\documentclass[11pt,a4paper]{article}
\pdfoutput=1

\usepackage{graphicx}
\usepackage{mathrsfs}
\usepackage{latexsym}
\usepackage{amsmath}
\usepackage{amssymb}
\usepackage{color}
\usepackage{subfig}
\usepackage{array}
\usepackage{multirow}
\usepackage{hhline}
\usepackage[table]{xcolor}
\usepackage{jheppub}
\usepackage{afterpage}

\usepackage[nohead,letterpaper]{geometry}
\usepackage{vmargin}
\setpapersize{USletter}
\setmargrb{2cm}{1cm}{2cm}{1.75cm}

\usepackage{xcolor}
\definecolor{plum}{rgb}{0.36078, 0.20784, 0.4}
\definecolor{chameleon}{rgb}{0.30588, 0.60392, 0.023529}
\definecolor{cornflower}{rgb}{0.12549, 0.29020, 0.52941}
\definecolor{scarlet}{rgb}{0.8, 0, 0}
\definecolor{brick}{rgb}{0.64314, 0, 0}

\newcommand{\ba}{\begin{eqnarray}}
\newcommand{\ea}{\end{eqnarray}}
\newcommand{\be}{\begin{equation}}
\newcommand{\ee}{\end{equation}}
\newcommand{\bd}{\begin{displaymath}}
\newcommand{\ed}{\end{displaymath}}

\numberwithin{equation}{section}

\usepackage[bf,medium,]{titlesec}

\titleformat{\section}
{\large\bfseries}{\thesection.}{.5em}{\large\bfseries}

\titleformat{\subsection}
{\bfseries}{\thesubsection}{.5em}{\bfseries}

\title{Deformations of Lifshitz holography with the Gauss-Bonnet term in ($n+1$) dimensions}

\author[a]{Miok Park}
\author[a,b]{and Robert B. Mann}

\affiliation[a]{Department of Physics, University of Waterloo,\\
Waterloo, Ontario N2L 3G1,Canada}
\affiliation[b]{Perimeter Institute for Theoretical Physics,\\
31 Caroline Street North, Waterloo, Ontario N2L 2Y5, Canada}

\emailAdd{m7park@sciborg.uwaterloo.ca}
\emailAdd{rbmann@sciborg.uwaterloo.ca}

\abstract{We investigate deformations of Gauss-Bonnet-Lifshitz holography in $(n+1)$ dimensional spacetime. Marginally relevant operators are dynamically generated by a momentum scale $\Lambda \sim 0$ and correspond to slightly deformed Gauss-Bonnet-Lifshitz spacetimes via a holographic picture. To admit (non-trivial) sub-leading orders of the asymptotic solution for the marginal mode, we find that the value of the dynamical critical exponent $z$ is restricted by $z= n-1-2(n-2) \tilde{\alpha}$, where $\tilde{\alpha}$ is the (rescaled) Gauss-Bonnet coupling constant. The generic black hole solution, which is characterized by the horizon flux of the vector field and $\tilde{\alpha}$, is obtained in the bulk, and  we explore its thermodynamic properties for various values of  $n$ and $\tilde{\alpha}$.}


\arxivnumber{arXiv:1305.5578}

\begin{document}
\maketitle



\newpage

\section{Introduction}
\label{sec:Intro}

Inspired by the AdS/CFT correspondence \cite{Maldacena:1997re}, new applications of the
holographic correspondence principle are being explored in many strongly coupled systems such as AdS/QCD duality {\cite{Forkel:2007cm,Gursoy:2007cb,Hoyos:2006gb}}, gravity/condensed matter theory duality (AdS/CMT) \cite{Bertoldi:2009dt, Braviner:2011kz, Hartnoll:2009sz, Horowitz:2010gk, Kachru:2008yh, Mann:2011hg,Ross:2011gu} or gravity/fluid duality {\cite{Bredberg:2010ky,Janik:2005zt, Rangamani:2009xk}}.

Since metallic systems that are strongly coupled can be engineered in laboratories, the AdS/CMT correspondence provides the possibility of direct experimental tests. One interesting avenue of study has emerged from quantum critical phenomena associated with continuous phase transitions, which shows unusual behaviour such as a vanishing characteristic energy scale and a diverging correlation length scale at the quantum critical point $g=g_c$. In the vicinity of the critical point, it is also observed that temporal correlations of the order parameter decay slower and slower as one approaches the critical point. Thus time scales differently from space in this limit, yielding a scaling symmetry
\begin{equation}
t \rightarrow \lambda^z t, \qquad \vec{x} \rightarrow \lambda \; \vec{x}
\end{equation}
where $z$ is the dynamical critical exponent; $z=1$ corresponds to conformal invariance, whereas $z \neq 1$ represents an anisotropic scaling invariance.

From the perspective of holographic correspondence, this anisotropic symmetry can be geometrically configured via
\begin{equation}
ds^2 = l^2 \bigg(- \frac{dt^2}{r^{2z}} + \frac{dr^2}{r^2} + \frac{dx_1^2 + dx_2^2}{r^2} \bigg) \label{nsmtr}
\end{equation}
which is called Lifshitz spacetime for $z \neq 1$ and obviously satisfies
\begin{equation}
t \rightarrow \lambda^z t, \qquad r \rightarrow \lambda \; r, \qquad \vec{x} \rightarrow \lambda \; \vec{x}. \label{nstrnsfmn}
\end{equation}
When $z \neq 1$, the non-trivial metric (\ref{nsmtr}) can be generated by having an anisotropic energy-momentum tensor that is supported by a massive vector field. Alternatively adding higher curvature terms such as $R^2$, Gauss-Bonnet, or Lovelock terms into the action can also engender an anisotropy of spacetime. Whereas when $z=1$, the metric (\ref{nsmtr}) with (\ref{nstrnsfmn}) restores the (asymptotic) AdS spacetime. In this paper, we consider the effects of both the massive vector field and the Gauss-Bonnet term in  ($n+1$) dimensional spacetime,  extending previous investigations along these lines {\cite{Cheng:2009df,Park:2012mn}}.

Gauss-Bonnet gravity generalizes Einstein gravity to include quadratic curvature terms. The equations of motion consist of the Einstein equations modified by additional quadratic curvature terms (with coupling constant $\alpha$) in such a way that the higher derivative terms  cancel out.  The additional terms in the equations of motion identically vanish in four dimensional spacetime, and so effects from such terms are manifest only in 5 dimensions or more. There are various motivations for considering such higher dimensional gravitational theories.

In gravity theory, considering higher curvature terms expands the array of black hole solutions in Einstein gravity to Gauss-Bonnet-Einstein black hole solutions. Examples include the Boulware-Deser solution for spherically symmetric spacetime {\cite{Boulware:1985wk}},{\cite{Wheeler:1985nh}} and the Gauss-Bonnet black hole solution in AdS spacetime {\cite{Cai:2001dz}},{\cite{Cho:2002hq}},{\cite{Cvetic:2001bk}},{\cite{Neupane:2002bf}}, and the properties of these black hole solutions have been studied {\cite{Jacobson:1993xs}},{\cite{Torii:2005xu}}. In supersting/M-theory the Gauss-Bonnet term naturally arises in effective low energy limit and leads to ghost-free nontrivial gravitational self interactions \cite{Zwiebach:1985uq}.  Also in cosmology, it provides one approach to understand the current acceleration of the universe in the context of the dark energy problem {\cite{Leith:2007bu}},{\cite{Nojiri:2005jg}}. In brane-world scenarios it yields additional interesting features {\cite{Charmousis:2002rc}},{\cite{Davis:2002gn}}. From the holographic correspondence point of view, involving  condensed matter systems, one expects that the higher curvature correction will induce interesting new effects {\cite{Gregory:2009fj}},{\cite{Kanno:2011cs}} or  provide a better explanation for physical phenomena shown in abnormal materials.

The main purpose of this paper is to find what role the Gauss-Bonnet coupling plays in deformed Gauss-Bonnet-Lifshitz (GB-Lifshitz) holography. In 5 dimensions this is the simplest extension that can yield interesting effects in a four-dimensional theory; we shall here examine any dimensionality. To do so, we assume  GB-Lifshitz spacetime in the ultraviolet (UV) energy regime and  Gauss-Bonnet-AdS (GB-AdS) spacetime in the infrared (IR) energy regime, and consider the special condition $z=(n-1)-2(n-2)\tilde{\alpha}$, where $z$ is the dynamic critical exponent and $\tilde{\alpha}$ is the rescaled Gauss-Bonnet coupling constant (defined below). This condition allows  us to describe the marginally relevant operator, when the operator is expanded to sub-leading orders. To generate the sub-leading orders of the gravitational solutions at the boundary, we introduce a momentum scale $\Lambda$, which is very small  ($\Lambda\sim 0$) and expand the solutions as a power series in this quantity. Then these sub-leading terms slightly deform the GB-Lifshitz spacetime, which lies in high energy scale $\Lambda^z/ T \rightarrow 0$, at the asymptotic region. We also consider the finite temperature theory by finding (planar) black hole solutions, which are expanded near the horizon. From the duality perspective, this configuration leads us to expect that the gravitational solution at the boundary of the deformed spacetime gives information about how marginally relevant operators behave near critical points at finite temperature in quantum critical theory. Thus we observe the behaviours of the physical quantities such as the free energy density or the energy density, which shows the properties of the marginally relevant operators, according to different values of $\tilde{\alpha}$ for given horizon flux of a massive vector field. At the zero temperature limit $\Lambda^z/T \rightarrow \infty$ (at fixed $\Lambda \sim 0$) the GB-AdS spacetime can emerge; it then might be possible to find the renormalization group flow between the deformed GB-Lifshitz and GB-AdS spacetime, as per the AdS case {\cite{Braviner:2011kz}},{\cite{Kachru:2008yh}}.

In section 2, the Gauss-Bonnet-Einstein action with   negative cosmological constant  coupled to a massive vector field is introduced, and its equations of motion are derived. In section 3, we consider  asymptotically Lifshitz spacetime  in the high energy regime, slightly deformed by a small value   $\Lambda$, and calculate its asymptotic solutions. In section 4, with these asymptotic solutions we perform   holographic renormalization by constructing the relevant counterterms so as to achieve a well-defined action principle. By doing so, we obtain a well-defined free energy density $\mathcal{F}$ and energy density $\mathcal{E}$. Moving  near the black hole horizon, in section 5, we derive the expanded black hole solutions in this region, and obtain the thermodynamic variables defined at the horizon. Also we check that  our analytic calculations agree with the integrated form of the first thermodynamic law. In section 6, we carry out numerical work  with the asymptotic solutions and the expanded black hole solutions. Choosing various values of $\tilde{\alpha} = \frac{1}{4}, \frac{1}{10}, 0, -\frac{1}{20}, -\frac{1}{2(n-2)}, -\frac{3}{10}$, we then fix the undetermined parameters ($\Lambda$, $f_0$, and $p_0$) appearing in  the metric for given $\tilde{\alpha}$ and $h_0$, which are variables characterizing black hole solution,  using numerical integration. Based on these fixed values, we explore the physical quantities, and finally plot the free energy density and the energy density depending $\log(\Lambda^z/T)$ and find their fitting functions.  In section 7, we summarize and discuss our work.

\section{Higher curvature gravity with a massive vector field}
\label{sec:CalDelta}

We start with $(n+1)$ dimensional gravitational action modified by higher curvature terms and coupled to a massive vector field
\begin{equation}
S = \int d^{n+1} x \sqrt{-g} \bigg( \frac{1}{2 \kappa_{n+1}^2} [R + 2 \tilde{\Lambda} + \alpha {\mathcal{L}_{GB}}] - \frac{1}{{g_v}^2} \bigg[ \frac{1}{4} H^2 + \frac{\gamma}{2} B^2 \bigg] \bigg) \label{ACTN0}
\end{equation}
where $\kappa_{n+1} = \sqrt{8 \pi G_{n+1}}$, $\tilde{\Lambda}$ is cosmological constant, $\alpha$ is the Gauss-Bonnet coupling constant and ${\mathcal{L}_{GB}} = R^2 - 4 R_{\mu \nu} R^{\mu \nu} + R_{\mu \nu \alpha \beta} R^{\mu \nu \alpha \beta}$. $H=dB$, and $g_{v}$ and $\gamma$ are the coupling constant and the squared mass of the vector field respectively.
This action yields equations of motion for the gravitational field
\begin{equation}
G_{\mu \nu} + \alpha L_{\mu \nu} - \tilde{\Lambda} g_{\mu \nu} = {\kappa_{n+1}^2} T_{\mu \nu} \label{EqGrvt}
\end{equation}
where
\begin{eqnarray}
&&G_{\mu \nu} = R_{\mu \nu} - \frac{1}{2} g_{\mu \nu} R, \\
&&L_{\mu \nu} = 2 \bigg( R R_{\mu \nu} - 2 R_{\mu \alpha} R^{\alpha}_{\; \; \nu} - 2 R^{\alpha \beta} R_{\mu \alpha \nu \beta} + R_{\mu}^{\; \; \alpha \beta \gamma} R_{\nu \alpha \beta \gamma} \bigg) - \frac{1}{2} g_{\mu \nu} {\mathcal{L}_{GB}}, \\
&&T_{\mu \nu} = \frac{1}{{g_v}^2} \bigg( H_{\mu \rho} H_{\nu}^{\; \rho} - \frac{1}{4} g_{\mu \nu} H^2 \bigg) + \frac{\gamma}{{g_v}^2} \bigg( B_{\mu}B_{\nu} - \frac{1}{2} g_{\mu \nu} B^2 \bigg),
\end{eqnarray}
and for the massive vector field
\begin{equation}
\nabla_{\mu} H^{\mu \nu} - \gamma B^{\nu} = 0. \label{EqFld}
\end{equation}
At the asymptotic region (i.e. $r \rightarrow 0$), for $z \neq 1$ we require  the equations of motion to admit the metric
\begin{equation}
ds^2 = l^2 \bigg(- \frac{dt^2}{r^{2z}} + \frac{dr^2}{r^2} + \frac{dx^2 + dy^2 + \cdots}{r^2} \bigg)
\end{equation}
which is supported by the vector potential described by
\begin{equation}
B = \frac{g_v l}{\kappa_{n+1}} \frac{q}{r^z} dt.
\end{equation}
These ansatz and boundary conditions fine-tune the cosmological constant to be
\begin{equation}
\tilde{\Lambda} = \frac{\chi_1}{2 l^2} - \tilde{\alpha} \frac{\chi_2}{l^2} \label{CsmlgcCsnt}
\end{equation}
where we replaced the coupling constant of the Gauss-Bonnet $\alpha$ with $\tilde{\alpha}$, which is
\begin{equation}
\tilde{\alpha} = \frac{\alpha(n-2)(n-3)}{l^2}, \label{cplgct}
\end{equation}
and the squared mass and the squared charge to be
\begin{equation}
\gamma = \frac{(n-1)z}{l^2}, \; \; \; \; \; \; \; q^2 = \frac{z-1}{z}(1-2 \tilde{\alpha})
\end{equation}
where the $\chi_1$ and $\chi_2$ are defined by
\begin{equation}
\chi_1 = z^2 + (n-2)z + (n-1)^2, \; \; \; \; \; \; \; \chi_2 = z^2 + (n-2)z + \frac{(n-1)(n-2)}{2}.
\end{equation}

For given $(n,z,\alpha)$ we require that   AdS$_{n+1}$ spacetime is also a solution to the equations
with cosmological constant (\ref{CsmlgcCsnt}).
In this case the metric is
\begin{equation}
ds^2_{AdS} = a \; l^2 \bigg(- \frac{dt^2}{r^2} + \frac{dr^2}{r^2} + \frac{dx^2 + dy^2 + \cdots}{r^2} \bigg) \label{AdSmtrc}
\end{equation}
where the $a$ is the scaling constant defined by
\begin{equation}\label{eqa}
a = \frac{n(n-1) + \sqrt{n(n-1)(1-2\tilde{\alpha})(n(n-1)- 4 \chi_2 \tilde{\alpha})}}{2(\chi_1 - 2 \chi_2 \tilde{\alpha})}
\end{equation}
appears because the cosmological constant (\ref{CsmlgcCsnt}) has been already fixed due to the Lifshitz boundary condition.

In order to configure the spacetime to induce  renormalization flow from UV Lifshitz spacetime to IR AdS spacetime at finite temperature (thereby recovering   isotropic scaling symmetry at low energy), we employ the ansatz
\begin{align}
ds^2 &= l^2 \bigg( -f(r) dt^2 + \frac{dr^2}{r^2} +p(r)(dx^2 + dy^2 + \cdots) \bigg), \label{nstzmtr}\\
B &= \frac{g_v l}{\kappa_{n+1}} h(r) dt, \label{nstzptn}
\end{align}
and these yield the solution for the GB-Lifshitz spacetime
\begin{equation}
\textrm{GB-Lifshitz :} \; \; \; \; f = \frac{1}{r^{2z}}, \; \; p = \frac{1}{r^2}, \; \; h = \frac{\sqrt{(z-1)(1-2\tilde{\alpha})}}{\sqrt{z}} \frac{1}{r^z}
\end{equation}
and for GB-AdS$_{n+1}$ spacetime
\begin{equation}
\textrm{Gb-AdS :} \; \; \; \; f = p = a r^{-2/\sqrt{a}}, \; \; h = 0.
\end{equation}

Plugging the ansatz (\ref{nstzmtr}) and (\ref{nstzptn}) into (\ref{EqGrvt}) and (\ref{EqFld})  yields
\begin{align}
&- \frac{2zh(r)^2}{f(r)} - \frac{r p'(r)}{p(r)} + \frac{r^2 f'(r)p'(r)}{2 f(r)p(r)} + \frac{r^2 p'(r)^2}{2 p(r)^2} - \frac{r^2 p''(r)}{p(r)} + \frac{\tilde{\alpha}}{2} \bigg( \frac{r^3 p'(r)^3}{p(r)}- \frac{r^4 f'(r)p'(r)^3}{2 f(r)p(r)^3} \nonumber\\
& - \frac{r^4 p'(r)^4}{2 p(r)^4} + \frac{r^4 p'(r)^2 p''(r)}{p(r)^3} \bigg)=0, \label{EqFGH1} \\
&2 \chi_1 + \frac{2(2n-3)z h(r)^2}{f(r)} - \frac{r f'(r)}{f(r)} + \frac{r^2 f'(r)^2}{2 f(r)^2} - \frac{(3n-5)r^2 f'(r)p'(r)}{2f(r)p(r)} - \frac{(n-2)^2 r^2 p'(r)^2}{2 p(r)^2}- \frac{r^2 f''(r)}{f(r)} \nonumber\\
&+ \tilde{\alpha} \bigg( -4 \chi_2 + \frac{3 r^3 f'(r)p'(r)^2}{2 f(r) p(r)^2} - \frac{r^4 f'(r)^2 p'(r)^2}{4 f(r)^2 p(r)^2} - \frac{r^3 p'(r)^3}{p(r)^3} + \frac{3(n-3)r^4 f'(r)p'(r)^3 }{4 f(r)p(r)^3} \nonumber\\
&+ \frac{(n^2 - 7 n +16)r^4 p'(r)^4}{8p(r)^4} + \frac{r^4 p'(r)^2 f''(r)}{2 f(r) p(r)^2} + \frac{r^4 f'(r)p'(r)p''(r)}{f(r)p(r)^2}- \frac{r^4 p'(r)^2 p''(r)}{p(r)^3} \bigg) = 0,  \label{EqFGH2}\\
&\chi_1 + \frac{(n-1)zh(r)^2}{f(r)} - \frac{r^2 h'(r)^2}{f(r)} - \frac{(n-1)r^2 f'(r)p'(r)}{2 f(r)p(r)}-\frac{(n-1)(n-2)r^2 p'(r)^2}{4 p(r)^2} \nonumber\\
& + \tilde{\alpha}  \bigg(-2 \chi_2 + \frac{(n-1)r^4 f'(r)p'(r)^3}{4 f(r)p(r)^3} + \frac{(n-1)(n-4)r^4 p'(r)^4}{16 p(r)^4} \bigg) =0. \label{EqFGH3}
\end{align}
For simplification we change variables via
\begin{equation}
p(r) = e^{\int^{r} \frac{q(s)}{s} ds}, \; \; \; \; f(r) = e^{\int^{r} \frac{m(s)}{s} ds}, \; \; \; \; h(r) = k(r) \sqrt{f(r)} \label{cv1}
\end{equation}
where   $m(r)$ is defined by using a new variable
\begin{align}
y(r) &= \bigg( 4 \chi_1 + 4(n-1)z k(r)^2 -2(n-1)m(r)q(r) - (n-1)(n-2)q(r)^2  \nonumber\\
&+ \tilde{\alpha} (-8 \chi_2 + (n-1)m(r)q(r)^3 + \frac{1}{4}(n-1)(n-4) q(r)^4) \bigg)^{1/2}, \label{cv2}
\end{align}
upon which (\ref{EqFGH1})-(\ref{EqFGH3}) are rewritten as
\begin{align}
r y'(r) =& -2(n-1) z k(r) - \frac{(n-1)}{2} q(r)y(r), \label{eqy}\\
r q'(r) =& \frac{\chi_1}{(n-1)} - z k(r)^2 - \frac{n}{4}q(r)^2 -\frac{y(r)^2}{4(n-1)} \nonumber\\
& + \frac{\tilde{\alpha}}{(2 - \tilde{\alpha} q(r)^2)} \bigg(- \frac{4 \chi_2}{(n-1)} + \frac{\chi_1 q(r)^2}{(n-1)} - z k(r)^2 q(r)^2 - \frac{n}{8} q(r)^4 - \frac{q(r)^2 y(r)^2}{4(n-1)} \bigg), \label{eqq}\\
r k'(r) =& - \frac{y(r)}{2} + \frac{k(r)y(r)^2}{4(n-1)q(r)} - \frac{z k(r)^3}{q(r)} + \frac{(n-2)}{4} k(r)q(r) - \frac{\chi_1 k(r)}{(n-1)q(r)} \nonumber\\
&+ \frac{\tilde{\alpha}}{(2 - \tilde{\alpha} q(r)^2)} \bigg( \frac{4 \chi_2 k(r)}{(n-1)q(r)} - \frac{\chi_1 k(r) q(r)}{(n-1)}- z k(r)^3 q(r) + \frac{n k(r)q(r)^3}{8} +\frac{k(r)q(r)y(r)^2}{4(n-1)}  \bigg). \label{eqk}
\end{align}
In terms of the new variables $k(r), q(r),$ and $y(r)$, the GB-Lifshitz spacetime is described by
\begin{equation}
\textrm{GB-Lifshitz : } q = -2, \; \; \; y = 2 \sqrt{z} \sqrt{(z-1)(1-2 \tilde{\alpha})} , \; \; \; k= \frac{\sqrt{(z-1)(1-2 \tilde{\alpha})}}{\sqrt{z}}, \label{prLshtzsltn}
\end{equation}
whereas for GB-AdS spacetime,
\begin{equation}
\textrm{GB-AdS : } q= - \frac{2}{\sqrt{a}}, \; \; \; \; \; \; \;y=0, \; \; \; \; \; \; \; k= 0.
\end{equation}

\section{Asymptotic behaviour}
\label{sec:Asmybh}

\begin{figure}[!b]
    \begin{center}
    \subfloat[$\tilde{\alpha}  \in \{-1, -1/2, -3/10, -1/4\}$]{\includegraphics[scale=0.7]{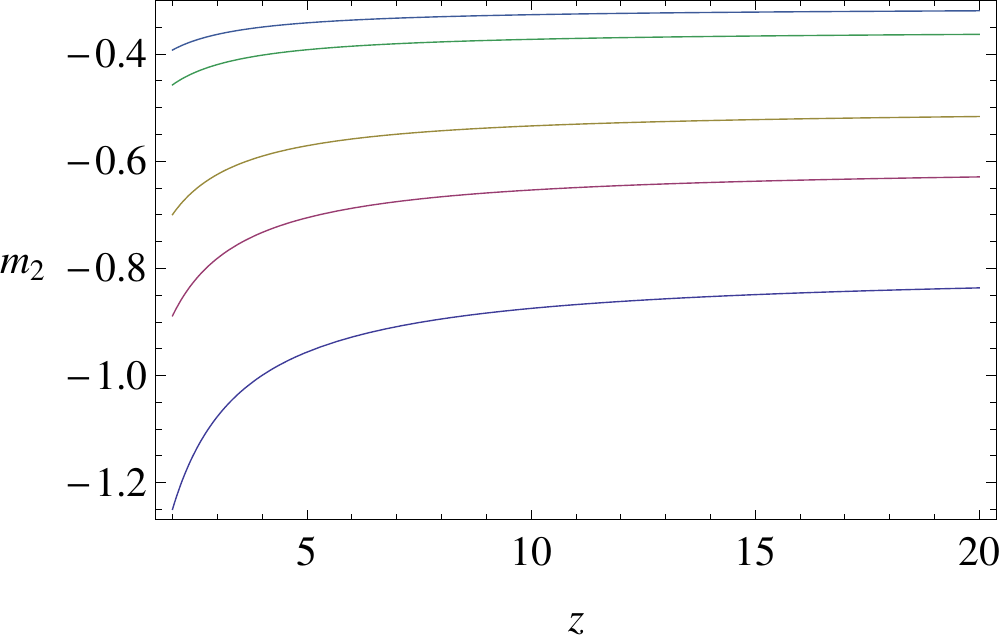}} \qquad \qquad \subfloat[$\tilde{\alpha}  \in \{1/7,1/6, 1/5,1/4\}$]{\includegraphics[scale=0.7]{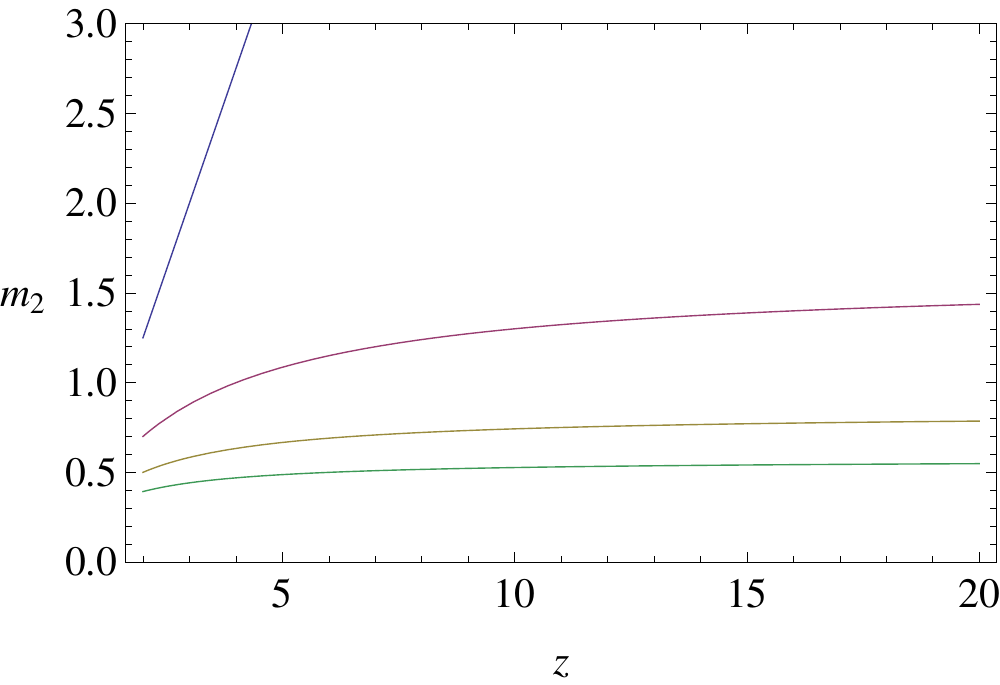}}
    \caption{$m_2$ versus $z$ for increasing values of $\tilde{\alpha}$ from bottom to top.}
    \label{fig:m2}
    \end{center}
\end{figure}

We next expand the marginal mode of the asymptotic solutions (\ref{prLshtzsltn}) to sub-leading orders, which are generated by a momentum scale $\Lambda \sim 0$, so that the Lifshitz spacetime becomes slightly deformed. The dimension of the marginal operator is $z+n-1$, and the non-trivial solution for the sub-leading orders, which become marginally relevant modes, are only allowed under the condition
\begin{equation}
z = (n-1) -2(n-2)\tilde{\alpha}. \label{malz}
\end{equation}
by equations of motion (\ref{eqy})-(\ref{eqk}). Thus, henceforth we deal with the case satisfying the  condition (\ref{malz}). Then the sub-leading orders admit the following form of the (Lifshitz spacetime) solution
\begin{align}
k(r)&= \frac{\sqrt{(z-1)(1 - 2 \tilde{\alpha})}}{\sqrt{z}} \bigg [ 1 + \frac{z+ (2z^2-5z-1)\tilde{\alpha}-2(z+1)(z-2)\tilde{\alpha}^2}{(z-1)^2 (z + (2-4z)\tilde{\alpha}) \log(r \Lambda)} + \frac{1}{2 (z-1)^4 (z+(2-4z)\tilde{\alpha})^2 \log^2(r \Lambda)}  \nonumber\\
&\bigg ( (z-1) \bigg(-3z^2 -4z^2(z-7)\tilde{\alpha} + (10 z^3-65z^2-20z+3)\tilde{\alpha}^2 - 2(z+1)(z^2 -26z+1)\tilde{\alpha}^3  \nonumber\\
&-4(z+1)^2 (z+2)\tilde{\alpha}^4 \bigg) + \frac{2}{(z+(2-4z)\tilde{\alpha})} \bigg( z^2(1-3z) -2z(4z^3 -25z^2 +13z -2)\tilde{\alpha} -(4z^5-87z^4 \nonumber\\
&+315z^3-189z^2+41z-4)\tilde{\alpha}^2 +2(7z^5 -144z^4 +433z^3-275z^2 +72z-13)\tilde{\alpha}^3-4(3z^5-90z^4 \nonumber\\
& +247z^3-145z^2 +34z-9)\tilde{\alpha}^4-8(z-2)(z+1)(2z^3 +9z^2-8z+1)\tilde{\alpha}^5 \bigg)\log(-\log(r \Lambda))  \bigg ) + \cdots \bigg]\nonumber\\
&+ (r \Lambda)^{2z +m_1} \log^{2-m_2}(r \Lambda) \bigg [\xi \bigg( 1+ \frac{\log(-\log(r \Lambda))}{(z-1)^2 (z-(z+1)\tilde{\alpha}) (z-(2-4z)\tilde{\alpha})^3 \log(r \Lambda)} \bigg( 2z^3(3z-1) \nonumber\\
& + 2z^2(2z^3-51z^2+33z-6)\tilde{\alpha} -2z(17z^4 -314z^3+273z^2-88z+12)\tilde{\alpha}^2
+2 (43z^5 -855z^4+833z^3 \nonumber\\
& -329z^2+76z-8)\tilde{\alpha}^3 -(58z^5 -207z^4+1892z^3-680z^2+194z-32)\tilde{\alpha}^4 -4(z+1)(17z^4+168z^3 \nonumber\\
& -180z^2+60z-9)\tilde{\alpha}^5 + 8(z+1)^2(2z^3+9z^2-8z+1)\tilde{\alpha}^6 \bigg) +\cdots \bigg) + \zeta \bigg( \frac{1}{\log(r \Lambda)}+ \cdots \bigg) \bigg ], \label{aympksol}
\end{align}
\begin{align}
q(r)&= -2 \bigg[ 1- \frac{z-(3z+1)\tilde{\alpha} + 2 (z+1)\tilde{\alpha}^2}{(z-1)(z+(2-4z)\tilde{\alpha}) \log(r \Lambda)}-\frac{1}{(z-1)^3(z+(2-4z)\tilde{\alpha})^2 \log^2 (r \Lambda)} \bigg( \frac{1}{2(z+2 \tilde{\alpha})} \nonumber\\
& \bigg( z^3 -2z^2(z^2+2z+2)\tilde{\alpha} + z(8z^3 +13z^2+14z+5)\tilde{\alpha}^2 -2(5z^4+15z^3+12z^2+7z+1)\tilde{\alpha}^3\nonumber\\
& +4(z+1)(z^3+6z^2+2z+1)\tilde{\alpha}^4 \bigg) + \frac{1}{(z+(2-4z)\tilde{\alpha})} \bigg(-z^2(3z-1) -2z(z^3-21z^2+12z-2)\tilde{\alpha} \nonumber\\
& + (11z^4-199z^3+137z^2-33z+4)\tilde{\alpha}^2 -2(10z^4-213z^3+157z^2-43z+9)\tilde{\alpha}^2+4(z^4 -97z^3 \nonumber\\
& +63z^2-11z+4)\tilde{\alpha}^4 \bigg)\log(-\log(r \Lambda)) \bigg) + \cdots \bigg] -\frac{2 \sqrt{z-1} \sqrt{z}}{(2z-1)\sqrt{1-2\tilde{\alpha}}} (r \Lambda)^{2z +m_1} \log^{2-m_2}(r \Lambda)  \nonumber\\
& \bigg[\xi \bigg( 1 + \frac{1}{(z-1)^2(2z-1)(z+(2-4z)\tilde{\alpha}) \log(r \Lambda)} \bigg( \frac{-1}{(2z-1)} \bigg(z(4z^2-7z+2)+(10z^3-21z^2+11z \nonumber\\
&-4)\tilde{\alpha}-2(z+1)\tilde{\alpha}^2 \bigg)+\frac{\log(-\log(r \Lambda))}{(z-(z+1)\tilde{\alpha})(z+(2-4z)\tilde{\alpha})^2 \log(r \Lambda)} \bigg( 2z^3(3z-1) +2z^2(2z^3-51z^2 \nonumber\\
&+33z-6)\tilde{\alpha}-2z(17z^4-314z^3+273z^2 -88z+12)\tilde{\alpha}^2 \bigg) \bigg) + \cdots \bigg) + \zeta \bigg( \frac{1}{\log(r \Lambda)}+ \cdots \bigg) \bigg],\label{aympqsol}
\end{align}
\begin{align}
y(r) &= 2 \sqrt{z} \sqrt{(z-1)(1-2 \tilde{\alpha})} \bigg[1 + \frac{z^2 - z(z+3)\tilde{\alpha} +2(z+1)\tilde{\alpha}^2}{(z-1)^2 (z+(2-4z)\tilde{\alpha}) \log(r \Lambda)}+ \frac{\log(-\log(r \Lambda))}{(z-1)^4 (z+(2-4z)\tilde{\alpha})^3 \log^2(r \Lambda)} \nonumber\\
& \bigg( z^3(1-3z)-2z^2(z^3 -18z^2 +8z-1)\tilde{\alpha} + z(7z^4 -123z^3 +21z^2+8z-1)\tilde{\alpha}^2 -2(3z^5 -79z^4 -63z^3 \nonumber\\
&+75z^2-20z +4) \tilde{\alpha}^3 -4(2 z^5 + 8z^4 +87z^3 -71z^2 +19z -5)\tilde{\alpha}^4 +8(z+1)(2z^3 +9z^2-8z+1)\tilde{\alpha}^5 \bigg) \nonumber\\
& + \cdots \bigg] - \frac{2z(z-2 \tilde{\alpha})}{(2z-1)(1-2 \tilde{\alpha})} (r \Lambda)^{2z +m_1} \log^{2-m_2}(r \Lambda) \bigg[\xi \bigg( 1 + \frac{1}{(z-1)^2(z+(2-4z)\tilde{\alpha}) \log(r \Lambda)} \nonumber\\
& \bigg( \frac{-z(4z-1)(z-1) + (14z^2-7z+5)(z-1)\tilde{\alpha} -2(2z+1)(z+1)(z-1)\tilde{\alpha}^2}{(2z-1)} \nonumber\\
& +\frac{(2z^3 (3z-1)+ 2z^2(2z^3 -51z^2+33z-6)\tilde{\alpha})\log(-\log(r \Lambda))}{(z-(z+1)\tilde{\alpha})(z+(z-4z) \tilde{\alpha})^2 }  \bigg)+ \cdots \bigg) + \zeta \bigg( \frac{1}{\log(r \Lambda)}+ \cdots \bigg) \bigg]. \label{aympysol}
\end{align}
where $m_1$ and $m_2$ become respectively
\begin{equation}
m_1(z, \tilde{\alpha}) = \frac{2(z-1)\tilde{\alpha}}{1-2\tilde{\alpha}}, \; \; \; \; \; \; m_2(z, \tilde{\alpha}) = -\frac{2 \tilde{\alpha} (z-(z+1)\tilde{\alpha})}{z+(2-4z)\tilde{\alpha}}. \label{m2}
\end{equation}
Here $2z+m_1$ is equivalent to $z+n-1$ by using (\ref{malz}) and determines the scaling dimension of the dual operator. Note that the denominator of $m_2$ must not be zero so as to ensure that the logarithmic term $\log^{2-m_2}$ remains finite; recall that we want the slightly perturbed metric near the origin. Since (\ref{m2}) becomes singular when $\tilde{\alpha}=\frac{z}{2(-1+2z)}$, where $\tilde{\alpha}$ decreases with increasing $z$, approaching $\frac{1}{4}$ as $z \rightarrow \infty$,  we require $\tilde{\alpha} \leq \frac{1}{4}$, where the equality is removed when $z = \infty$. This ensures $m_2$ is finite for any   $z \geq 2$. In Fig.\ref{fig:m2}, we depict $m_2$ versus $z$  by varying $\tilde{\alpha}$; we find that the value of $m_2$ does not exceed $-\frac{5}{4}$ for $|\tilde{\alpha}| < 1$ and any positive $z$.

In (\ref{aympksol})-(\ref{aympysol}), a new momentum scale $\Lambda$ appears. We take it to be very small compared to the energy scale, where the Lifshitz spacetime lies. Thus $\Lambda$ slightly deforms the Lifshitz spacetime and by doing so generates the marginally relevant mode;  the pure Lifshitz spacetime solutions (\ref{prLshtzsltn}) are recovered when $\Lambda \rightarrow 0$. Furthermore we see from (\ref{aympksol})-(\ref{aympysol}) that these solutions contain different modes due to the arbitrary parameters $\xi$ and $\zeta$. This introduces an ambiguity related to defining the scale of $\Lambda$ since the solution is invariant under the transformation
\begin{equation}
F(\Lambda r; \zeta, \xi ; \lambda) = F(e^{\lambda'/z} \Lambda r; e^{-2 \lambda'} (\zeta - \lambda' \xi), e^{-2 \lambda'} \xi; \lambda + \lambda')
\end{equation}
where $F$ stands for $k$, $q$, and $y$ functions. We fix $\lambda=0$ in (\ref{aympksol})-(\ref{aympysol}) as  in
previous work where $\tilde{\alpha}=0$ {\cite{Cheng:2009df}} and {\cite{Park:2012bv}}.

By substituting (\ref{aympksol})-(\ref{aympysol}) into the relation (\ref{cv1}) and (\ref{cv2}), the original functions $f$ and $p$ are obtained as
\begin{align}
\frac{f}{l^2} &= F_{0}^2 \frac{\log^{-2\frac{m_2}{m_1}z}(r \Lambda)}{(r \Lambda)^{2z}} \bigg(1 -\frac{1}{(z-1)^3 (z+(2-4z)\tilde{\alpha})^3 \log(r \Lambda)} \bigg( z^3 (7z-4) + 2z^2 (4z^3 -49z^2 +37z-7) \tilde{\alpha}  \nonumber\\
& -z (64z^4 -487z^3 +406z^2-111z+8)\tilde{\alpha}^2 + 2(89z^5 -539z^4+450z^3-131z^2 +7z+4)\tilde{\alpha}^3 -4(39z^5 \nonumber\\
&  -217z^4 +128z^3+3z^2-21z+8)\tilde{\alpha}^4+ 8z^2 (z+1)(2z^2-19z+11)\tilde{\alpha}^5 + \bigg(2z^3(3z-1)+4z^2(z^3 \nonumber\\
&-21z^2+12z-2)\tilde{\alpha} -2z (11z^4 -199z^3+137z^2-33z+4)\tilde{\alpha}^2 +4(10z^4-213z^3+157z^2-43z+9)\tilde{\alpha}^3 \nonumber\\
&-8(z^4 -97z^3+63z^2-11z+4)\tilde{\alpha}^4-16z(2z^3+9z^2-8z+1)\tilde{\alpha}^5 \bigg) \log(-\log(r\Lambda))\bigg) + \cdots\bigg), \label{aympfsol}\\
\frac{p}{l^2} & =P_{0}^2 \frac{\log^{2\frac{m_2}{m_1}}(r \Lambda)}{(r \Lambda)^2}
\bigg(1 + \frac{1}{(z-1)^3(z+(2-4z)\tilde{\alpha})^3 \log(r \Lambda)} \bigg( z^2(5z-2) + 2z(3z^3 - 38z^2 +24z\nonumber\\
&-4)\tilde{\alpha}
-(38z^4 -377z^3 +280 z^2-69z +8)\tilde{\alpha}^2 + (82z^4 -818z^3 +640z^2 -178z+34)\tilde{\alpha}^3-(54z^4 \nonumber\\
&-732z^3+504z^2-92z+28)\tilde{\alpha}^4 - 8(z+1)(2z^3 +17z^2 -15z+2)\tilde{\alpha}^5 + \bigg( 2z^2(3z-1)+4z(z^3\nonumber\\
&-21z^2+12z-2)\tilde{\alpha} - (22z^4 -398z^3 +274z^2-66z+8)\tilde{\alpha}^2 +4(10z^4 -213z^3+157z^2 -43z+9)\tilde{\alpha}^3 \nonumber\\
&-8(z^4-97z^3+63z^2 -11z+4)\tilde{\alpha}^4 -16(z+1)(2z^3 +9z^2-8z+1)\tilde{\alpha}^5 \bigg)\log(-\log(r \Lambda)) \bigg) + \cdots \bigg) \label{aymppsol}
\end{align}
where $F_0$ and $P_0$ are  integration constants. As we are assuming the Lifshitz spacetime in high energy regime, we introduce an arbitrary scale $\mu$
\begin{equation}
\log(r \Lambda) = \log(r \mu) - \log \frac{\mu}{\Lambda},
\end{equation}
and expand the asymptotic solutions into UV regions where $\mu \gg \Lambda$
\begin{equation}
\bigg| \frac{1}{\log \frac{\mu}{\Lambda}} \bigg|, \qquad \bigg|\frac{\log(r \mu)}{\log \frac{\mu}{\Lambda}} \bigg| \ll 1, \label{expfctr}
\end{equation}
and then (\ref{aympfsol}) and (\ref{aymppsol}) become
\begin{align}
f =& \frac{1}{r^{2z}} \bigg[1 + \frac{1}{(z-1)^3(z-(2-4z) \tilde{\alpha})^3 \log(\frac{\mu}{\Lambda})} \bigg(z^3 \bigg(7z-4+2z(z-1)^2 \log(r \mu) + 2(3z-1) \log(\log(\frac{\mu}{\Lambda})) \bigg) \nonumber\\
&+ 2z^2 \bigg( 4z^3 -49z^2 +37z-7-(11z-3)z(z-1)^2 \log(r \mu) + 2(z^3 -21z^2 +12z-z) \log(\log(\frac{\mu}{\Lambda})) \bigg) \tilde{\alpha} \label{expdf}\nonumber\\
&- z \bigg(64 z^4 - 487 z^3 + 406 z^2 - 111 z + 8 - 12 z^2 (z-1)^2 (7z-3) \log(r \mu) + \cdots \bigg)\tilde{\alpha}^2 + \cdots \bigg)  + \cdots \bigg], \\
p=& \frac{1}{r^2} \bigg[1 - \frac{(1 - 2 \tilde{\alpha})}{(z-1)^3(z-(2-4z) \tilde{\alpha})^3 \log(\frac{\mu}{\Lambda})} \bigg(z^2 \bigg( 5z-2 + 2z(z-1)^2 \log(r \mu) + 2(3z-1) \log(\log(\frac{\mu}{\Lambda})) \bigg) \nonumber\\
& + 2z \bigg(3z^3 - 33z^2 + 22z - 4 - 3 z (z-1)^2 (3 z-1) \log(r \mu) + 2 (z^3-18 z^2+ 11 z -2) \log(\log(\frac{\mu}{\Lambda})) \bigg) \tilde{\alpha}  \nonumber\\
& -\bigg(26 z^4 -245 z^3 +192 z^2 -53 z + 8 - 24 z^2 (z-1)^2 (2 z-1) \log(r \mu)  + \cdots \bigg)\tilde{\alpha}^2 + \cdots \bigg) + \cdots  \bigg] \label{expdp}
\end{align}
where the $t$ and $x$ coordinates are rescaled by
\begin{align}
t \rightarrow \bigg( \Lambda \log^{\frac{m_2}{m_1}} \bigg(\frac{\mu}{\Lambda}\bigg) \bigg)^z\frac{1}{F_{0}} t, \; \; \; \; \; \; x \rightarrow \frac{\Lambda}{\log^{\frac{m_2}{m_1}} (\frac{\mu}{\Lambda})} \frac{1}{P_{0}} x.
\end{align}

\section{Holographic Renormalization}
\label{sec:HoloRn}

As we are considering the spacetime at finite temperature, the action is related to the free energy via
\begin{equation}
F = - T \log{\mathcal{Z}} = T S_{\epsilon}(g_{*})
\end{equation}
where $S_{\epsilon}$ and $g_{*}$ indicate the Euclidean action and metric respectively. At the asymptotic boundary near $r \rightarrow 0$, the action includes a boundary term to yield the bulk equations of motion
\begin{equation}
S_{\epsilon} = S_{\textrm{bulk},\epsilon} + S_{\textrm{boundary}, \epsilon} \label{ecldS1}
\end{equation}
where each term is
\begin{align}
S_{\textrm{bulk},\epsilon} &= \int d^{n+1} x \sqrt{g} \bigg( \frac{1}{2 \kappa_{n+1}^2}[R + 2 \tilde{\Lambda} + \alpha \mathcal{L}_{GB}] - \frac{1}{{g_{v}}^2} \bigg[\frac{1}{4} H^2 + \frac{\gamma}{2} B^2 \bigg] \bigg), \nonumber\\
S_{\textrm{boundary},\epsilon} &= \frac{1}{\kappa_{n+1}^2} \int_{\Sigma} d^n x \sqrt{\gamma} \bigg( K + 2 \alpha (J - 2 \hat{G}^{ab} K_{ab}) \bigg) \label{ecldS2}
\end{align}
and where $\hat{G}^{ab}$ is the $n$-dimensional Einstein tensor on $\Sigma$ corresponding to $\gamma_{ab}$ and $J = J_{ab} \gamma^{ab}$, with
\begin{equation}
J_{ab} = \frac{1}{3} (2 K K_{ac} K^{c}_{\; \;b} + K_{cd} K^{cd} K_{ab} -2 K_{ac} K^{cd}K_{db} - K^2 K_{ab}). \label{ecldSsub}
\end{equation}
Defining $F = \int d^{n-1}x \; \mathcal{F}$, the free energy density $\mathcal{F}$, which is the free energy density per unit $(n-1)$-dimensional spatial volume, is
\begin{align}
{\mathcal{F}}_{\textrm{bulk}} &= -\frac{l^{n-1}}{2 \kappa_{n+1}^2}  \lim_{r \rightarrow 0} r \sqrt{f(r)} p'(r) p(r)^{\frac{n-3}{2}} \bigg( 1 - \alpha \frac{(n-2)r^2}{l^2} \frac{p'(r)}{p(r)} \bigg( \frac{(n-3)}{2} \frac{p'(r)}{p(r)} + \frac{f'(r)}{f(r)} \bigg) \bigg), \nonumber\\
{\mathcal{F}}_{\textrm{boundary}} &= \frac{l^{n-1}}{\kappa_{n+1}^2}  \lim_{r \rightarrow 0} \bigg( r \bigg(\sqrt{f(r)} p(r)^{\frac{n-1}{2}} \bigg)' - \alpha \frac{ (n-1)(n-2)r^3}{6 l^2} \frac{p'(r)^2}{f(r)p(r)} \bigg( \sqrt[3]{f(r)} p(r)^{\frac{n-3}{2}} \bigg)' \bigg).
\end{align}
Considering the boundary of the spacetime near $r \rightarrow 0$, these quantities are divergent upon insertion of (\ref{expdf}) - (\ref{expdp}), and so   counterterms are required to render them finite. These counterterms also should be constructed so as to preserve the covariance of the action at the boundary and to yield a well-defined variational principle. To meet the first condition, one convenient option would be to employ $B^2 =  B^\mu B_\mu$ and to consider a functional of this quantity on the boundary, which could be any form if canceling divergences were the only criterion. We take the counterterms to be the combination   $C_0 + C_1 B^2 + C_2 B^4 + \cdots$ and then replace $B^2$ with $- \frac{\kappa^2_{n+1}}{g^2_v}B^2 - \frac{(z-1)(1-2 \tilde{\alpha})}{z}$, because the counterterms should vanish for the pure Lifshitz spacetime.

The counterterms are precisely expressed by
\begin{align}
{\mathcal{F}}_{\textrm{c.t}} &= \frac{l^{n-1}}{2 \kappa_{n+1}^2} \lim_{r \rightarrow 0} \sqrt{f(r)}p(r)^{\frac{n-1}{2}} \bigg( \sum^{2}_{j=0} C_{j} \bigg( - \frac{\kappa^2_{n+1}}{g^2_v}B^2 - \frac{(z-1)(1-2 \tilde{\alpha})}{z} \bigg)^j \bigg) \nonumber\\
&= \frac{l^{n-1}}{2 \kappa_{n+1}^2} \lim_{r \rightarrow 0} \sqrt{f(r)}p(r)^{\frac{n-1}{2}} \bigg( \sum^{2}_{j=0} C_{j} \bigg( k(r)^2 - \frac{(z-1)(1-2 \tilde{\alpha})}{z} \bigg)^j \bigg) \label{ct}
\end{align}
where the $C_j$ are not constants but rather series in $1/\log(r \Lambda)$, which expand similarly to (\ref{aympksol}) - (\ref{aympysol}). Thus the free energy density is finally written as
\begin{align}
{\mathcal{F}} =& {\mathcal{F}}_{\textrm{bulk}} + {\mathcal{F}}_{\textrm{boundary}} + {\mathcal{F}}_{\textrm{c.t}} \nonumber\\
=& \frac{l^{n-1}}{2 \kappa_{n+1}^2} \lim_{r \rightarrow 0} \sqrt{f(r)} p(r)^{\frac{n-1}{2}} \bigg( \frac{(n-2) r p'(r)}{p(r)} + \frac{r f'(r)}{f(r)} + \tilde{\alpha} \bigg( - \frac{r^3}{2} \frac{f'(r)}{f(r)} \frac{p'(r)^2}{p(r)^2} - \frac{(n-4)}{6} \frac{r^3 p'(r)^3}{p(r)^3}  \bigg)  \nonumber\\
& + \sum^{2}_{j=0} C_{j} \bigg( k(r)^2 - \frac{(z-1)(1-2 \tilde{\alpha})}{z} \bigg)^j \bigg). \label{fedfn}
\end{align}

The counterterms also yield a well-defined variational principle of the on-shell action; this process defines the boundary stress tensor, which yields the conserved quantities. Here we obtain some physical quantities such as energy density $\mathcal{E}$, pressure $\mathcal{P}$, and flow $\mathcal{J}^{i}$ by finding the boundary stress tensor. To do this we start with the variation of the action known as
\begin{equation}
\delta {\mathcal{F}} = \frac{\sqrt{\gamma}}{2} \tau^{ab} \delta \gamma_{ab} + \mathcal{J}^a \delta B_a \label{delF}
\end{equation}
where $\tau^{ab}$ contributes to yield conserved quantities via
\begin{equation}
Q = - \int d^{n-1} x \sqrt{\sigma} \xi_{a} k_{b} \tau^{ab}
\end{equation}
where $\sqrt{\sigma} = \sqrt{\sigma_{x_1 x_1} \cdots \sigma_{x_{n-1} x_{n-1}}}$ is the spatial volume element,   $\xi_{a}$ are boundary Killing fields, and $k_{a}$ is the unit vector normal to the boundary surface. In the presence of the non-scalar fields, the boundary stress tensor $\tau^{ab}$ should be modified to include the contribution of the massive vector fields. To do so, we employ the vielbein frame at the boundary of the metric
\begin{equation}
\gamma_{ab} = \eta_{\hat{a} \hat{b}} e^{\hat{a}}_{a} e^{\hat{b}}_{b}, \; \; \; \eta = \textrm{diag}(\pm1,1,1, \cdots)
\end{equation}
where
\begin{equation}
e^{\hat{t}} = e^{\hat{t}}_{a} dx^a = \sqrt{f} d \tau, \; \; \; \; e^{\hat{x_i}} = \sqrt{p} \; dx_{i}.
\end{equation}
Then (\ref{delF}) is rewritten with the new boundary stress tensor $\mathcal{T}^{ab}$
\begin{equation}
\delta \mathcal{F} = \sqrt{\gamma} \; \mathcal{T}^{a}_{\; \; \hat{a}} \delta e^{\hat{a}}_{a} + \mathcal{J}^{\hat{a}} \delta B_{\hat{a}}
\end{equation}
where
\begin{equation}
\mathcal{T}^{ab} = T^{a}_{\; \; \hat{a}} e^{b \hat{a}}, \qquad \mathcal{T}^{ab} = \tau^{ab} + \frac{1}{\sqrt{\gamma}} \mathcal{J}^{(a} B^{b)}.
\end{equation}
The energy density is expressed as
\begin{equation}
\mathcal{E} = \sqrt{\sigma} k_{a} \xi_{b} T^{ab} = \sqrt{\gamma} \tau^{t}_{\; \; t} + \mathcal{J}^{t} B_{t}
\end{equation}
and the pressure is given by
\begin{equation}
\mathcal{P} = -\sqrt{\gamma} \tau^{x}_{\; \; x}.
\end{equation}
From the variation of the action (\ref{ecldS1})-(\ref{ecldSsub}) with respect to   $\gamma_{ab}$, we obtain for  $\tau^{ab}$ the expression
\begin{align}
&\tau^{ab} = \frac{2}{\sqrt{\gamma}} \frac{\delta {\mathcal{F}}}{\delta \gamma_{ab}} = \frac{1}{\kappa_{n+1}^2} \bigg[ (K \gamma^{ab} - K^{ab}) +2 \alpha (J \gamma^{ab} - 3 J^{ab} - 2 \hat{P}^{acdb} K_{cd})  \nonumber\\
&+\frac{1}{2 l } \sum^{2}_{j=0} C_{j} \bigg( \gamma^{ab} \bigg(-\frac{\kappa_{n+1}^2}{{g_v}^2} {B}^2 -\frac{(z-1)(1 - 2 \tilde{\alpha})}{z} \bigg)^j  + \frac{ 2 j {\kappa_{n+1}}^2}{{g_v}^2} {B}^{a} {B}^{b} \bigg( -\frac{\kappa_{n+1}^2}{{g_v}^2} {B}^2 - \frac{(z-1)(1 - 2 \tilde{\alpha})}{z} \bigg)^{j-1} \bigg) \bigg]
\end{align}
where
\begin{equation}
\hat{P}_{abcd} = \hat{R}_{abcd} + 2 \hat{R}_{b[c}g_{d]a} - 2 \hat{R}_{a[c}g_{d]b} + \hat{R} g_{a[c}g_{d]b},
\end{equation}
in which the $\hat{R}$'s are various contractions of Riemann tensors defined on the $n$-dimensional hypersurface associated with $\gamma_{ab}$. From the variation of the action with respect to the vector field strength $B_{t}$, ${\mathcal{J}}^{\hat{t}}$ we obtain
\begin{align}
{\mathcal{J}}^{\hat{t}} =& \sqrt{f(r)} \frac{\delta S }{\delta B_{t}},\nonumber\\
=& \frac{l^{n-2}}{g_v \kappa_{n+1}} \lim_{r \rightarrow 0} \sqrt{f(r)} p(r)^{\frac{n-1}{2}} \bigg(\frac{r(k(r) \sqrt{f(r)})'}{\sqrt{f(r)}} + k(r) \sum^{2}_{j=0} j C_{j} \bigg( k(r)^2 - \frac{(z-1)(1-2 \tilde{\alpha})}{z} \bigg)^{j-1} \bigg) \nonumber\\
=& \frac{l^{n-2}}{g_v \kappa_{n+1}} \lim_{r \rightarrow 0} \sqrt{f(r)} p(r)^{\frac{n-1}{2}} \bigg(- \frac{1}{2}y(r) + k(r) \sum^{2}_{j=0} j C_{j} \bigg( k(r)^2 - \frac{(z-1)(1-2 \tilde{\alpha})}{z} \bigg)^{j-1} \bigg) \label{flfn}
\end{align}
where other components of ${\mathcal{J}}^{\hat{a}}$ vanish. Then the energy density becomes
\begin{align}
\mathcal{E} =& \frac{l^{n-1}}{2 \kappa_{n+1}^2 } \lim_{r \rightarrow 0} \sqrt{f(r)} p(r)^{\frac{n-1}{2}} \bigg[ (n-1) \frac{r p'(r)}{p(r)} - k(r) y(r) + \tilde{\alpha} \bigg( - \frac{(n-1)}{6} \frac{r^3 p'(r)^3}{p(r)^3} \bigg) \nonumber\\
& + \sum^{2}_{j=0} C_{j} \bigg( k(r)^2 - \frac{(z-1)(1-2 \tilde{\alpha})}{z} \bigg)^j \bigg], \label{edfn}
\end{align}
and the pressure takes the same form of the free energy density with the opposite sign
\begin{equation}
\mathcal{P} = - \mathcal{F}. \label{prss}
\end{equation}

To ensure a finite value of the action and a well-defined variational principle $\delta S = 0$, we expect our counterterms (\ref{ct}) to render the free energy density (\ref{fedfn}), the flow (\ref{flfn}), and the energy density (\ref{edfn}) finite when $r \rightarrow 0$. Expanding these functions with the asymptotic solution (\ref{aympksol})-(\ref{aympysol}), they  take the form
\begin{equation}
F \sim \frac{1}{r^{z+n-1}} \bigg( F_1[\frac{1}{\log(r \Lambda)}] + \cdots (r \Lambda)^{z+n-1} \log^{2-m_2}(r \Lambda) \bigg( F_2[\frac{1}{\log(r \Lambda)}] + \cdots \bigg) \bigg) + \textrm{counterterms} \label{cmmnf}
\end{equation}
where $F$ stands for $\mathcal{F}$, $\mathcal{E}$, and $\mathcal{J}^{\hat{t}}$, and $F_1$ and $F_2$ indicate the parts involving series of $\log (r \Lambda)$ in the expression of $\mathcal{F}$, $\mathcal{E}$, and $\mathcal{J}^{\hat{t}}$. As shown in (\ref{cmmnf}),  divergences occur as $ r \rightarrow 0$. The divergences due to $\frac{1}{r^{z+n-1}}$ are eliminated by the two coefficients $C_0$ and $C_1$,  expanded as a series in $1/\log(r \Lambda)$. To prevent  divergences due to $\log^{2-m_2}(r \Lambda)$, another coefficient $C_2$   is required. Then $C_0$, $C_1$, and $C_2$ become
\begin{align}
C_{0} =& -\frac{2 (-3 + 6 z + (6 - 14 z) \tilde{\alpha} + 4 (-2 + 3 z) \tilde{\alpha}^2)}{-3 + 6 \tilde{\alpha}} +  \frac{2(z - 2
\tilde{\alpha})(1 + 2 (-2 + z) \tilde{\alpha}) (-z + (1 + z) \tilde{\alpha})^3}{(z-1)^3 (2 z-1)(z + (2 - 4 z)\tilde{\alpha})^2 \log^2(r \Lambda)} \nonumber\\
&+\frac{(z - (1 + z) \tilde{\alpha})^2}{3(2z-1)^2 (z-1)^5 (z + (2 - 4 z) \tilde{\alpha}^3) \log^3(r \Lambda)} \bigg( 3 z (1 - 2 z - 3 z^2 + 2 z^3 + 4 z^4) + (-3 - 21 z \nonumber\\
&+ 109 z^2 + 33 z^3 - 146 z^4 - 48 z^5 + 16 z^6) \tilde{\alpha} + \cdots \bigg) + \cdots,\label{C0} \\
C_{1} =& z - \frac{2 z (z - 2 \tilde{\alpha}) (z - (1 + z)\tilde{\alpha})^2}{(z-1)^2 (2 z-1)(2 \tilde{\alpha}-1) (z + (2 - 4 z)\tilde{\alpha})\log(r \Lambda)} \nonumber\\
& -\frac{z (-z + (1 + z) \tilde{\alpha})}{2 (2 z-1)^2 (z-1)^4 (2 \tilde{\alpha}-1) (z + (2 - 4 z) \tilde{\alpha})^2 \log^2(r \Lambda)} \bigg(z (3 - 10 z - z^2 + 14 z^3) + (-3 - 9 z \nonumber\\
& + 113 z^2 - 103 z^3 - 70 z^4 + 24 z^5) \tilde{\alpha} + \cdots \bigg) + \cdots, \label{C1} \\
C_{2} =& \frac{z^2 (1 - 3 z + 4 z \tilde{\alpha})}{4 (z-1) (2 z-1) (2 \tilde{\alpha}-1)^2} + \frac{z^2}{4(2z-1)^2 (z-1)^3 (2 \tilde{\alpha}-1)^2 (z + (2 - 4 z) \tilde{\alpha}) \log(r \Lambda)} \bigg( z (3 - 14 z \nonumber\\
& + 15 z^2) + (-3 + 3 z + 65 z^2 - 119 z^3 + 30 z^4) \tilde{\alpha} + \cdots \bigg) + \frac{\Delta}{\log^2(r \Lambda)} \nonumber\\
&+ \frac{z^2 \log(-\log(r \Lambda))}{4 (1 - 2 z)^2 (-1 + z)^5 (1 - 2 \tilde{\alpha}^2 (-z + (1 + z) \tilde{\alpha}) (-z + (-2 + 4 z) \tilde{\alpha})^3  \log^2(r \Lambda)} \nonumber\\
& \times \bigg( (1 - 3 z)^2 z^3 (-3 + 5 z) + z^2 (-9 + 110 z - 532 z^2 + 1154 z^3 - 955 z^4 + 120 z^5) \tilde{\alpha} + \cdots \bigg). \label{C2}
\end{align}
where $C_0$ and $C_1$ keep expanding with infinite series of $1/\log(r \Lambda)$ as the asymptotic solutions expand. However $C_2$ just consists of a finite number of terms, because the $\log^{2-m_2} (r \Lambda)$ quantity in (\ref{cmmnf}) is divided by $1/ \log^{i}(r \Lambda)$  (where $i=1,2,\cdots$) from $F_2$ and so when $i$ becomes $2-m_2$ divergences due to $\log^{2-m_2} (r \Lambda)$ vanish. These solutions, however, have an ambiguity $\Delta$ as shown in (\ref{C2}). It turns out that $\Delta$ does not affect  numerical work for the free energy density and the energy density, as we shall show in the next section.

Inserting the solution (\ref{C0})-(\ref{C2}) into (\ref{fedfn}), (\ref{flfn}), and (\ref{edfn}), we find
\begin{align}
\mathcal{F} =& \frac{l^{n-1}}{\kappa^2_{n+1}} \frac{\sqrt{z}}{\sqrt{(z-1)(1-2 \tilde{\alpha})}} \bigg( (z-(z+1) \tilde{\alpha}) \zeta + \frac{X}{(2z-1)(z-1)^2(z-(-2+4z)\tilde{\alpha})^3} \xi \bigg), \label{fdwtcc} \\
\mathcal{E} =& - \frac{l^{n-1}}{\kappa^2_{n+1}} \frac{(z-2 \tilde{\alpha})}{\sqrt{(z-1)z}(2z-1)(1-2 \tilde{\alpha})^{3/2}} \bigg( (z-(z+1) \tilde{\alpha}) \zeta - \frac{Y}{(2z-1)(z-1)^2 (z-(-2+4z) \tilde{\alpha})^3} \xi \bigg), \label{edwtcc} \\
\mathcal{J}^{\hat{t}} =& \frac{1}{g_v} \frac{l^{n-2}}{\kappa_{n+1}} \frac{1}{(1-2z)^2 (1- 2 \tilde{\alpha})^2} \bigg( -(z-1)(z-2 \tilde{\alpha}) \tilde{\alpha} \; \zeta + \frac{4(z-1)(1-2 \alpha)^3 \Delta}{z} \xi \nonumber\\
& + \frac{Z}{2(z-1)^4 (1-2z)(z-(1+z) \tilde{\alpha})(z-(-2+4z) \tilde{\alpha})^3} \xi \bigg), \label{flwtcc}
\end{align}
where $X$, $Y$, and $Z$ are functions of $z$ and $\tilde{\alpha}$, and are displayed in (\ref{XYZ}). This result implies for the pure Lifshitz spacetime to be
\begin{equation}
\mathcal{F} = \mathcal{E} = \mathcal{J}^{\hat{t}} = 0,
\end{equation}
since $\zeta=\xi=0$ for the pure Lifshitz solution.

\section{Finite Temperature}
\label{sec:FiniteT}

We have so far worked near boundary of the spacetime. We now move to the near-horizon region of a  black hole. In this section, we first obtain (planar) black hole solutions expanded near the horizon, and find some thermodynamic quantities such as temperature $T$, entropy density $s$, horizon flux density of the massive vector field $\phi$ at near horizon $r=r_+$. Then we prove the integrated first law by constructing an RG-invariant quantity $\bar{K}$. We find analytic expressions
for $\frac{\mathcal{F}_0}{Ts}$ and $\frac{\mathcal{E}_0}{Ts}$ when the marginally relevant modes are turned off. In subsequent sections we check these relationships numerically and find agreement in all dimensions we study.

\subsection{Expansion and Physical quantities near the horizon}
\label{sec:NearHrz}

Let us consider a black hole solution defined by $f(r_+)=0$, and expand the solution near horizon $r=r_+$. The expansions in  powers of $(1-\frac{r}{r_+})$ are determined by equation of motion (\ref{EqFGH1})-(\ref{EqFGH3}). We find
\begin{align}
f(r) &= f_{0} \bigg( \bigg(1- \frac{r}{r_{+}} \bigg)^2 + \bigg( 1- \frac{r}{r_{+}} \bigg)^3 + \frac{1}{12(1-2 \tilde{\alpha})^2(z-2 \tilde{\alpha})^2} \bigg( z (8 h_0^2 (-2 + 3 z) + z (7 + 14 z - 6 z^2)) \nonumber\\
&+ 2 (32 {h_0}^4 + 16 {h_0}^2 (1 + z - 6 z^2) + z (-18 - 30 z - 35 z^2 + 29 z^3)) \tilde{\alpha} + \cdots \bigg) \bigg(1- \frac{r}{r_{+}} \bigg)^4 + \cdots \bigg),\label{hrsnslf} \\
p(r) &= p_{0} \bigg(1 + \frac{(4 {h_0}^2 + z - 3 z^2 - (16 {h_0}^2 + z - 11 z^2) \tilde{\alpha} + 2 (-1 + 8 {h_0}^2 + z - 8 z^2) \tilde{\alpha}^2 + 8 z^2 \tilde{\alpha}^3)}{2 (z - 2 \tilde{\alpha}) (-1 + 2 \tilde{\alpha})} \bigg(1- \frac{r}{r_{+}} \bigg)^2  \nonumber\\
&+ \frac{(4 {h_0}^2 + z - 3 z^2 - (16 {h_0}^2 + z - 11 z^2) \tilde{\alpha} + 2 (-1 + 8 {h_0}^2 + z - 8 z^2) \tilde{\alpha}^2 + 8 z^2 \tilde{\alpha}^3)}{2 (z - 2 \tilde{\alpha}) (-1 + 2 \tilde{\alpha})} \bigg(1- \frac{r}{r_{+}} \bigg)^3 + \cdots \bigg), \label{hrsnslp}\\
h(r) &= \sqrt{f_{0}} h_0 \bigg( \bigg(1- \frac{r}{r_{+}} \bigg)^2 +\bigg(1- \frac{r}{r_{+}} \bigg)^3 + \frac{1}{24 (1 - 2 \tilde{\alpha})^2 (z - 2 \tilde{\alpha})^2} \bigg(z (z (20 + 10 z - 9 z^2)+8 {h_0}^2(-1 + 3 z)) \nonumber\\
&+ 4 (8 {h_0}^4 - 4 {h_0}^2 (-1 + 2 z + 9 z^2) + z (-21 - 27 z - 5 z^2 + 14 z^3)) \tilde{\alpha} + \cdots \bigg) \bigg(1- \frac{r}{r_{+}} \bigg)^4 + \cdots  \bigg)\label{hrsnslh}
\end{align}
where $f_0$ and $p_0$ are constants, which are not determined by equations of motion, and control the scaling of the metric by redefining the coordinates so that $dt'^2 = f^2_0 dt^2$ and $dx'^2_i= p^2_0 d x^2_i$. In the next section, we tune $f_0$ and $p_0$ with $\log(\frac{\Lambda}{\mu})$ shown in (\ref{expdf}) and (\ref{expdp}) at the energy scale   $\mu = \frac{1}{r_+}$ by matching near horizon solutions (numerically integrated towards asymptotics) and the asymptotic solutions. If $f_0$ and $p_0$ are fixed by doing so, the expanded black hole solutions (\ref{hrsnslf})-(\ref{hrsnslh}) are characterized by two parameters $h_0$ and $\tilde{\alpha}$. Namely, (\ref{hrsnslf})-(\ref{hrsnslh}) describe a two-parameter family $\{h_0, \tilde{\alpha} \}$ of black hole solutions.

Thermodynamic properties of (\ref{hrsnslf})-(\ref{hrsnslh}) are found by calculating temperature $T$ and entropy density $s$ near horizon. The temperature is determined by demanding the periodicity of the imaginary time coordinate $\tau$ so that the spacetime becomes regular at $r=r_+$, and the entropy is defined by $S=\frac{A}{4 G_{n+1}}$ which is proportional to the area of the black hole. Using our metric (\ref{nstzmtr}), the $T$ and $s$ can be written as
\begin{align}
T &= \frac{r_+}{2 \pi} \sqrt{\frac{1}{2} \frac{d^2 f(r)}{dr^2}} \bigg|_{r=r_+}, \label{tmprtr}\\
s &= 2 \pi \frac{l^{n-1}}{\kappa^2_{n+1}} p(r_+)^{\frac{n-1}{2}} \label{ntrpdnst}
\end{align}
where $S=\int s \; d^{n-1} x$. We also can calculate the horizon flux $\Phi$ of the massive vector field defined by
\begin{equation}
\Phi = \oint \sqrt{\gamma} \vec{H} \cdot d \vec{A} = \oint \phi \; d^{n-1} x \
\end{equation}
where
\begin{equation}
\phi = \frac{l^{n-2}g_v r_+}{\kappa_{n+1}} \bigg( \frac{p(r)^{\frac{n-1}{2}}}{\sqrt{f(r)}} \frac{d h(r)}{dr} \bigg) \bigg|_{r=r_+} \label{flux}
\end{equation}
is the horizon flux density. Plugging near horizon solutions (\ref{hrsnslf})-(\ref{hrsnslp}) into (\ref{flfn})--(\ref{tmprtr}), each quantity becomes
\begin{equation}
T = \frac{\sqrt{f_0}}{2 \pi}, \qquad s = 2 \pi p^{\frac{n-1}{2}}_{0} \frac{l^{n-1}}{\kappa^2_{n+1}}, \qquad \phi = 2 h_0 p^{\frac{n-1}{2}}_{0} \frac{l^{n-2} g_v}{\kappa_{n+1}}. \label{thrmdnmqtt}
\end{equation}

\subsection{Integrated First Law of Thermodynamics}
\label{sec:IntFirLawThermo}

So far we obtained the asymptotic solutions (\ref{aympksol})-(\ref{aympysol}), and calculated the free energy density (\ref{fedfn}) and the energy density (\ref{edfn}). Then we derived the near horizon solutions (\ref{hrsnslf})-(\ref{hrsnslp}), and computed thermodynamic quantities such as temperature $T$ and the entropy density $s$ near horizon in (\ref{thrmdnmqtt}). Now we check the consistency of our calculations by proving the integrated first law of thermodynamics.

To connect the physical variables of the asymptotic region and near horizon region, we construct an RG-invariant quantity $\bar{K}$ to be a constant at both regions. First, by using the asymptotic solution, we combine the functions $q, k, y$ and $m$ in an invariant combination. Then $\bar{K}$ takes the form
\begin{align}
\bar{K} =& \frac{\sqrt{f(r)} p(r)^{\frac{n-1}{2}}}{2} \bigg[ q(r) - m(r) - k(r) y(r) - \tilde{\alpha} \bigg( \frac{1}{2}q(r)^3 - \frac{1}{2} m(r) q(r)^2 \bigg) \bigg] \\
=& \frac{\sqrt{f(r)} p(r)^{\frac{n-1}{2}}}{4 q(r)} \bigg[\frac{1}{(n-1)} y(r)^2 + n q(r)^2 - 4z k(r)^2 - 2 k(r) q(x) y(x) - \frac{4 \chi_1}{(n-1)} \nonumber\\
& +\tilde{\alpha} \bigg(- \frac{n}{4} q(r)^4 + \frac{8 \chi_2}{(n-1)} \bigg) \bigg]
\end{align}
which produces with (\ref{aympksol})-(\ref{aympysol})
\begin{align}
\bar{K} =& \frac{2(z-(1+z)\tilde{\alpha})}{\sqrt{z-1}\sqrt{z}(1-2z)(1-2 \tilde{\alpha})^{\frac{3}{2}}} \bigg( (z-(1+z)\tilde{\alpha})\zeta - \frac{\xi}{(1-2z)(z-1)^2(z+(2-4z) \tilde{\alpha})^3} \nonumber\\
& \times \bigg(-z^3 (1 - 3 z + z^2) + z^2 (-6 + 28 z - 48 z^2 + 21 z^3 - 4 z^4) \tilde{\alpha} + z (-12 + 95 z - 287 z^2  \nonumber\\
&+ 424 z^3 - 262 z^4 + 72 z^5) \tilde{\alpha}^2 + (-8 + 116 z - 514 z^2 + 1193 z^3 - 1630 z^4 + 1125 z^5 - 322 z^6) \tilde{\alpha}^3  \nonumber\\
&+ (28- 231 z + 685 z^2 - 1283 z^3 + 1887 z^4 - 1558 z^5 + 472 z^6) \tilde{\alpha}^4 - 2 (7 - 27 z - 62 z^2 + 246 z^3 \nonumber\\
&- 47 z^4- 231 z^5 + 90 z^6) \tilde{\alpha}^5 + 4 (1 + z)^2 (2 - 17 z + 41 z^2 - 32 z^3 + 4 z^4) \tilde{\alpha}^6 \bigg) \bigg).
\end{align}
Employing the expression for the free energy density (\ref{fedfn}) and the energy density(\ref{edfn}), we obtain the
  algebraic relation
\begin{align}
&\frac{1}{2} \sqrt{f(r)} p(r)^{\frac{n-1}{2}} \bigg[ \frac{(n-1)r p'(r)}{p(r)} - k(r) y(r) + \tilde{\alpha} \bigg(- \frac{(n-1)}{6} \frac{r^3 p'(r)^3}{p(r)^3} \bigg) + \sum^{2}_{j=0} C_{j} \bigg( k(r)^2 - \frac{(z-1)(1-2\tilde{\alpha})}{z} \bigg)^j \bigg] \nonumber\\
&=\frac{1}{2} \sqrt{f(r)} p(r)^{\frac{n-1}{2}} \bigg[\frac{(n-2)r p'(r)}{p(r)} + \frac{r f'(r)}{f(r)} + \tilde{\alpha} \bigg( - \frac{r^3}{2} \frac{f'(r)}{f(r)} \frac{p'(r)^2}{p(r)^2} - \frac{(n-4)}{6} \frac{r p'(r)^3}{p(r)^3} \bigg) \nonumber\\
& \; \; \; \; + \sum^{2}_{j=0} C_{j} \bigg(k(r)^2 - \frac{(z-1)(1-2 \tilde{\alpha})}{z} \bigg)^j \bigg] + \bar{K}
\end{align}
or simply
\begin{equation}
\mathcal{E} = \mathcal{F} + \frac{l^{n-1}}{\kappa^2_{n+1}} \bar{K}. \label{Kbar1}
\end{equation}

Next, let us calculate $\bar{K}$ with (\ref{hrsnslf})-(\ref{hrsnslp}) at $r=r_+$. This yields for $\bar{K}$
\begin{equation}
\bar{K} = \sqrt{f_0} P^{\frac{n-1}{2}}_{0} = T s \frac{\kappa^2_{n+1}}{l^{n-1}}
\end{equation}
and by plugging the above into (\ref{Kbar1}), we find
\begin{equation}
\mathcal{F} = \mathcal{E} - Ts \label{frslwthrm}
\end{equation}
which is the integrated form of the first law of thermodynamics. We have therefore proved that our analytic calculation agrees with  the first law of thermodynamics.  We shall make use of (\ref{frslwthrm})   for checking our numerical calculations, which are performed with the aim of observing the behaviour of the marginally relevant modes where $\Lambda \sim 0$, in section 6.

For $\Lambda = 0$, in \cite{Park:2012bv} it was expected regardless of dimensionality that
\begin{equation}
\mathcal{F}_0 = - \mathcal{E}_0 = - \frac{1}{2} T s \label{feqe}
\end{equation}
which is derived by using $\mathcal{F}_0 = - \mathcal{E}_0$ from the trace Ward identity in \cite{Taylor:2008tg}, and (\ref{prss}). This fact was also proved by numerical calculation in \cite{Park:2012bv}. More generally, using
 (\ref{prss})  we  expect $\mathcal{P}_0 = \frac{z}{n-1} \mathcal{E}_0$ implying
\begin{equation}
\mathcal{F}_0 = - \frac{z}{n-1} \mathcal{E}_0 \label{tWI}
\end{equation}
in any dimensionality.  In the Gauss-Bonnet case we are considering, $z=n-1-2(n-2) \tilde{\alpha}$ leads us to
expect
\begin{equation}
\mathcal{F}_0 = -\eta(n, \tilde{\alpha})\mathcal{E}_0 \qquad \textrm{where} \; \; \; \eta(n, \tilde{\alpha}) = 1-\frac{2(n-2)}{(n-1)} \tilde{\alpha}, \label{eta}
\end{equation}
and
\begin{equation}
\frac{\mathcal{F}_0}{Ts}= - \frac{n-1-2 \tilde{\alpha}(n-2)}{2(n-1)-2\tilde{\alpha}(n-2)} , \qquad \qquad \frac{\mathcal{E}_0}{Ts} = \frac{n-1}{2(n-1)-2 \tilde{\alpha} (n-2)}. \label{tW2}
\end{equation}
In the next section we shall numerically compute  $\eta(n, \tilde{\alpha})$ function for $n=4,5,\ldots,9$ for all values
of $ \tilde{\alpha}$ we consider.  We will see that
our results are consistent with (\ref{eta}) and (\ref{tW2}) in all dimensions.

The lack of equality of the magnitudes of the free energy density and the energy density is also seen in (\ref{fdwtcc})-(\ref{edwtcc}) for $\Lambda \sim 0$, where $\mathcal{F} \neq - \mathcal{E}$ when $\tilde{\alpha} \neq 0$ even if   $\xi =0$. This fact stands in contrast to the results for $\tilde{\alpha} = 0$ {\cite{Park:2012bv}}, in which $\mathcal{F} = - \mathcal{E}$ holds for $\Lambda \sim 0$ if $\beta = 0$, where $\beta$ plays the same role as $\xi$ in this paper.

\section{Exploring Near the Quantum Critical Point}
\label{sec:FiniteT}

In holographic duality picture at zero temperature, we can imagine that the matter fields residing on the boundary of the spacetime become those of the conformal field theory, which is dual to the AdS spacetime in IR regimes, or the Lifshitz-like field theory (or quantum critical theory), which is dual to the Lifshitz spacetime in UV regimes. Our interests are to consider this configuration at finite temperature in the presence of the massive vector field and the Gauss-Bonnet terms in the gravity action, and to investigate the thermodynamic behaviour of the deformed GB-Lifshitz spacetime with a black hole, which from the duality perspective corresponds to the marginally relevant operators near the quantum critical point. To generate the marginally relevant modes, the momentum scale $\Lambda$ is introduced into the asymptotic Lifshitz metric, and it is assumed that $\Lambda \sim 0$, which is much smaller than the background temperature $T$ ($\Lambda^z \ll T$). Hence it slightly deforms the Lifshitz spacetime, recovering the pure Lifshitz spacetime for $\Lambda=0$.

In this section, our aims are to show the implications of the Gauss-Bonnet terms for holographic renormalization flow and to observe the behaviour of physical quantities such as the free energy density $\mathcal{F}/Ts$ and the energy density $\mathcal{E}/Ts$ as functions of $\log(\frac{\Lambda^z}{T})$.

To do these, we first fix the undetermined parameters $\Lambda$, $f_0$, and $p_0$ (in the numerical calculation we consider the quantities  $\Lambda r_+$, $\frac{f}{f_0} \frac{r^z}{r^z_+}$, and $\frac{p}{p_0} \frac{r^2}{r^2_+}$). This is performed by numerically integrating the near horizon solution towards the boundary and reading off the matching values of $\Lambda$, $f_0$, and $p_0$ in sections 6.1.1 and 6.1.2. Based on these values the thermodynamic functions $\mathcal{F}$ and $\mathcal{E}$ are numerically integrated from near horizon to the boundary, and their numerical values are found for each $h_0$ in section 6.1.3. In section 6.2 we plot the data for $\mathcal{F}$ and $\mathcal{E}$ in a function of $\log(\frac{\Lambda^z}{T})$.

\subsection{Integrating towards the Lifshitz Boundary}
\label{sec:NmrIntgr}

As mentioned in section 5.1, we have  scaling ambiguities in the metric due to the undetermined constants $f_0$ and $p_0$. We here find values of $\Lambda$ for each $h_0$ by matching the numerical integration of near horizon solutions with the asymptotic solutions at the middle regions, and tune $f_0$ and $p_0$ with the matching value of $\Lambda$ for the same $h_0$ by using the same method. For the numerical calculation, we set up the arbitrary scale $\mu = r^{-1}_+$ and use dimensionless quantities:  $\frac{r}{r_+}$, $\Lambda r_+$, $\frac{\mathcal{F}}{Ts}$, $\frac{\mathcal{E}}{Ts}$ and $\frac{\Lambda^z}{T}$.

Before performing the numerical work, we need to find the appropriate range of $\tilde{\alpha}$ and  then choose specific values of $\tilde{\alpha}$, as (\ref{hrsnslf})-(\ref{hrsnslh}) describe a two-parameter family $\{h_0, \tilde{\alpha}\}$ of  black hole solutions. The values of $h_0$ will be computed for each $\tilde{\alpha}$ and will be discussed further in the following subsection.

Analyzing the asymptotic solutions (\ref{aympksol})-(\ref{aympysol}) in section 2, recall that $m_2$ yields the restriction $\tilde{\alpha} \leq \frac{1}{4}$, where   equality is not attained when $z = \infty$. We expect  that $|\tilde{\alpha}| < 1$ since  the Gauss-Bonnet term associated with the coupling constant $\tilde{\alpha}$ via (\ref{cplgct}) is considered as a small correction to Einsteinian gravity. Then the range of $\tilde{\alpha}$ becomes
\begin{equation}
-1 < \tilde{\alpha} \leq \frac{1}{4},
\end{equation}
and this corresponds to the range
\begin{equation}
\frac{n}{2} \leq z < 3n-5
\end{equation}
upon using $z=n-1-2(n-2)\tilde{\alpha}$ from (\ref{malz}).
To observe the $\tilde{\alpha}$-dependence for a broad range of values, we sparsely choose the values of $\tilde{\alpha}$, which are the two positive values $\frac{1}{4}$, $\frac{1}{10}$, and three negative values $-\frac{1}{20}$ (bigger than $-\frac{1}{2(n-2)}$), $-\frac{1}{2(n-2)}$, $-\frac{3}{10}$ (smaller than $-\frac{1}{2(n-2)}$); we also consider $\tilde{\alpha}=0$ {\cite{Park:2012bv}} for comparison. Table \ref{table:vlz} lists the cases that we consider.

\begin{table}[!h]
\begin{center}
  \begin{tabular}{| c || >{\centering}m{1.8cm} | >{\centering}m{1.8cm} | >{\centering}m{1.8cm} | >{\centering}m{1.8cm} | >{\centering}m{2.1cm} | >{\centering}m{1.8cm} m{0cm}|} \hline
  \multirow{2}{*}{} & \multicolumn{6}{c}{\bf{$z$}} & \tabularnewline [10pt] \hhline{~-------}
       & $\tilde{\alpha} = \frac{1}{4}$ & $\tilde{\alpha} = \frac{1}{10}$ &  $\tilde{\alpha} = 0$ & $\tilde{\alpha} = - \frac{1}{20}$ & $\tilde{\alpha} = - \frac{1}{2(n-2)}$ & $\tilde{\alpha} = -\frac{3}{10}$ & \tabularnewline [10pt] \hline \hline
    $n=4$ & 2   & 2.6 & 3 & 3.2 & 4 & 4.2 & \tabularnewline [10pt] \hline
    $n=5$ & 2.5 & 3.4 & 4 & 4.3 & 5 & 5.8 & \tabularnewline [10pt] \hline
    $n=6$ & 3   & 4.2 & 5 & 5.4 & 6 & 7.4 & \tabularnewline [10pt] \hline
    $n=7$ & 3.5 & 5   & 6 & 6.5 & 7 & 9   & \tabularnewline [10pt] \hline
    $n=8$ & 4   & 5.8 & 7 & 7.6 & 8 & 10.6& \tabularnewline [10pt] \hline
    $n=9$ & 4.5 & 6.6 & 8 & 8.7 & 9 & 12.2&\tabularnewline [10pt] \hline
    \end{tabular}
    \caption{values of $z$ according to $n$ and $\tilde{\alpha}$}
    \label{table:vlz}
\end{center}
\end{table}

\subsubsection{Matching $\Lambda$}
\label{sec:MtchLamb}

In the asymptotic solutions (\ref{aympksol})-(\ref{aympysol}), $\Lambda$ is the only unfixed quantity. To fix $\Lambda$, we expand these solutions into the high energy regimes by introducing the arbitrary scale $\mu$ and using the condition (\ref{expfctr}), which assumed that $\mu \gg \Lambda$. Then we set $\mu = r_+$, which means that our asymptotic solutions are described at the energy scale $r_+$, and fit the asymptotic solutions towards the horizon.

Near the horizon the expanded black hole solutions (\ref{hrsnslf})-(\ref{hrsnslh}) should be changed to  the functions  $k$, $q$, and $y$ by using the relations (\ref{expfctr}) so as to avoid the scale ambiguities of $f_0$ and $p_0$ for now. Then we numerically integrate the solutions towards the boundary by obeying the equations of the motion (\ref{eqy})-(\ref{eqk}).

\begin{figure}
        \subfloat[For $\tilde{\alpha}=1/4$ and $n=4$, $h_0=0.75120$ corresponds to $\log(\Lambda r_+) = -23874.4$]{\includegraphics[scale=0.7]{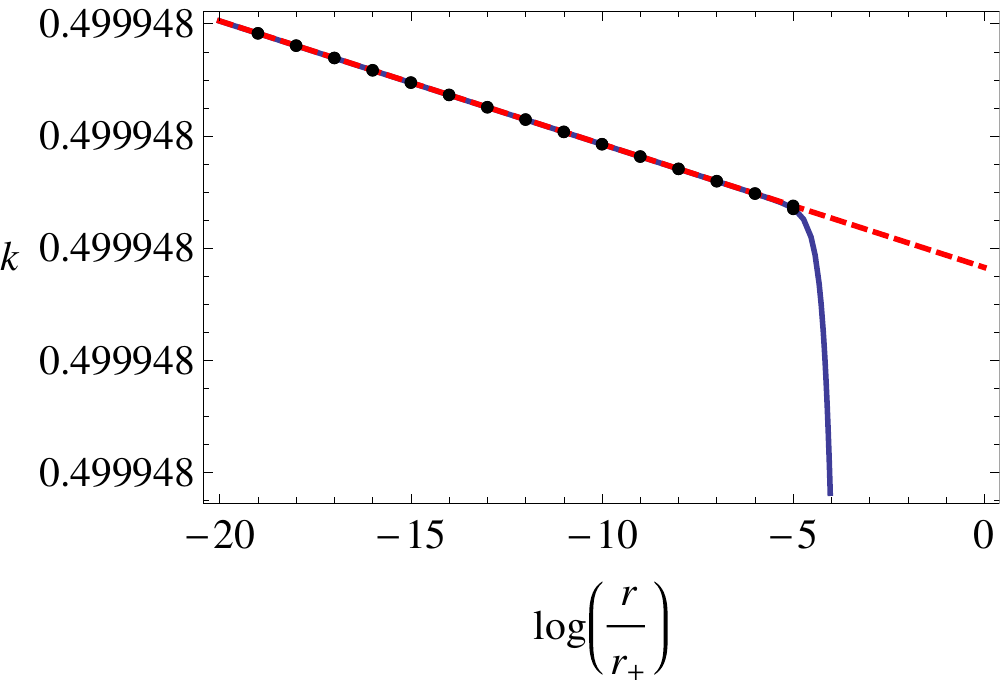}} \; \; \qquad \subfloat[For $\tilde{\alpha}=1/10$ and $n=4$, $h_0=1.26160$ corresponds to $\log(\Lambda r_+) =-8611.8$]{\includegraphics[scale=0.7]{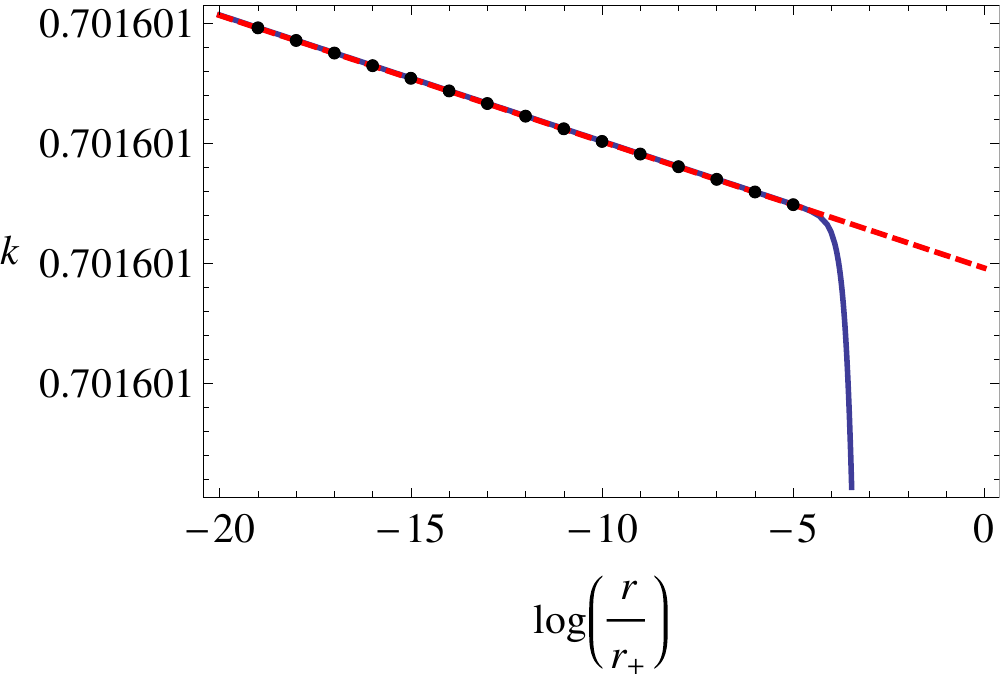}} \\
        \vspace{0.5cm}
        \subfloat[For $\tilde{\alpha}=0$ and $n=4$, $h_0=1.63430$ corresponds to $\log(\Lambda r_+) = -5167.9$]{\includegraphics[scale=0.7]{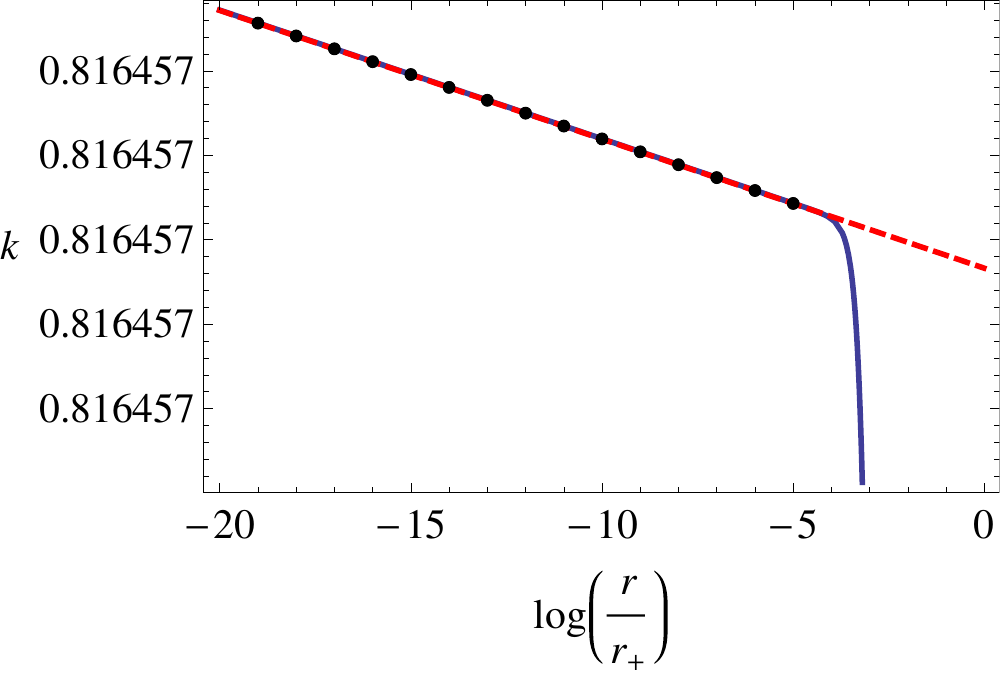}} \; \; \qquad \subfloat[For $\tilde{\alpha}=-1/20$ and $n=4$, $h_0=1.82980$ corresponds to $\log(\Lambda r_+) =-3422$]{\includegraphics[scale=0.7]{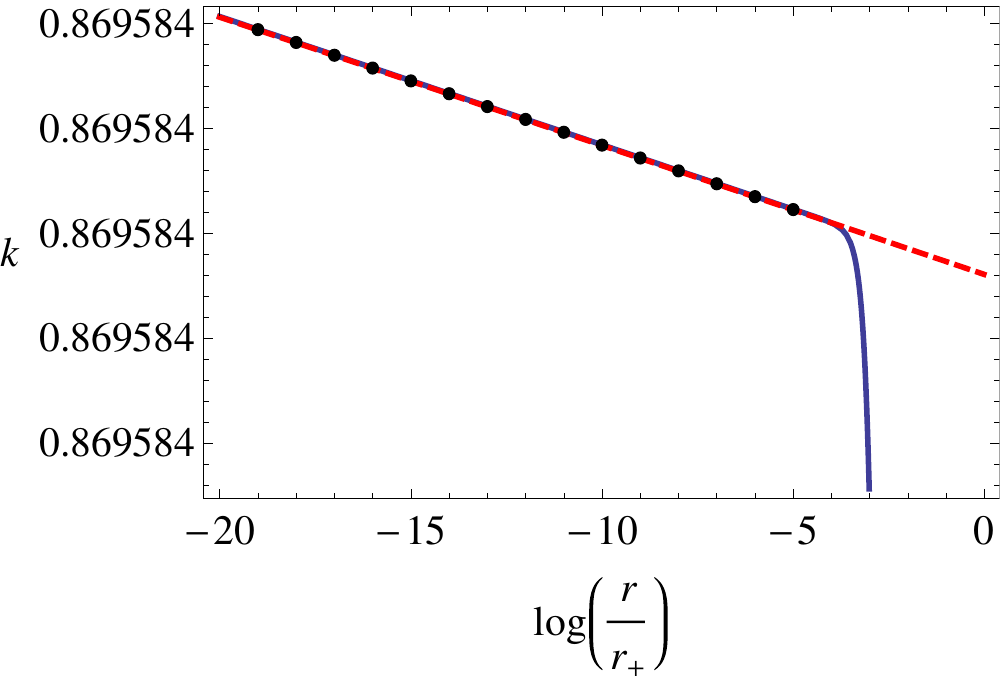}} \\
        \vspace{0.5cm}
        \subfloat[For $\tilde{\alpha}=-1/4$ and $n=4$, $h_0=2.66868$ corresponds to $\log(\Lambda r_+) = -4137.2$]{\includegraphics[scale=0.7]{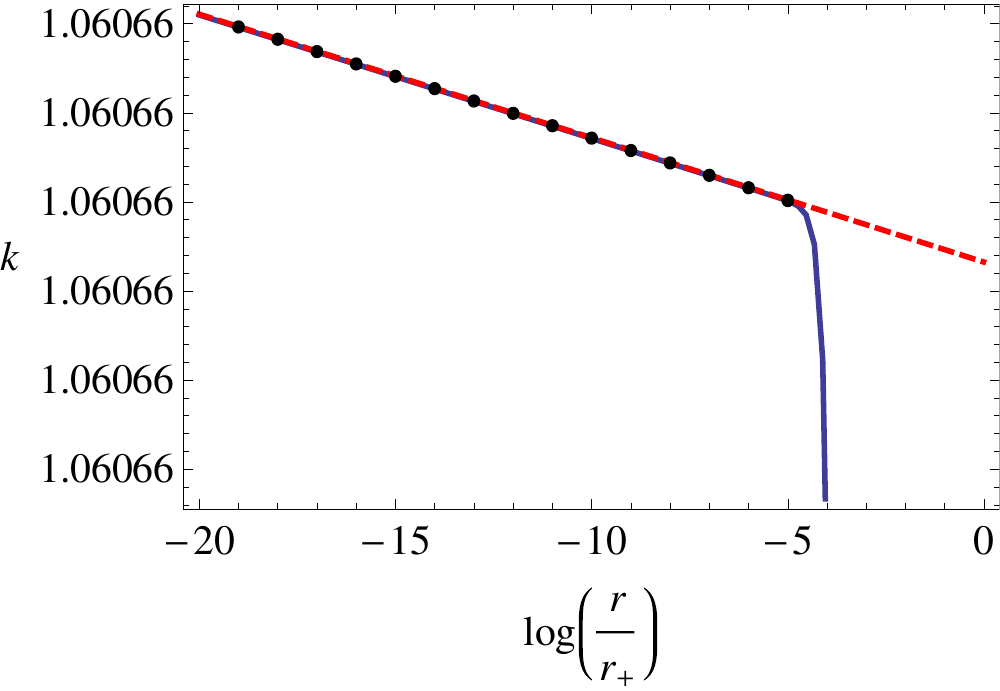}} \; \; \qquad \subfloat[For $\tilde{\alpha}=-3/10$ and $n=4$, $h_0=2.89170$ corresponds to $\log(\Lambda r_+) =-2110.4$]{\includegraphics[scale=0.7]{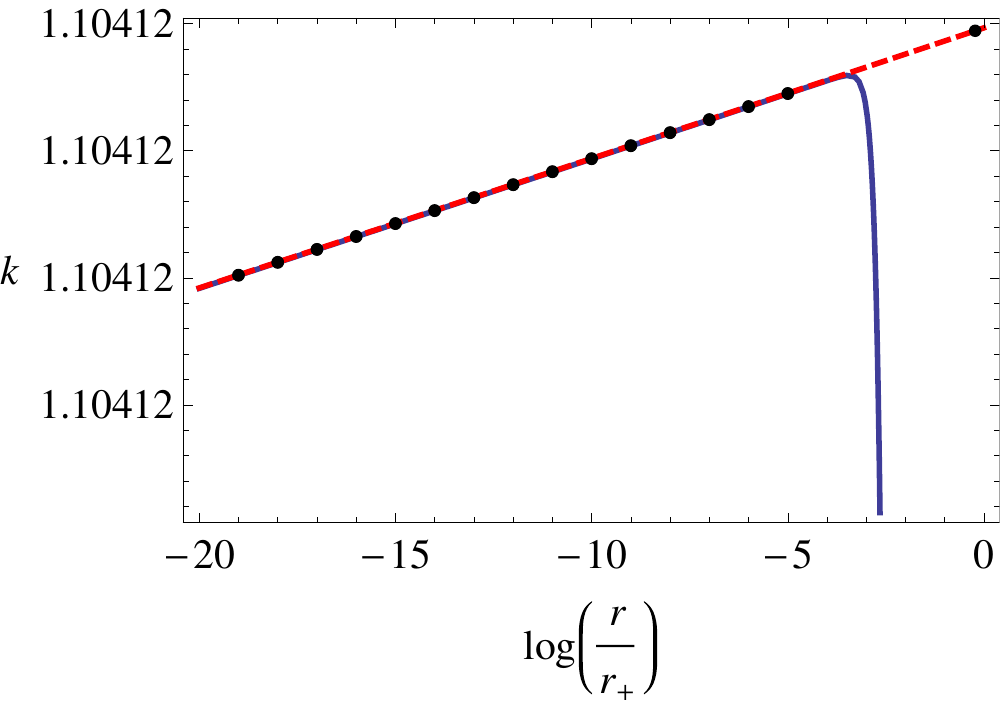}} \\
        \vspace{0.5cm}
        \caption{Extracting $\log( \Lambda r_{+})$}
        \label{fig:mtchk}
\end{figure}

At the middle region, we match the fitting of the asymptotic solutions with the numerical integration for given $h_0$ by controlling the value of $\log \Lambda r_+$ as shown in Fig. {\ref{fig:mtchk}}, which displays the  function $k$ for $n=4$ and for each value of $\tilde{\alpha}$ in Table {\ref{table:vlz}}. The red dashed line is the fitted asymptotic solution, that is the one including $\log \Lambda r_+$, and the blue solid line is the result of  numerical integration of the horizon solution. In our numerical work, the integration is performed from $\log(r/r_+) \sim -0.15$ to $\log(r/r_+) \sim - 10^4$.

Note that in Fig. {\ref{fig:mtchk}}, the graph (f) shows the different behaviour from (a)-(e) by having a positive slope, which means that the $k$ function decreases as  the boundary is approached. This positive-slope pattern is observed only for the $k$ function in the range $\tilde{\alpha} < -1/(2(n-2))$, which is equivalent to $z > n$; for $\tilde{\alpha} \geq-1/(2(n-2))$ or $z \leq n$, the slope of $k$ is negative, meaning it increases as the boundary is approached. The signs of the slopes of $q$ and $y$ versus $\log(r/r_+)$ are
negative  for all values of $\tilde{\alpha}$.  Note that $k$ is defined via $h=k\sqrt{f}$ from (\ref{cv1}), where $f$ is the $-g_{tt}$-component of the metric and $h$ is the vector field potential. Hence $k$ only has information about the charge of the vector field and the Gauss-Bonnet coupling constant $\tilde{\alpha}$. Fig. {\ref{fig:mtchk}}-(f) illustrates that when $z > n$ the weight for the charge of the vector field and $\tilde{\alpha}$ near the horizon become larger than at the boundary.

This process is performed by varying $h_0$. Beyond the maximum value of $\log \Lambda r_+$, the numerical integration of the near horizon solution exponentially grows as  the boundary is approached, and so does not match the asymptotic solution.
This means that if the horizon flux $\phi$ in (\ref{thrmdnmqtt}) is too large, which corresponds to high temperature, the deformation of the spacetime is not allowed any more and becomes the asymptotically pure Gauss-Bonnet-Lifshitz spacetime (having a black hole).  Hence  there exists a maximum value of $h_0$.

\begin{figure}[h]
    \begin{center}
    \includegraphics{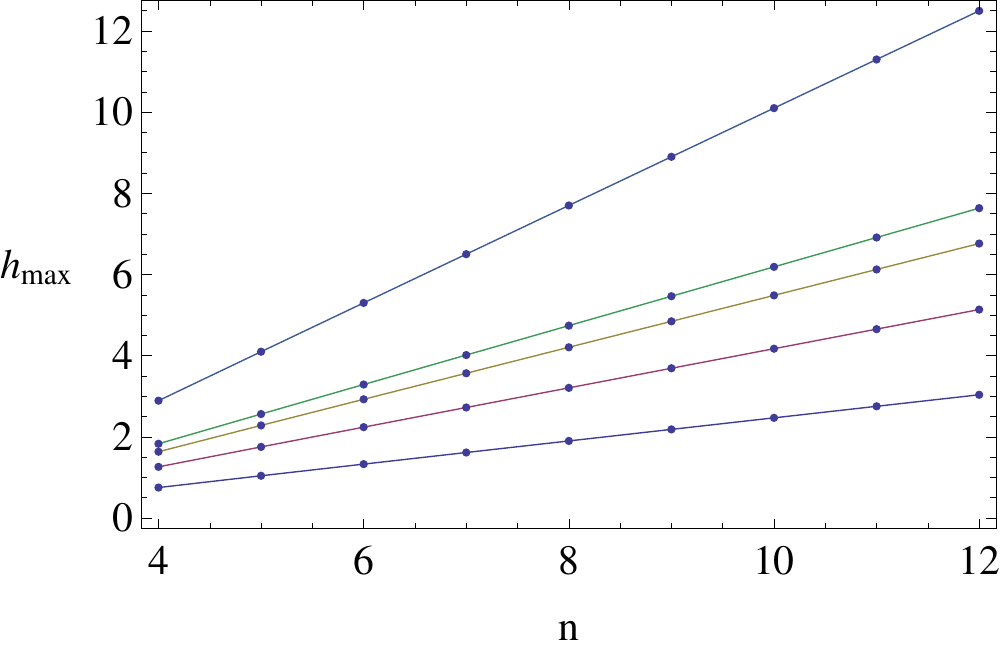}
    \caption{Dots are the maximum value of $h_0$ for given $n$ and $\tilde{\alpha}$, and dots sharing the same $\tilde{\alpha}$ are connected with the same color of line to distinguish them. From the bottom, each different color of line respectively indicates $\tilde{\alpha} = \frac{1}{4}, \frac{1}{10}, 0 , -\frac{1}{20}$, and $-\frac{3}{10}$.}
    \label{fig:hmaxvsn}
    \vspace{0.2cm}
    \end{center}
\end{figure}

\begin{figure}[!h]
    \begin{center}
    \subfloat[The maximum value of $h_0$ on $z$]{\includegraphics[scale=0.7]{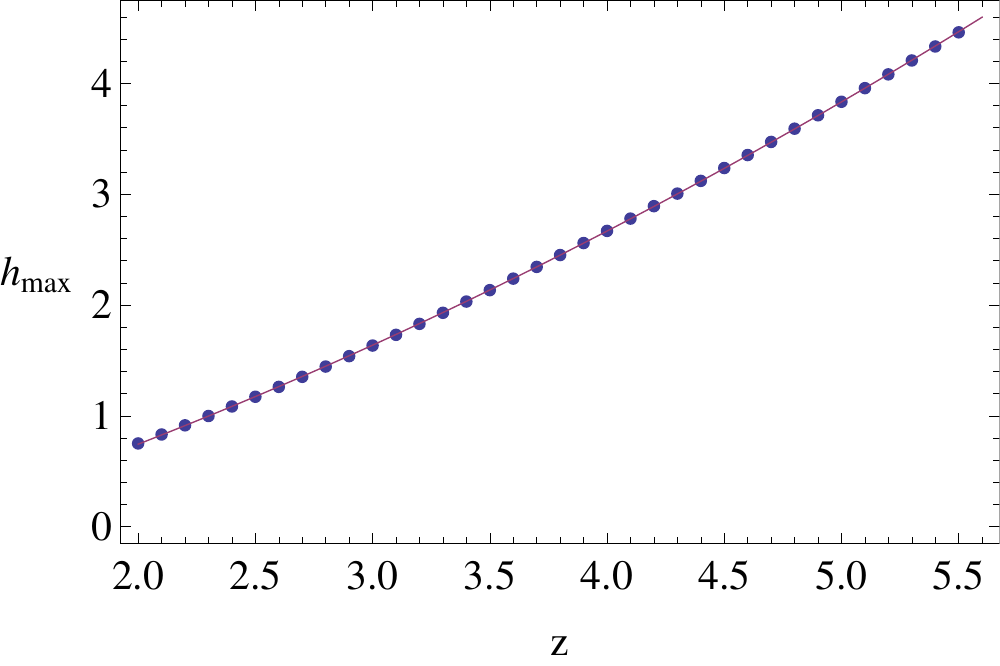}} \; \; \qquad \subfloat[The maximum value of $h_0$ on $\tilde{\alpha}$]{\includegraphics[scale=0.7]{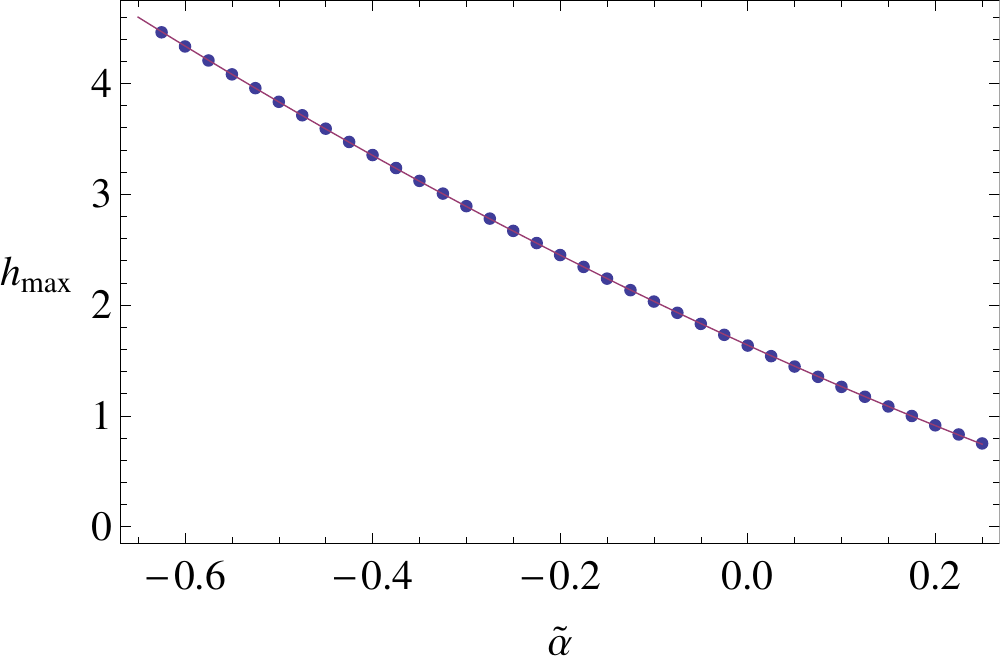}}
    \caption{Dependence of $z$ or $\tilde{\alpha}$ on $h_{max}$ when $n=4$.}
    \label{fig:hmaxvsza}
    \end{center}
\end{figure}

The maximum values of $h_0$ are recorded in Table {\ref{table:hmax}}, and show linearly increasing dependence on the dimensionality of spacetime for the same $\tilde{\alpha}$. For decreasing $\tilde{\alpha}$ the slope gets larger  as shown in Fig. {\ref{fig:hmaxvsn}}, where dots are the numerical data of the maximum value of $h_0$ and the data obtained with the same value of $\tilde{\alpha}$ are connected with the line. For $n=4$, the dependence of $\tilde{\alpha}$ (or $z$) on the maximum values of $h_0$ is also found in Fig. {\ref{fig:hmaxvsza}}, where the dots correspond to the maximum values of $h_0$, and the lines are the fitting functions obtained from the data. These are not linear; rather  $h_{max}\sim-0.633878 + 0.553198 z + 0.0680033 z^2$ for Fig. {\ref{fig:hmaxvsza}}(a) and $h_{max}\sim1.63775- 3.84487 \tilde{\alpha} + 1.08805 \tilde{\alpha}^2$ for Fig. {\ref{fig:hmaxvsza}}(b). These results indicate that $h_0$ obviously depends on $\tilde{\alpha}$.

A minimum value of the flux also exists; its value becomes zero at the horizon. In this case, the influence of the massive vector field vanishes, and so the spacetime is described completely by an asymptotically  Gauss-Bonnet AdS black hole.

Very small values of $h_0$ correspond to the zero temperature limit $\Lambda^z/T \rightarrow \infty$; in this limit the AdS spacetime emerges in the IR regime. Hence for fixed $\Lambda \sim 0$, the slightly deformed UV Lifshitz spacetime and IR AdS spacetime with Gauss-Bonnet corrections emerge instead. Thus we might expect  holographic renormalization group flow at zero temperature
similar to the $\tilde{\alpha}=0$ case {\cite{Braviner:2011kz}},{\cite{Kachru:2008yh}}. The presence and nature of the renormalization flow interpolating both spacetimes for $\tilde{\alpha}\neq 0$ has not yet been studied yet;  it remains for future work.

\begin{table}[!hb]
\begin{center}
  \begin{tabular}{| c || >{\centering}m{2cm} | >{\centering}m{2cm} | >{\centering}m{2cm} | >{\centering}m{2cm}| >{\centering}m{2.1cm} | >{\centering}m{1.9cm} m{0cm}|} \hline
    \multirow{2}{*}{} & \multicolumn{6}{c}{\bf{$h_{max}$}} & \tabularnewline [10pt] \hhline{~-------}
        & $\tilde{\alpha} = \frac{1}{4}$ & $\tilde{\alpha} = \frac{1}{10}$ &  $\tilde{\alpha} = 0$ & $\tilde{\alpha} = - \frac{1}{20}$ & $\tilde{\alpha} = -\frac{1}{2(n-2)}$ & $\tilde{\alpha} = -\frac{3}{10}$ & \tabularnewline [10pt] \hline \hline
    $n=4$  & 0.7512 & 1.2616 & 1.6343 & 1.8298 & 2.6687 & 2.8917 & \tabularnewline [10pt] \hline
    $n=5$  & 1.0416 & 1.7528 & 2.2822 & 2.5622 & 3.2519 & 4.0981 & \tabularnewline [10pt] \hline
    $n=6$  & 1.3288 & 2.2394 & 2.9255 & 3.2900 & 3.8614 & 5.3005 & \tabularnewline [10pt] \hline
    $n=7$  & 1.6147 & 2.7239 & 3.5668 & 4.0158 & 4.4818 & 6.5012 & \tabularnewline [10pt] \hline
    $n=8$  & 1.8998 & 3.2073 & 4.2070 & 4.7406 & 5.1079 & 7.7011 & \tabularnewline [10pt] \hline
    $n=9$  & 2.1844 & 3.6900 & 4.8465 & 5.4647 & 5.7372 & 8.9004 & \tabularnewline [10pt] \hline
    $n=10$ & 2.4688 & 4.1723 & 5.4856 & 6.1884 & 6.3686 & 10.0992& \tabularnewline [10pt] \hline
    $n=11$ & 2.7530 & 4.6543 & 6.1244 & 6.9118 & 7.0013 & 11.2977& \tabularnewline [10pt] \hline
    $n=12$ & 3.0371 & 5.1361 & 6.7629 & 7.6346 & 7.6349 & 12.4958&\tabularnewline [10pt] \hline
    $\vdots$ & $\vdots$ & $\vdots$ & $\vdots$ & $\vdots$ & $\vdots$ & $\vdots$ &\tabularnewline [10pt] \hline
    \end{tabular}
    \caption{The maximum values of $h_{0}$ for each values of $n$ and $\tilde{\alpha}$}
    \label{table:hmax}
\end{center}
\end{table}

\subsubsection{Matching $f_0$ and $p_0$}
\label{sec:Mtchfandp}

We have other variables left to be determined, which are $f_0$ and $p_0$ in (\ref{hrsnslf})-(\ref{hrsnslh}). Since $\Lambda$ is fixed, we tune $f_0$ and $p_0$ based on the value of $\Lambda$ for each set $\{h_0, \tilde{\alpha}\}$ by using the same method applied for $\Lambda$.

\begin{figure}
        \subfloat[For $\tilde{\alpha}=1/4$ ($z=2$) and $h_0=0.75120$, which corresponds to $\log(\Lambda r_+)=-23874.4$, the red dashed line is matched to the blue solid line at $f_0 r^{2z}_+ =19.66$ (left) and $p_0 r^2_+ = 2.2942$ (right). The dots correspond to the values for finding the matching condition.]{\includegraphics[scale=0.7]{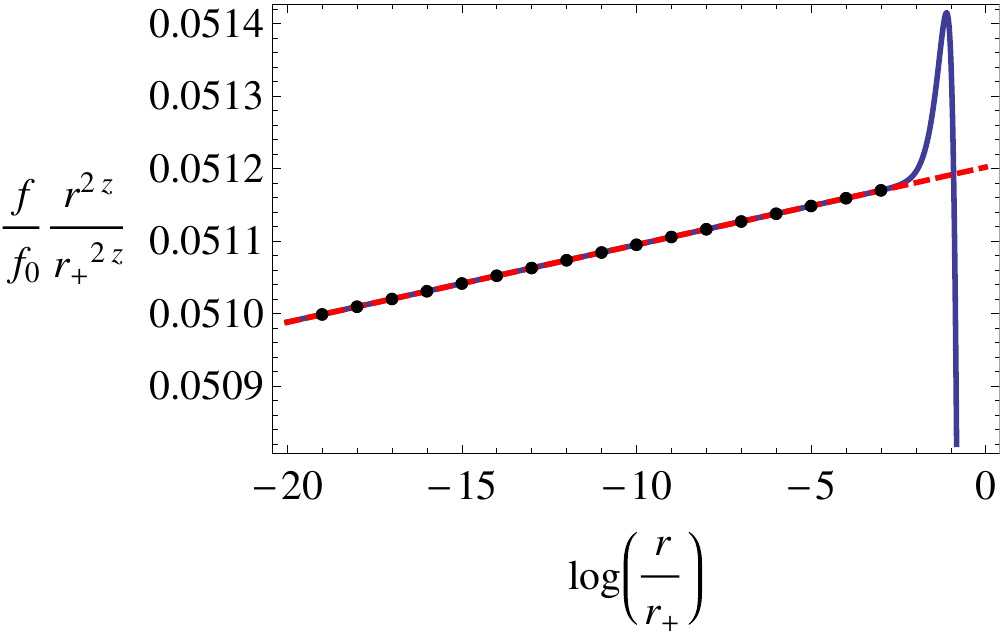} \; \; \; \; \; \includegraphics[scale=0.7]{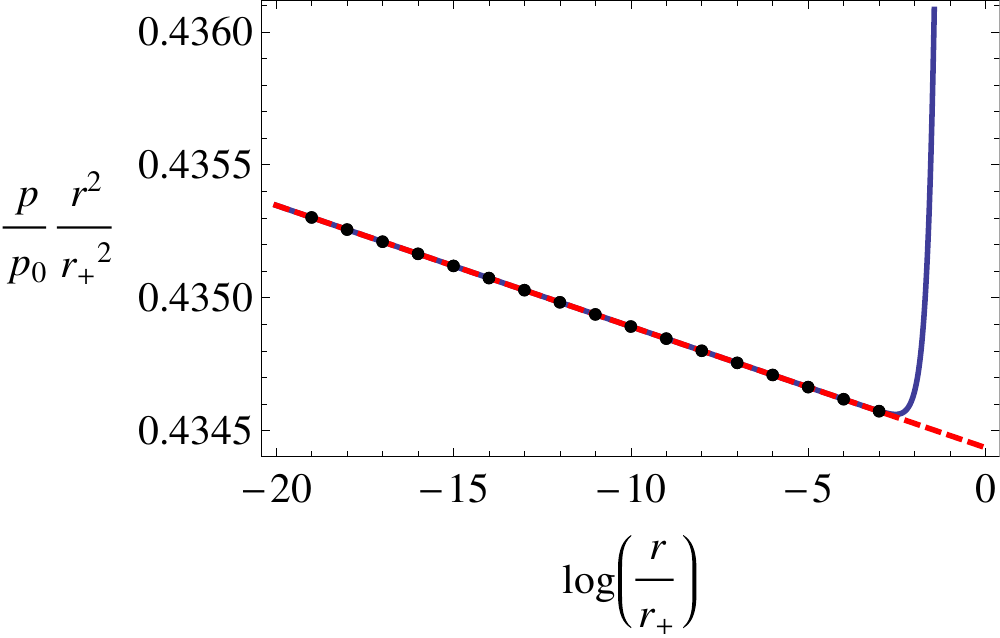}} \\
        \vspace{0.8cm}\\
        \subfloat[For $\tilde{\alpha}=1/10$ (z=2.6) and $h_0=1.26160$, which corresponds to $\log(\Lambda r_+)=-8611.8$, the red dashed line is matched to the blue solid line at $f_0 r^{2z}_+ =30.66$ (left) and $p_0 r^2_+ = 1.9725$ (right). The dots correspond to the values for finding the matching condition.]{\includegraphics[scale=0.7]{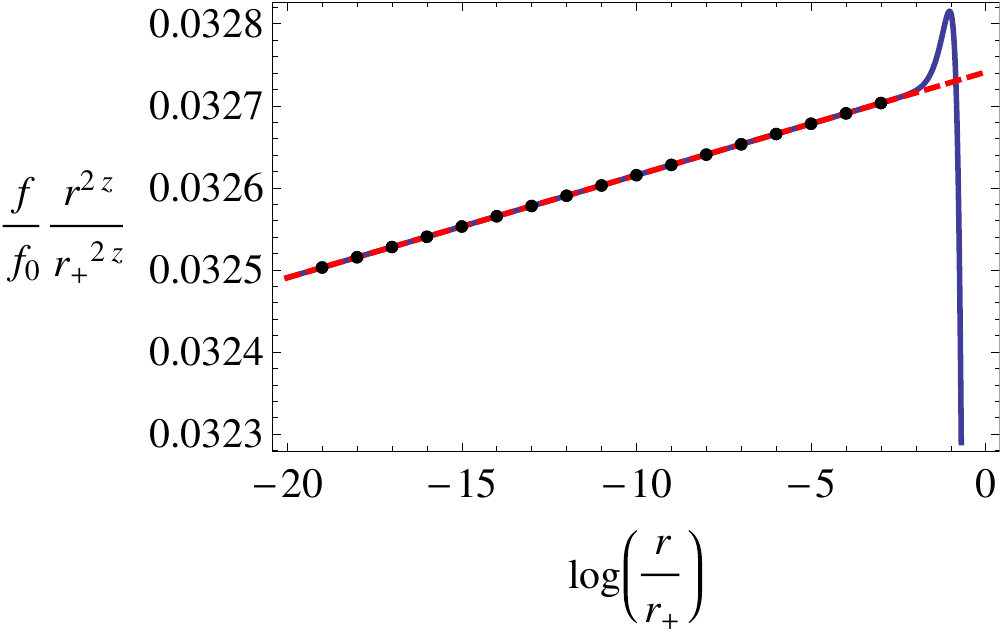} \; \; \; \; \; \includegraphics[scale=0.7]{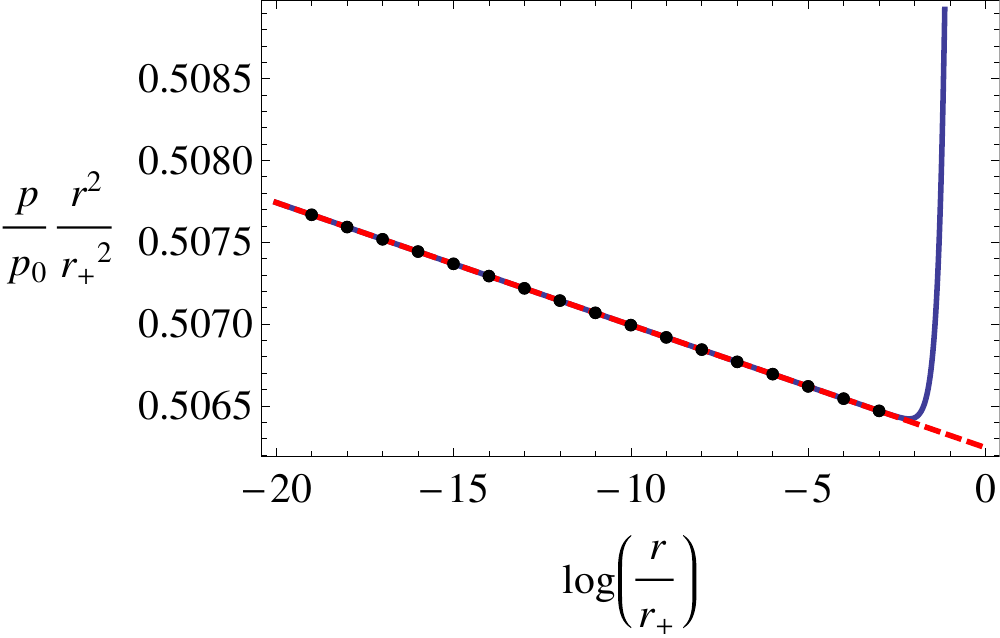}} \\
        \vspace{0.8cm}\\
        \subfloat[For $\tilde{\alpha}=0$ (z=3) and $h_0=1.63430$, which corresponds to $\log(\Lambda r_+)=5167.9$, the red dashed line is matched to the blue solid line at $f_0 r^{2z}_+ =39.00$ (left) and $p_0 r^2_+ = 1.8303$ (right). The dots correspond to the values for finding the matching condition.]{\includegraphics[scale=0.7]{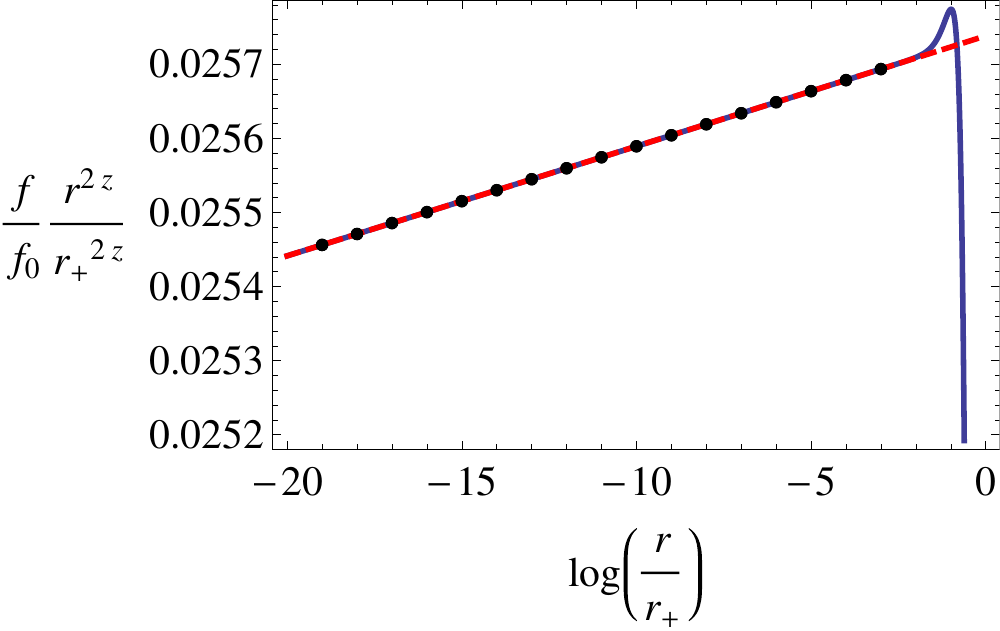} \; \; \; \; \; \includegraphics[scale=0.7]{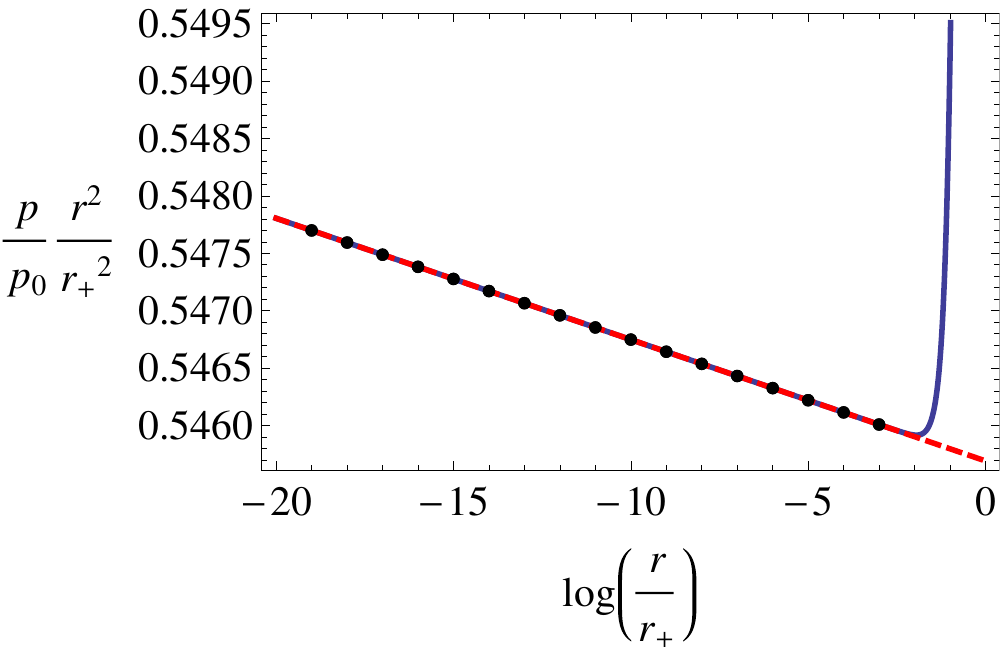}} \\
        \caption{Extracting $f_{0}$ and $p_{0}$ for positive and zero $\tilde{\alpha}$ for $n=4$ }
        \label{fig:mtchfp1}
\end{figure}

\begin{figure}
        \subfloat[For $\tilde{\alpha}=-1/20$ ($z=3.2$) and $h_0=1.82980$, which corresponds to $\log(\Lambda r_+)=-3422$, the red dashed line is matched to the blue solid line at $f_0 r^{2z}_+ =43.53$ (left) and $p_0 r^2_+ =1.7727$ (right). The dots correspond to the values for finding the matching condition.]{\includegraphics[scale=0.7]{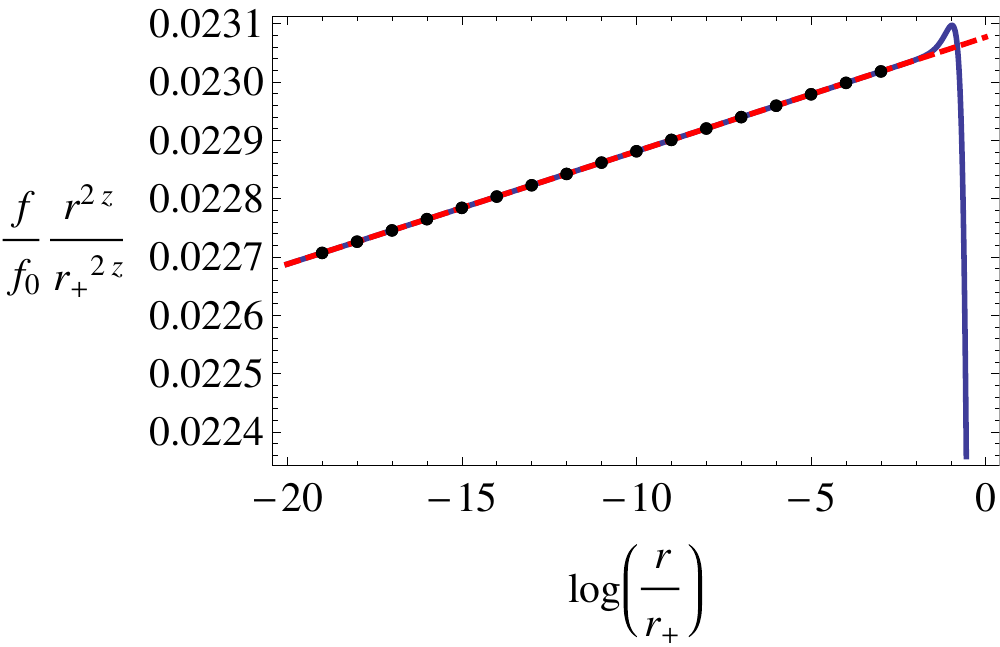} \; \; \; \; \; \includegraphics[scale=0.7]{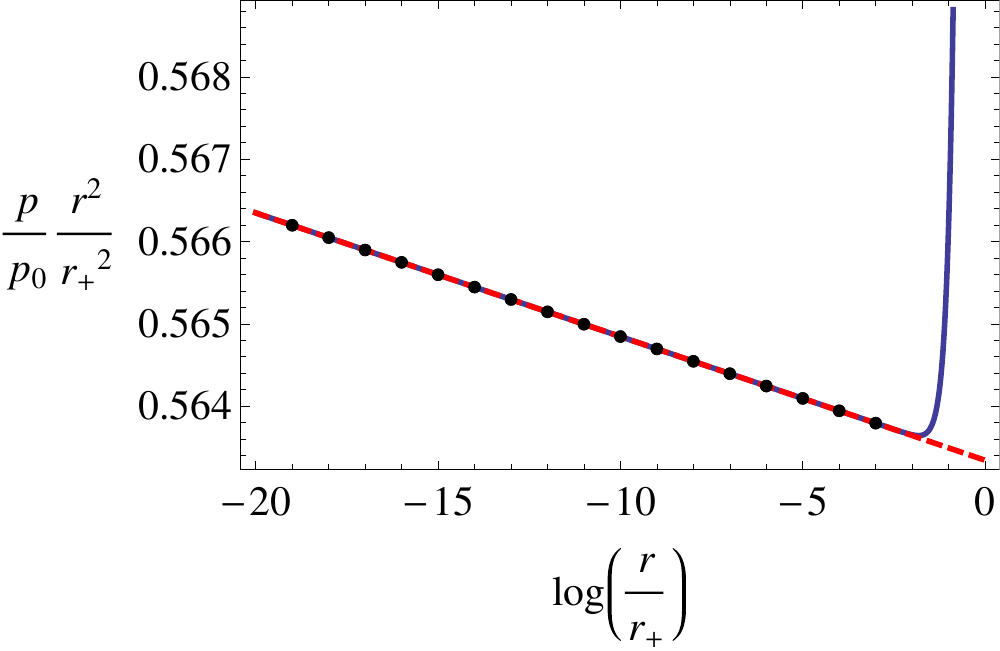}} \\
        \vspace{0.8cm}\\
        \subfloat[For $\tilde{\alpha}=-1/4$ ($z=4$) and $h_0=2.66868$, which corresponds to $\log(\Lambda r_+)=-4137.2$, the red dashed line is matched to the blue solid line at $f_0 r^{2z}_+ =63.88$ (left) and $p_0 r^2_+ =1.6042$ (right). The dots correspond to the values for finding the matching condition.]{\includegraphics[scale=0.7]{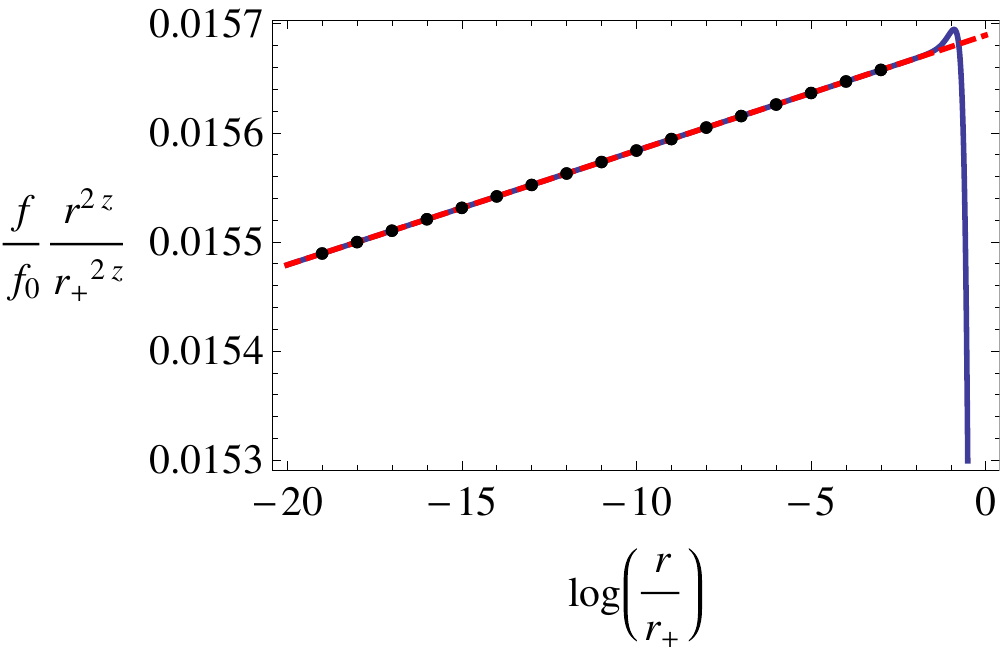} \; \; \; \; \; \includegraphics[scale=0.7]{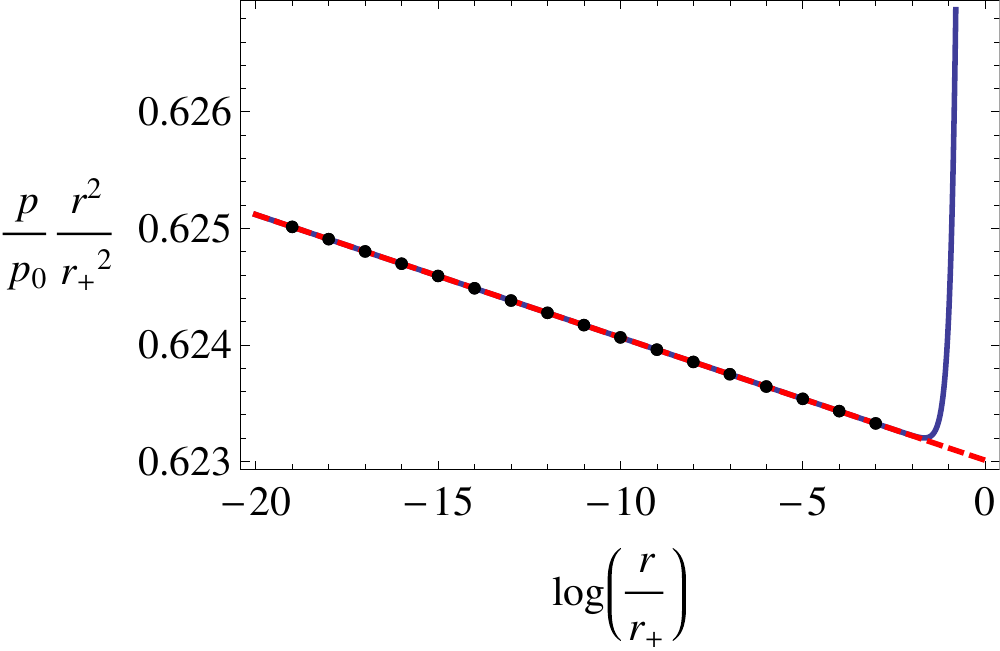}} \\
        \vspace{0.8cm}\\
        \subfloat[For $\tilde{\alpha}=-3/10$ ($z=4.2$) and $h_0=2.89170$, which corresponds to $\log(\Lambda r_+)=-2110.4$, the red dashed line is matched to the blue solid line at $f_0 r^{2z}_+ =69.76$ (left) and $p_0 r^2_+=1.5719$ (right). The dots correspond to the values for finding the matching condition.]{\includegraphics[scale=0.7]{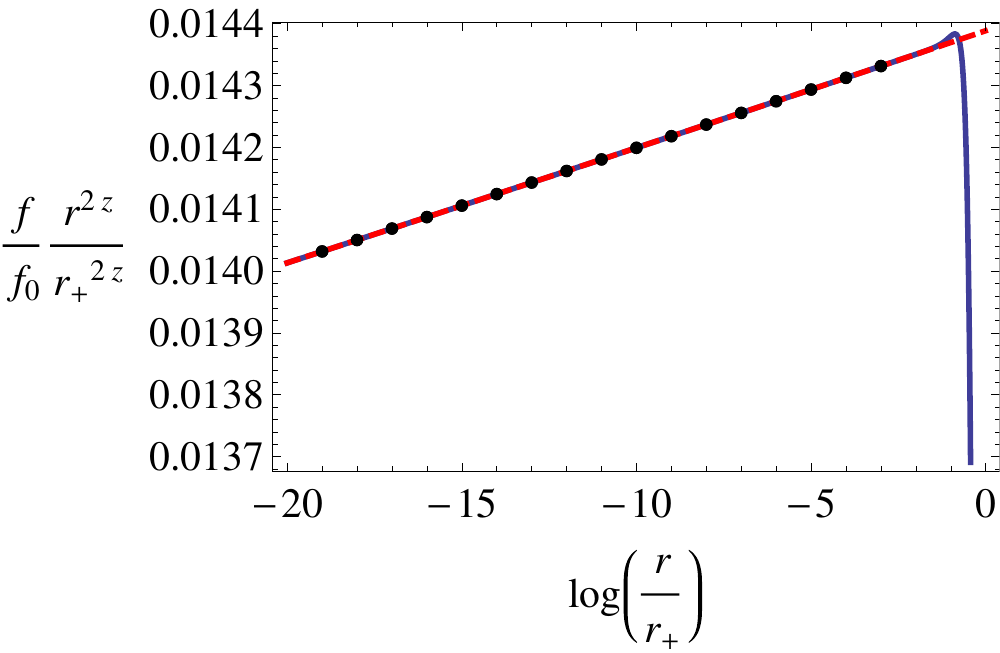} \; \; \; \; \; \includegraphics[scale=0.7]{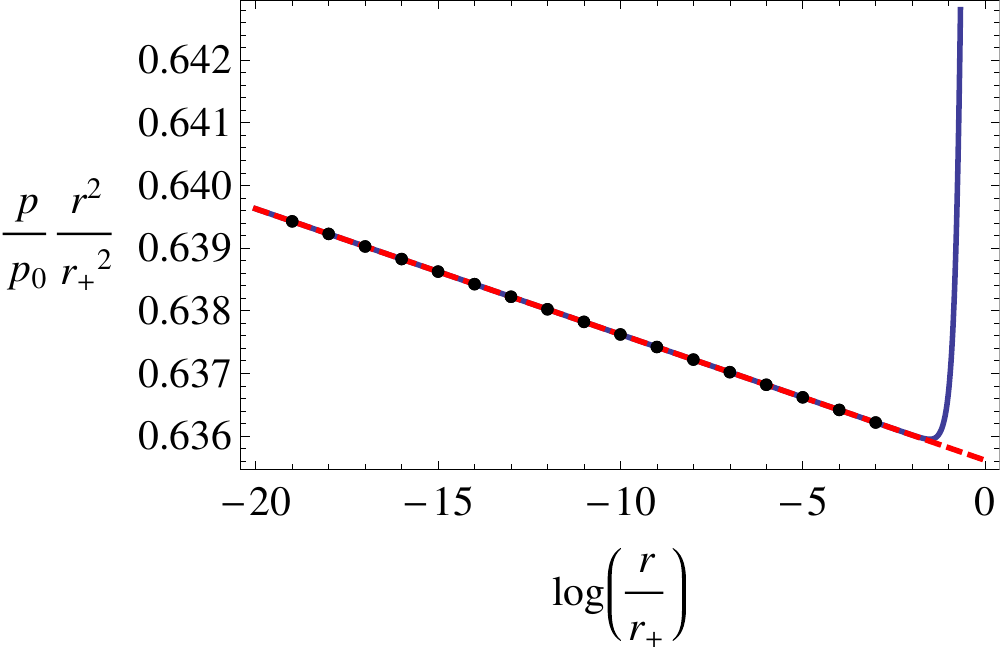}}
        \caption{Extracting $f_{0}$ and $p_{0}$ for negative $\tilde{\alpha}$ for $n=4$ }
        \label{fig:mtchfp2}
\end{figure}

We bring the asymptotic solutions (\ref{expdf})-(\ref{expdp}) expanded in the high energy regime in terms of $\mu$, where $\mu=r^{-1}_+$, and fit the solutions with the determined value of $\Lambda$ (given $\{h_0, \tilde{\alpha}\}$) to the horizon. Near the horizon, we numerically integrate the  the equations of motion (\ref{EqFGH1})-(\ref{EqFGH3})  towards the boundary using
(\ref{hrsnslf})-(\ref{hrsnslh}) as initial conditions. At the middle region, we adjust $f_0 r^{2z}_+$ or $p_0 r^{2}_+$ and read off the values for which both solutions agree. This process for $f$ and $p$ functions is depicted in Figures {\ref{fig:mtchfp1}} and {\ref{fig:mtchfp2}} for $\tilde{\alpha}=1/4, 1/10,0,-1/20,-1/4,-3/10$ for $n=4$. The red dashed line is the fit of the boundary solution and the blue solid line is the numerical result of the near horizon solution.  For $n=5,6,7,8,9$ higher-dimensional cases we find similar patterns for the same values of $\tilde{\alpha}$.


Since we have now fixed  $\Lambda$, $f_0$, and $p_0$, the thermodynamic quantities calculated in (\ref{thrmdnmqtt}) can be determined. The temperature is given by
\begin{equation}
\log \bigg(\frac{\Lambda^z}{T} \bigg) = z \log(\Lambda r_+)  + \log (2 \pi) - \frac{1}{2} \log (f_0 r^{2z}_+) \label{fxtmprtr}
\end{equation}
which is computed from the known values of $\Lambda$ and $f_0$. In next sections, we collect all data on the free energy density $\mathcal{F}/Ts$ and the energy density $\mathcal{E}/Ts$ for given $\{h_0, \tilde{\alpha}\}$, where the specific value of $h_0$ corresponds to each set of values $\{ \Lambda, f_0, p_0 \}$, and fit their results to a function of $\log \Lambda^z/T$.  We consider the dimensionless quantities  $\mathcal{F}/Ts$ and $\mathcal{E}/Ts$.  For the entropy density $s$ from (\ref{thrmdnmqtt}), we consider the dimensionless quantity $s/T^{\frac{n-1}{z}}$ as a function of $\log(\Lambda^z/T)$, plotted in Fig. {\ref{fig:ntrp}}.

\subsubsection{Free Energy Density and Energy Density}
\label{sec:EDandFEDonr}

\begin{figure}
       \subfloat[For $\tilde{\alpha}=1/4$ ($z=2$) and $h_0=0.75120$, which corresponds to $\log(\Lambda r_+)=-23874.4$. Blue line is the numerical results of the free energy density over $Ts$ (left) and the energy density over $Ts$ (right), and red dashed line is reading-off the constant value in the intermediate regime.]{\includegraphics[scale=0.7]{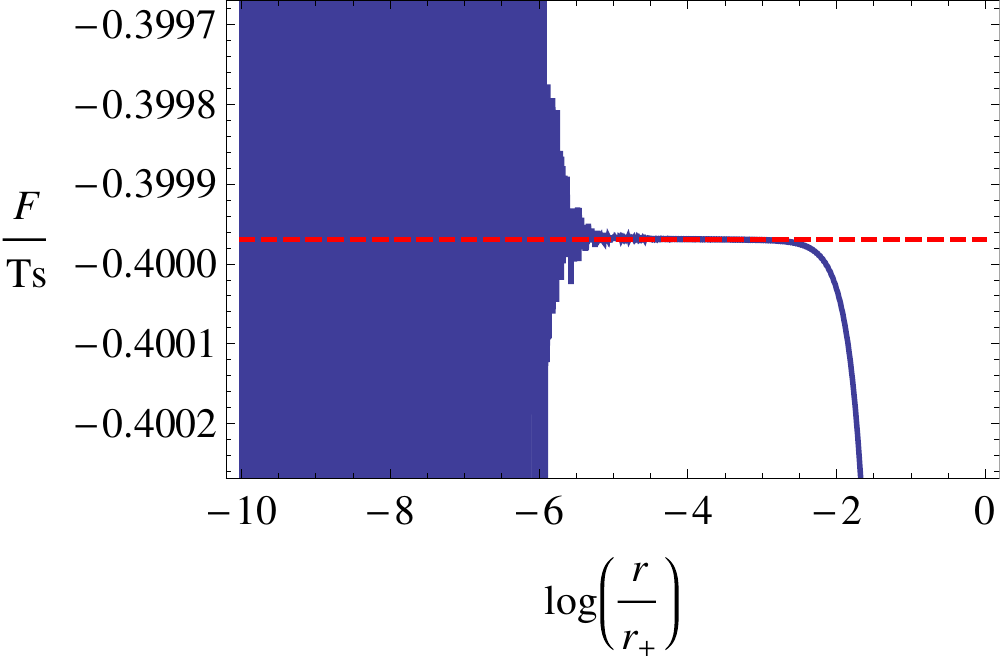} \qquad \qquad \includegraphics[scale=0.7]{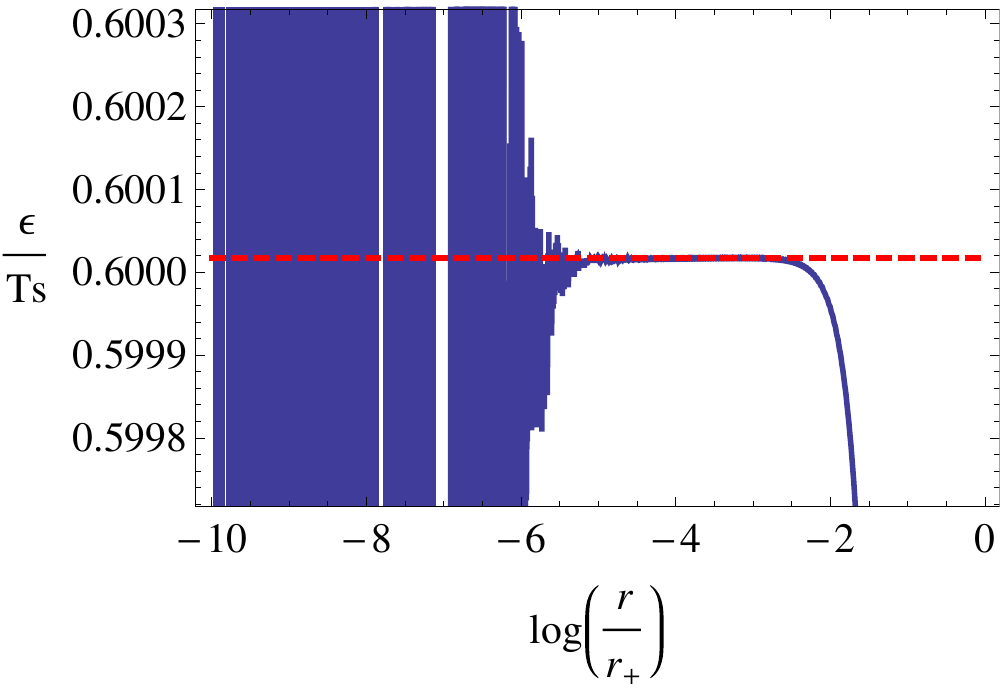}}\\
       \vspace{0.3cm}\\
       \subfloat[For $\tilde{\alpha}=1/10$ ($z=2.6$) and $h_0=1.26156$, which corresponds to $\log(\Lambda r_+)=-5648.3$. Blue line is the numerical results of the free energy density over $Ts$ (left) and the energy density over $Ts$ (right), and red dashed line is reading-off the constant value in the intermediate regime.]{\includegraphics[scale=0.7]{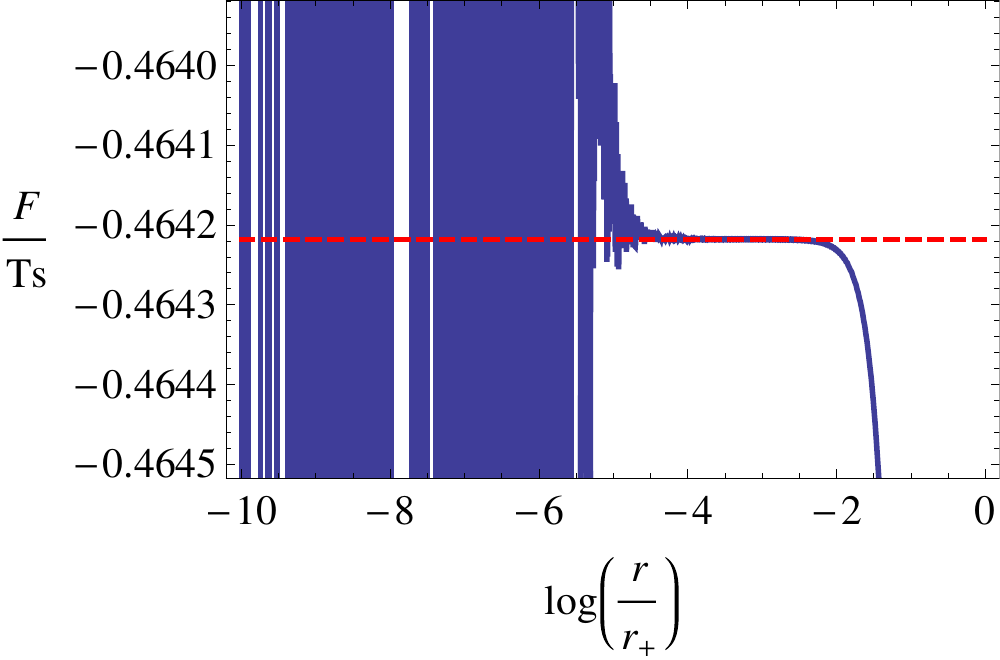} \qquad \qquad \includegraphics[scale=0.7]{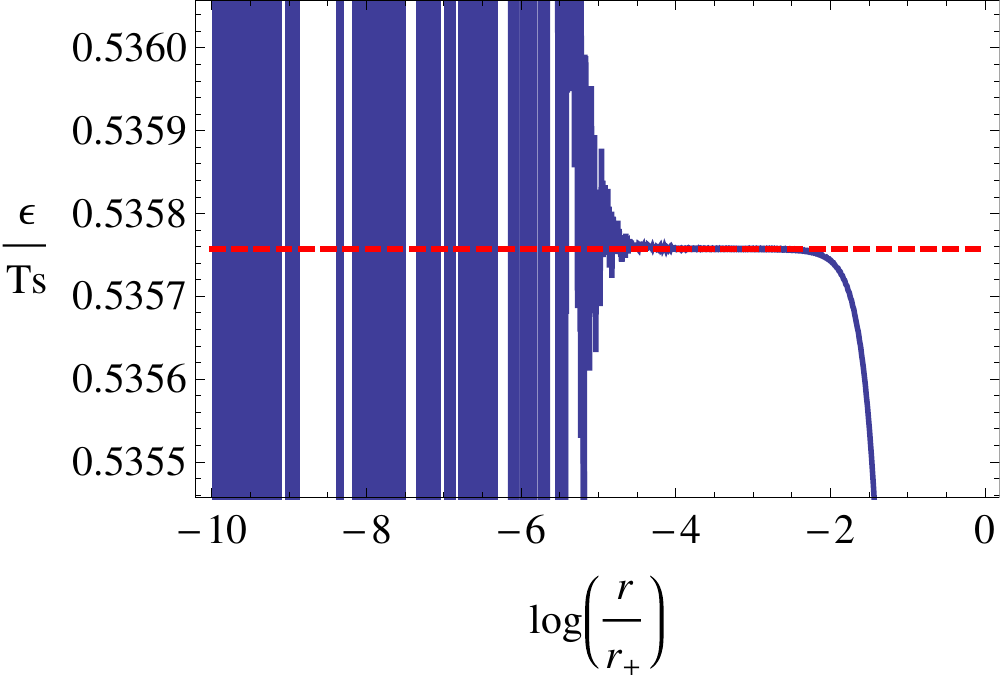}}\\
       \vspace{0.3cm}\\
       \subfloat[For $\tilde{\alpha}=0$ (z=3) and $h_0=1.63430$, which corresponds to $\log(\Lambda r_+)=5167.9$. Blue line is the numerical results of the free energy density over $Ts$ (left) and the energy density over $Ts$ (right), and red dashed line is reading-off the constant value in the intermediate regime.]{\includegraphics[scale=0.7]{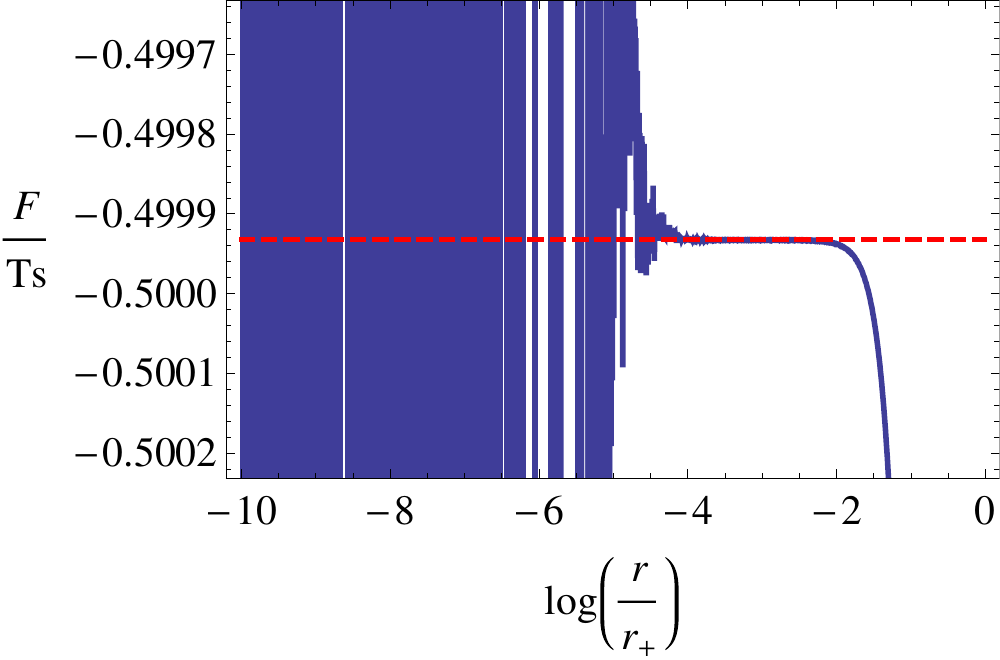} \qquad \qquad \includegraphics[scale=0.7]{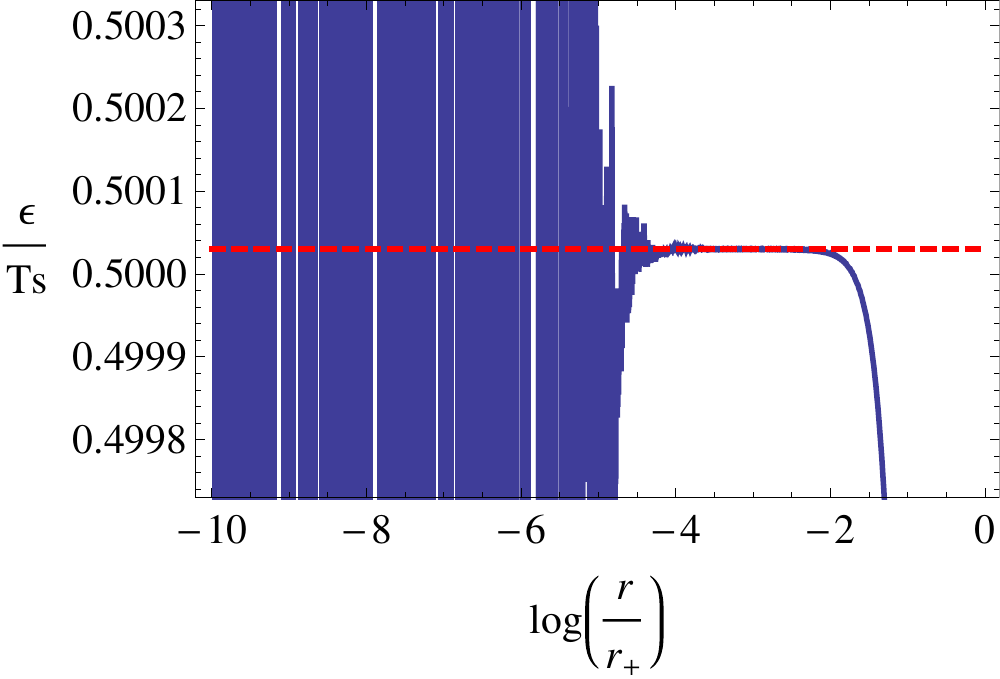}}\\
        \caption{${\mathcal{F}}/Ts$ and ${\mathcal{E}}/Ts$ versus $\log(\frac{r}{r_+})$ for positive and zero $\tilde{\alpha}$ for $n=4$}
        \label{fig:frednst1}
\end{figure}

\begin{figure}[p!]
    \subfloat[For $\tilde{\alpha}=-1/10$ ($z=3.2$) and $h_0=1.82976$, which corresponds to $\log(\Lambda r_+)=-2523.1$.]{\includegraphics[scale=0.67]{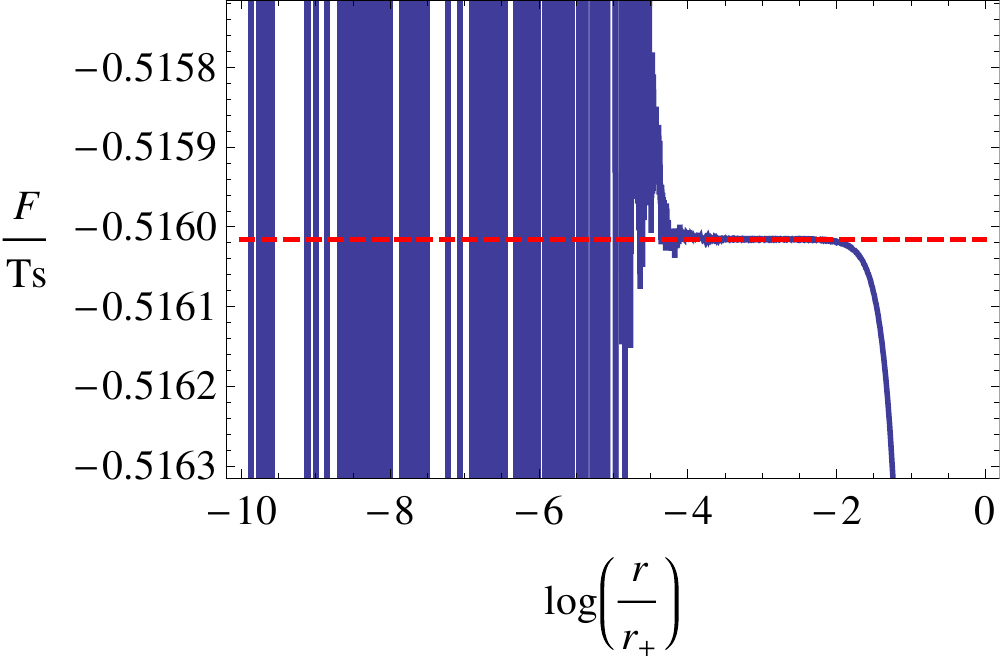} \qquad \qquad \includegraphics[scale=0.67]{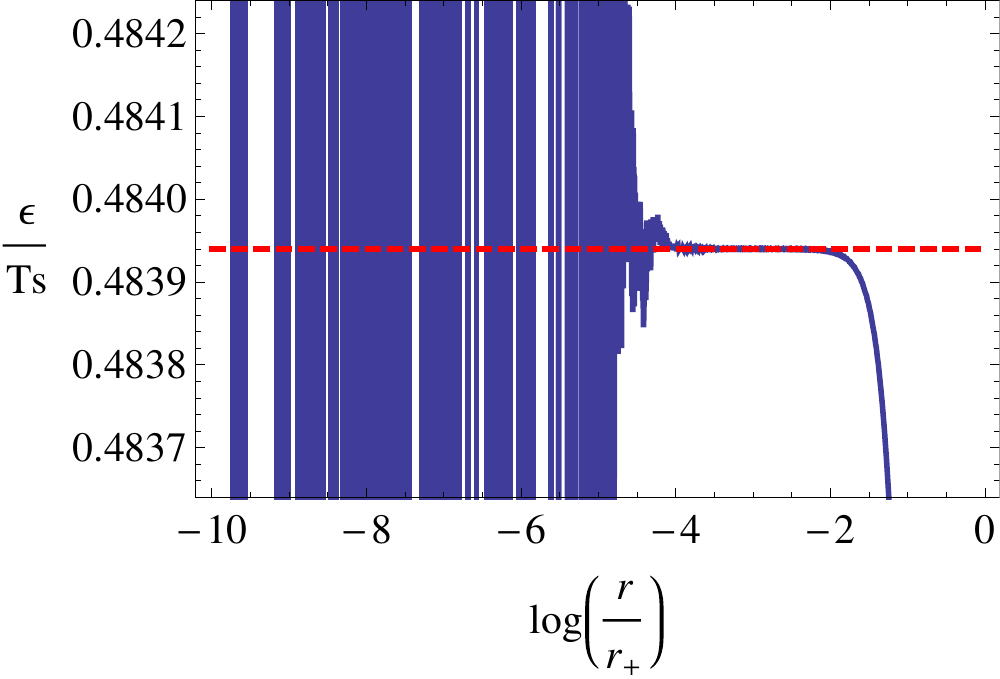}}\\
    \subfloat[For $\tilde{\alpha}=-1/4$ ($z=4$) and $h_0=2.66868$, which corresponds to $\log(\Lambda r_+)=-4137.2$.]{\includegraphics[scale=0.67]{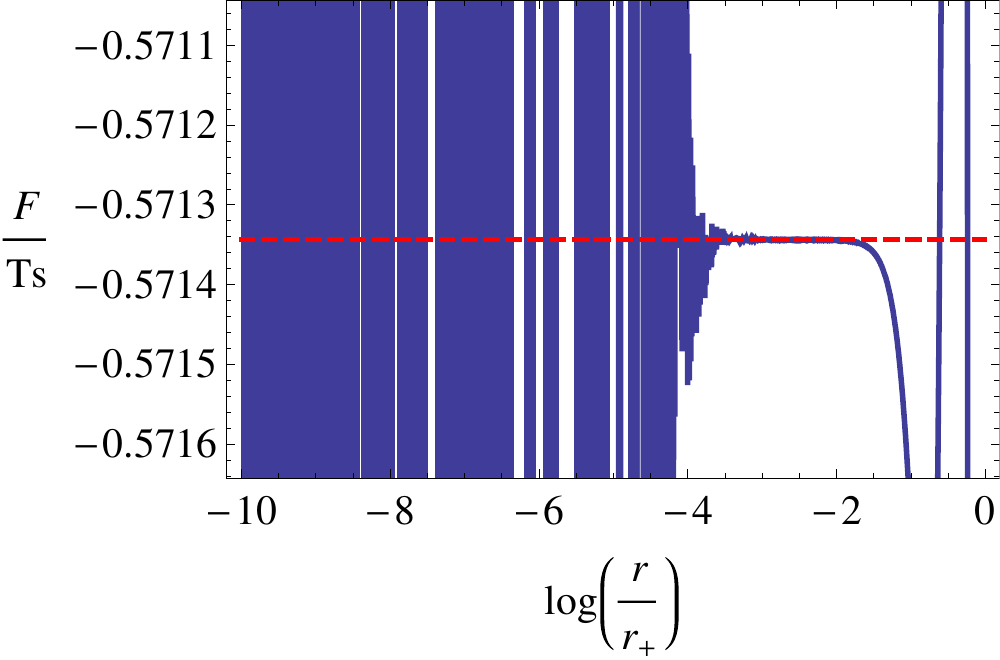} \qquad \qquad \includegraphics[scale=0.67]{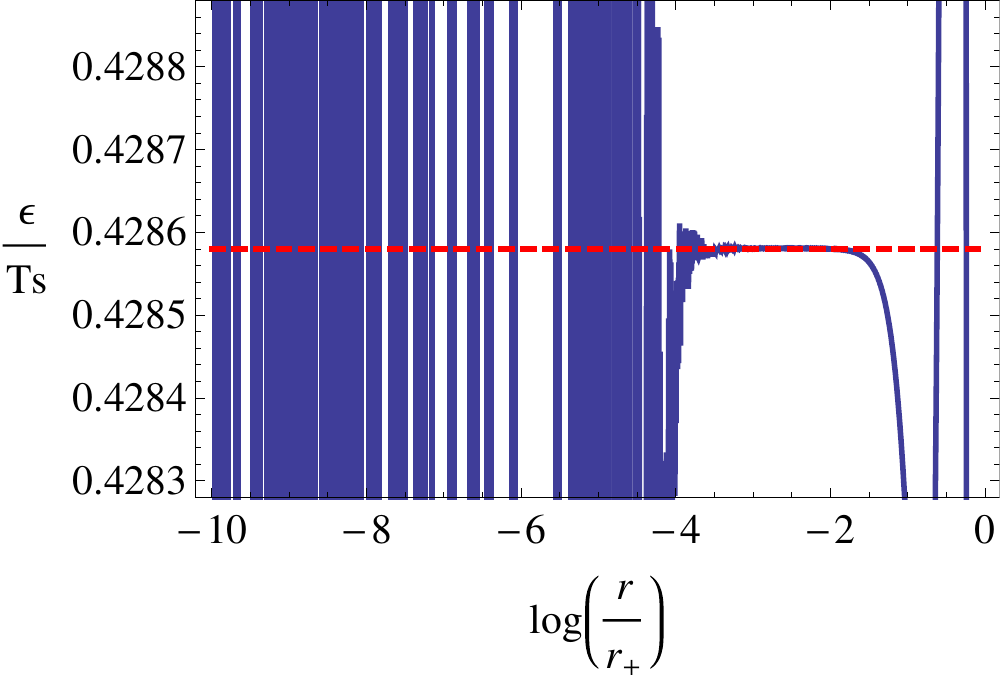}}\\
    \subfloat[For $\tilde{\alpha}=-3/10$ ($z=4.2$) and $h_0=2.89170$, which corresponds to $\log(\Lambda r_+)=-2110.4$.]{\includegraphics[scale=0.67]{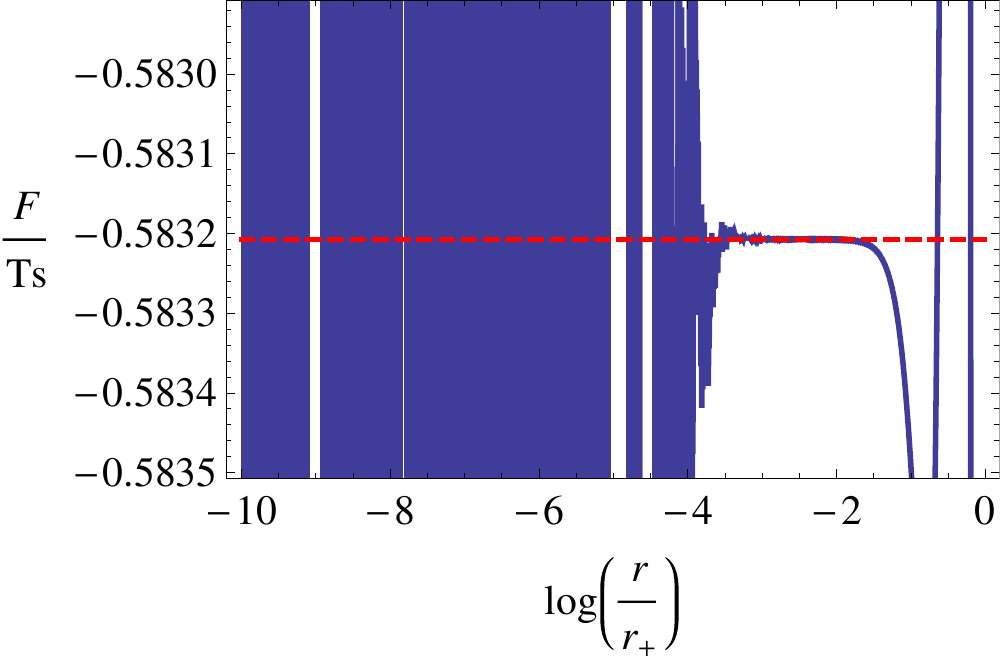} \qquad \qquad \includegraphics[scale=0.67]{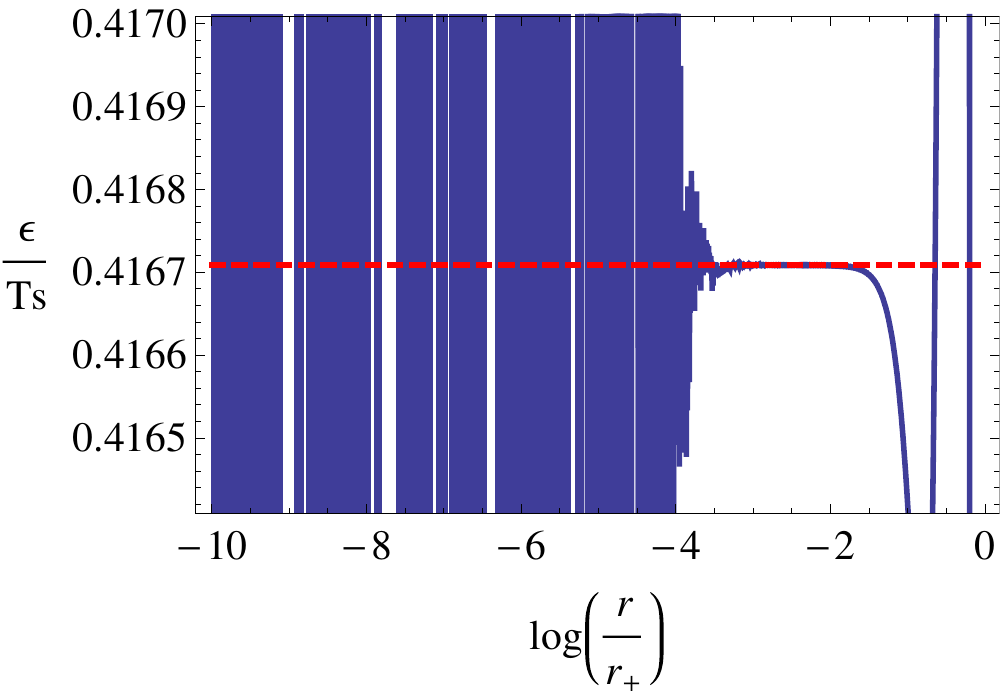}}\\
    \subfloat[enlarged graphs of the above (c)]{\includegraphics[scale=0.67]{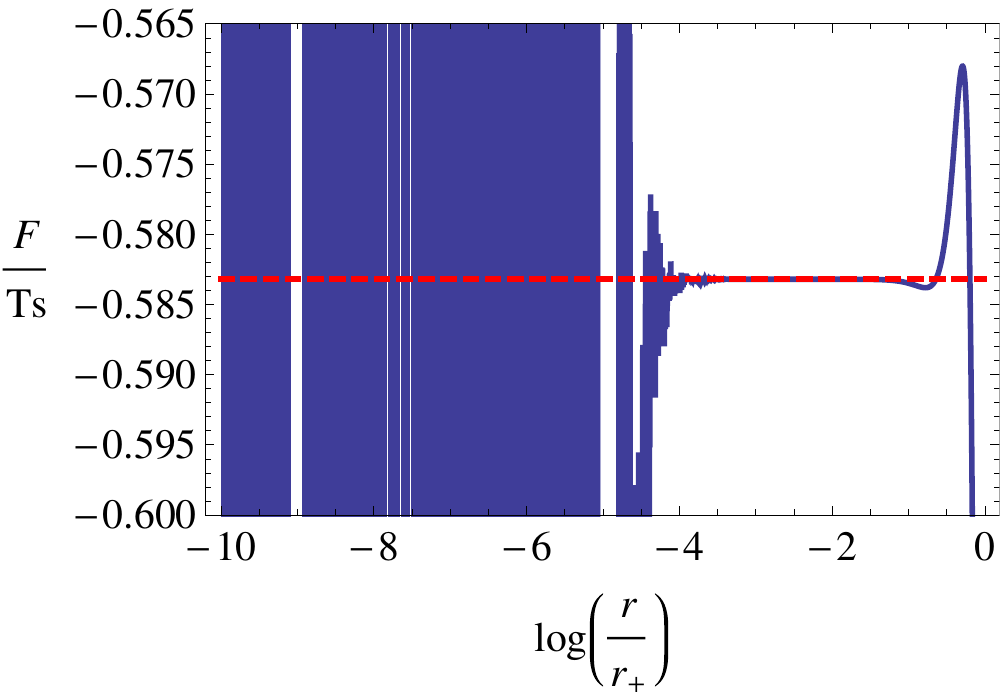} \qquad \qquad \includegraphics[scale=0.67]{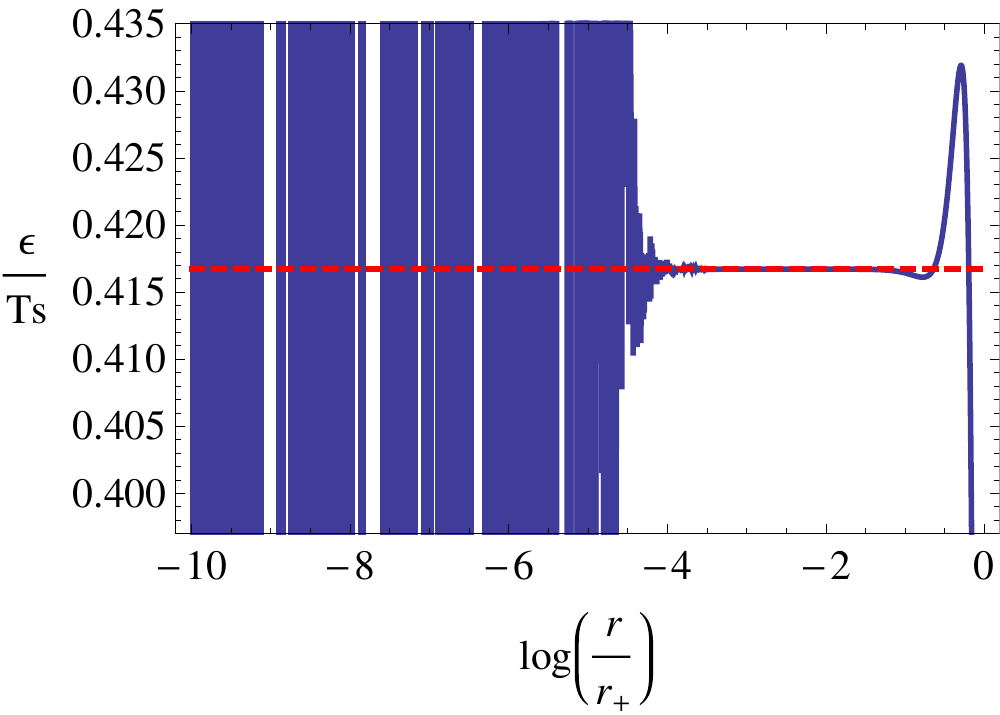}}
    \caption{${\mathcal{F}}/Ts$ and ${\mathcal{E}}/Ts$ versus $\log(\frac{r}{r_+})$ for negative $\tilde{\alpha}$ for $n=4$}
    \label{fig:frednst2}
\end{figure}

The dimensionless quantities $\mathcal{F}/Ts$ and $\mathcal{E}/Ts$, respectively the free energy density   (\ref{fedfn}) and the energy density  (\ref{edfn}), are numerically integrated from the horizon towards the boundary for given $\{h_0, \tilde{\alpha}\}$. The results are depicted in Figures {\ref{fig:frednst1}} and {\ref{fig:frednst2}}, where the blue line is the numerical result and the red dashed line is read  off from the stable value in the flat region. As shown in Figures {\ref{fig:frednst1}} and {\ref{fig:frednst2}}, $\mathcal{F}/Ts$ and $\mathcal{E}/Ts$ yield   constant values over some finite range of $\log(r/r_+)$, but oscillate infinitely upon approaching the boundary. The reason for these diverging and oscillating behaviours can be found in the holographic renormalization process in section 3. Recall  that the free energy density and the energy density diverge as $r \rightarrow 0$, and so the counterterms  (\ref{C0})-(\ref{C2}) were constructed to render them finite. As mentioned, $C_0$ and $C_1$ expand to infinite series in $\log (\Lambda r_+)$; however only terms up to $1/\log^4(\Lambda r)$ (15 terms for each of $C_0$ and $C_1$) are computed due to limitations of the analytic calculation. This finite number  of terms is employed in the numerical calculation of $\mathcal{F}/Ts$ and $\mathcal{E}/Ts$. On the other hand, the numerical work includes higher orders (i.e. higher than $1/\log^4 \Lambda r$), and so the diverging properties are not totally eliminated by our analytic solutions using counterterms.

In Figures {\ref{fig:frednst2}} (b) and (c), which corresponds to cases of $z=4$ and $z=4.2$ respectively, unstable behaviours are observed near the horizon. To see the whole pattern the scales are zoomed out in Figure  {\ref{fig:frednst2}} (d). This pattern of having a sharp peak near the horizon  commonly appears for the  range $z \gtrsim n$ for any $n$ and $h_0$. The unstable behaviour of $\mathcal{F}/Ts$ and $\mathcal{E}/Ts$ for $z \gtrsim n$ can be understood via the same rationale for  the different pattern of the $k$-function for $z > n$. That is, as $z$ gets bigger and approaches $z \sim n$ due to the decreasing value of $\tilde{\alpha}$, the effect of both the charge of the vector field and $\tilde{\alpha}$ gets bigger  near the horizon than at the boundary, causing unstable behaviour in both $\mathcal{F}/Ts$ and $\mathcal{E}/Ts$.

\afterpage{\clearpage}

\subsection{Behaviours of $\mathcal{F}$ and $\mathcal{E}$ near Quantum Critical Regimes}
\label{sec:EDandFEDonTmp}

\begin{figure}[!h]
        \subfloat[For $\alpha=\frac{1}{4}$ (or $z=2$). Dots are numerical results running $h_0$ from $0.75120$ to $0.75038$, which corresponds to $\log \Lambda^z/T$ from $-47748.5$ to $-3945.48$. Solid line is the fitting function denoted in Table {\ref{table:fitfuncn4}}.]{\includegraphics[scale=0.55]{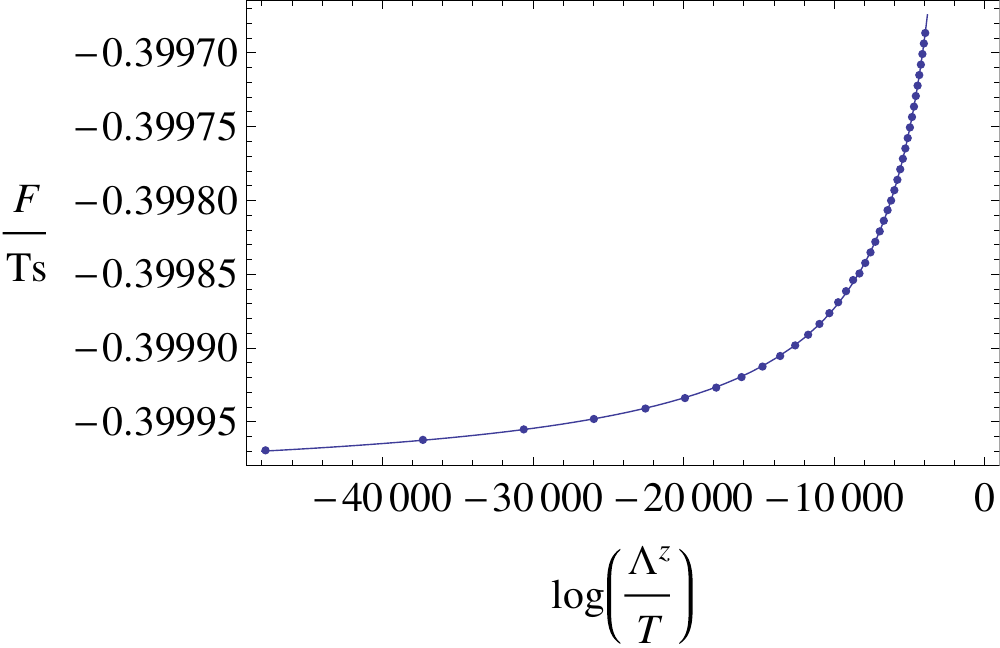} \includegraphics[scale=0.55]{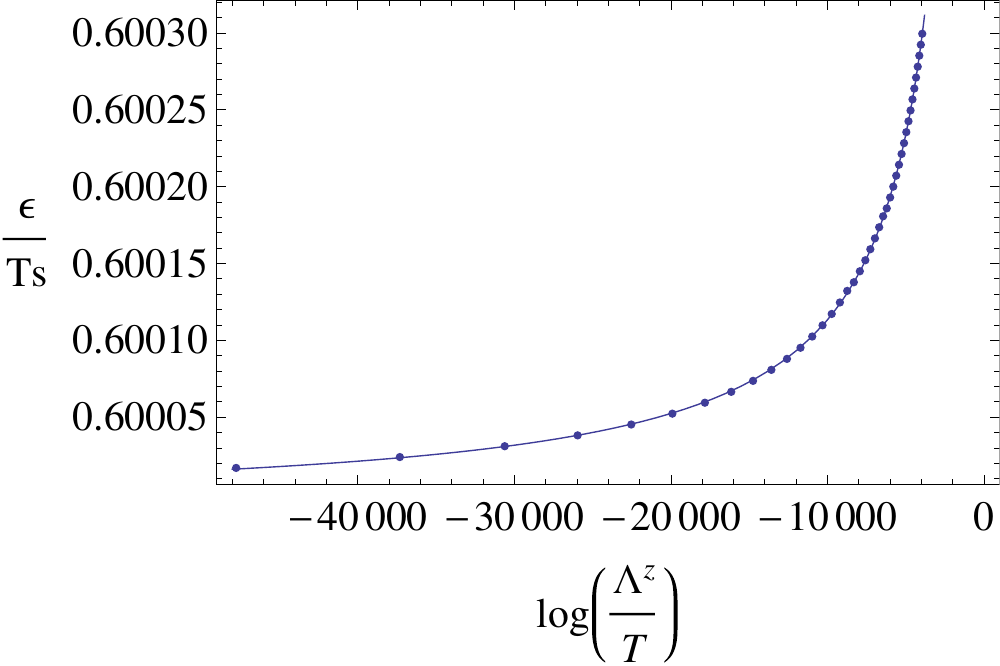}   \includegraphics[scale=0.55]{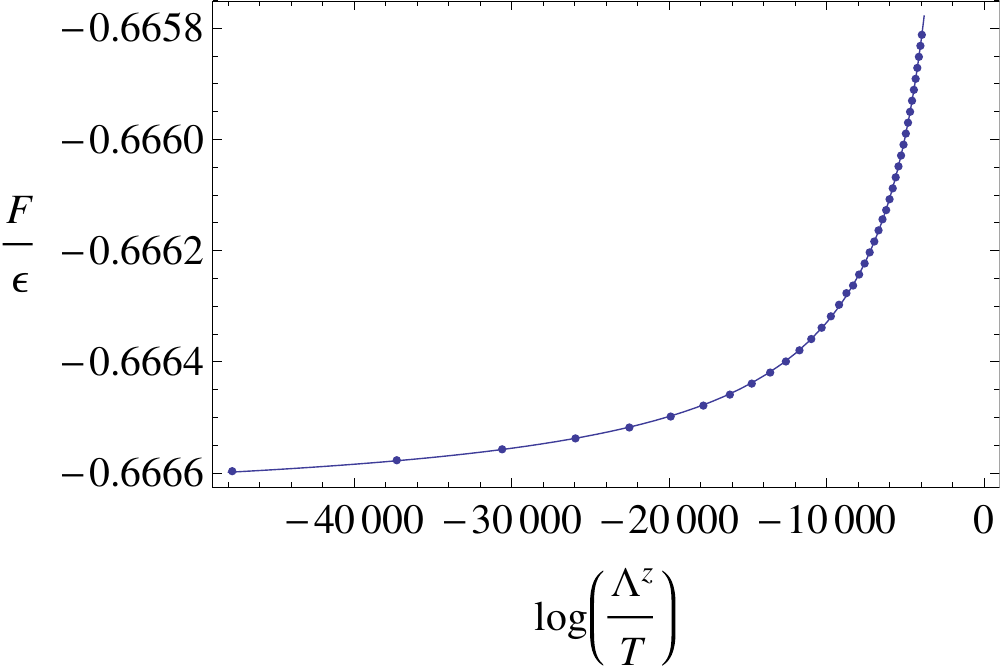}} \\
        \vspace{0.3cm}\\
        \subfloat[For $\alpha=\frac{1}{10}$ (or $z=2.6$). Dots are numerical results running $h_0$ from $1.26160$ to $1.26078$, which corresponds to $\log \Lambda^z/T$ from $-22390.6$ to $-1960.29$. Solid line is the fitting function denoted in Table {\ref{table:fitfuncn4}}.]{\includegraphics[scale=0.55]{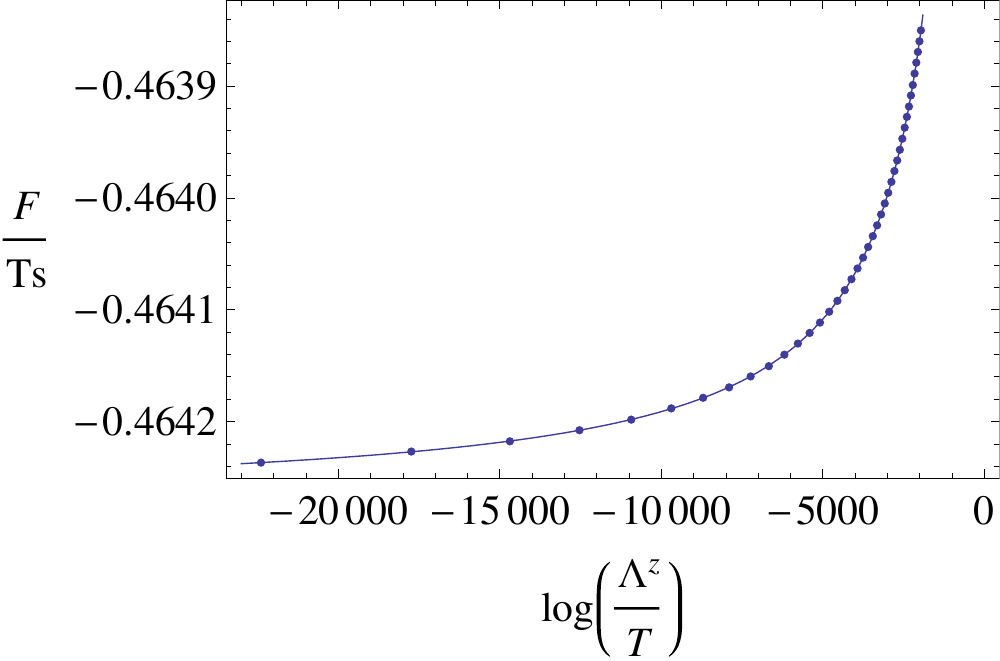} \includegraphics[scale=0.55]{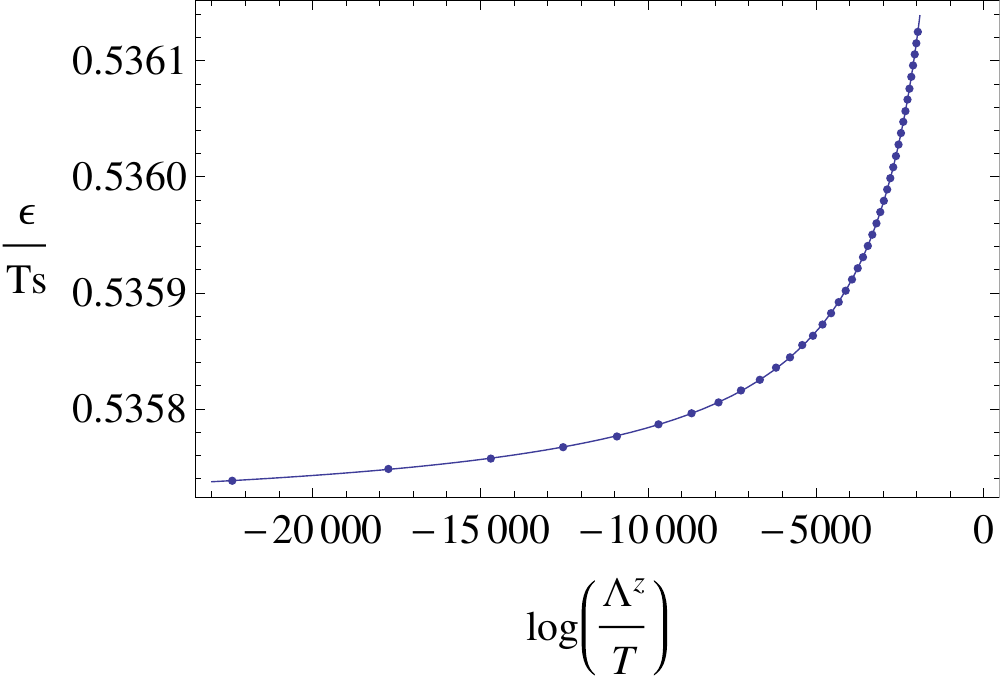}  \includegraphics[scale=0.55]{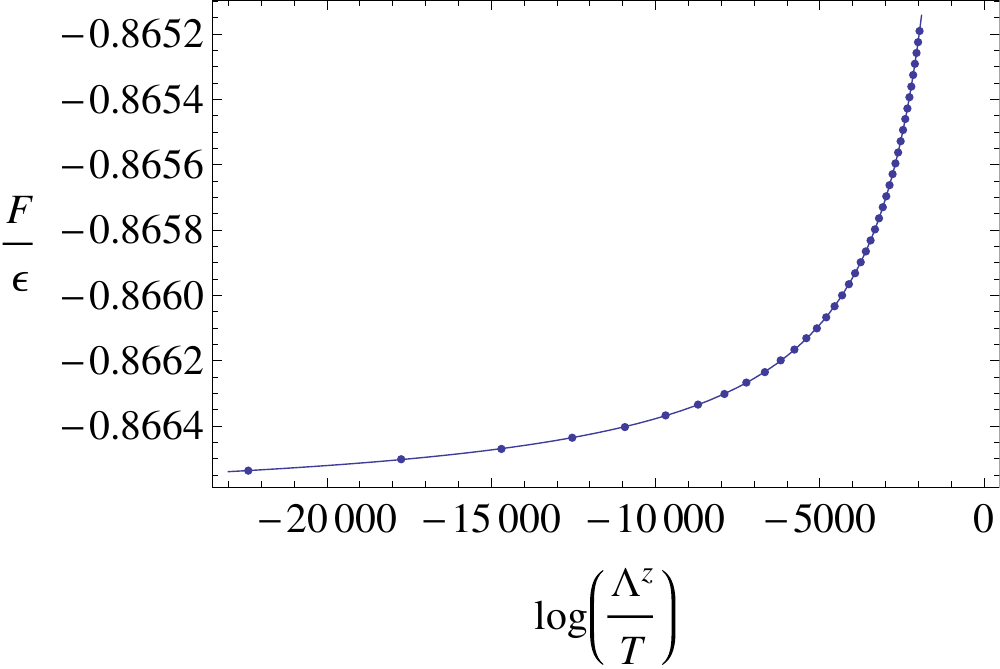}} \\
        \vspace{0.3cm}\\
        \subfloat[For $\alpha=0$ (or $z=3$). Dots are numerical results running $h_0$ from $1.63430$ to $1.63348$, which corresponds to $\log \Lambda^z/T$ from $-15503.7$ to $-15522.51$. Solid line is the fitting function denoted in Table {\ref{table:fitfuncn4}}.]{\includegraphics[scale=0.55]{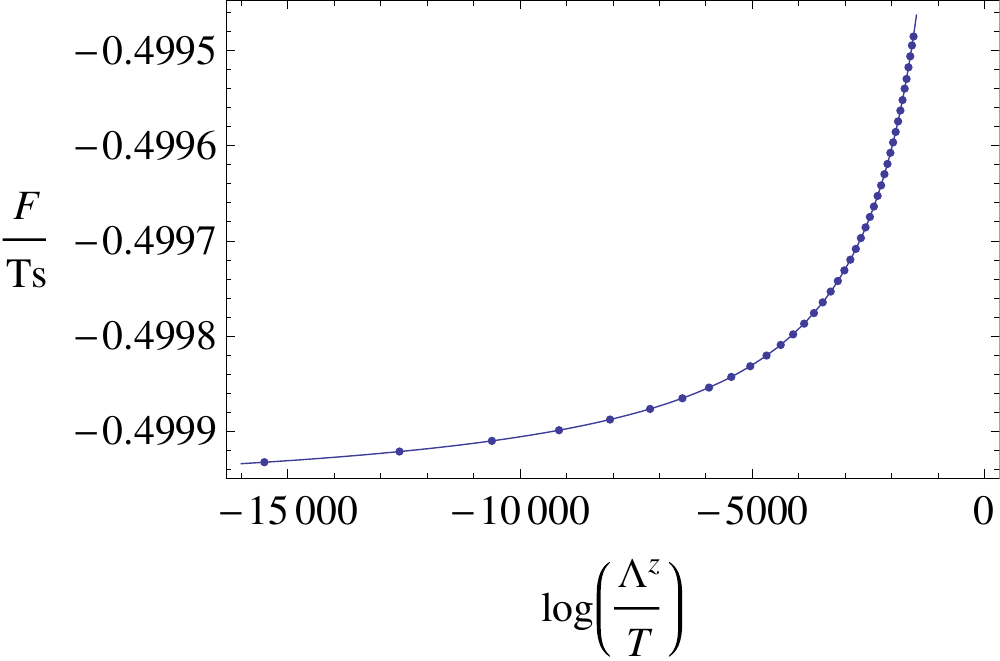} \includegraphics[scale=0.55]{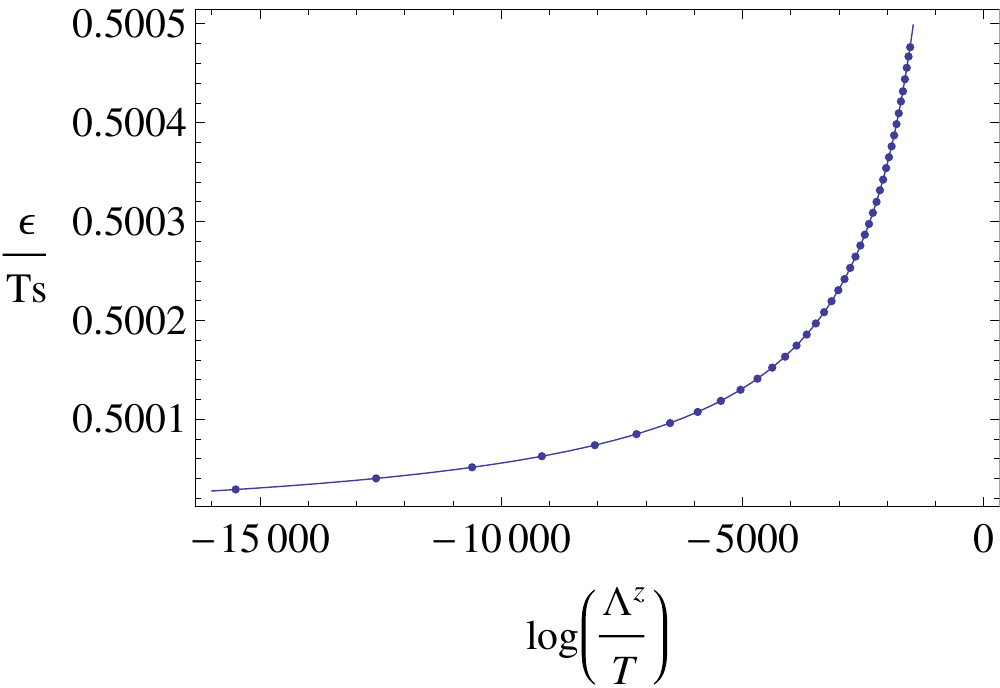} \includegraphics[scale=0.55]{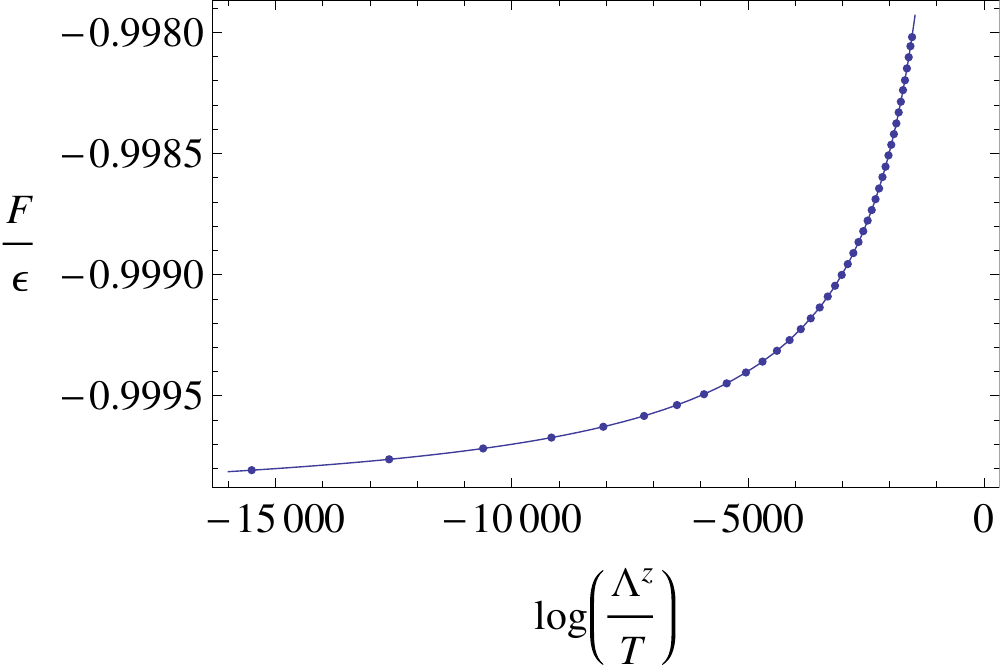}}
        \caption{Plots of ${\mathcal{F}}/Ts$ , ${\mathcal{E}}/Ts$ and ${\mathcal{F}}/{\mathcal{E}}$ versus $\log(\Lambda^z/T)$ for positive and zero $\tilde{\alpha}$ for $n=4$}
        \label{fig:frvsT1}
\end{figure}

\begin{figure}[!h]
        \subfloat[For $\alpha=-\frac{1}{20}$ (or $z=3.2$). Dots are numerical results running $h_0$ from $1.82980$ to $1.82898$, which corresponds to $\log \Lambda^z/T$ from $-10950.4$ to $-1354.62$. Solid line is the fitting function denoted in Table {\ref{table:fitfuncn4}}.]{\includegraphics[scale=0.55]{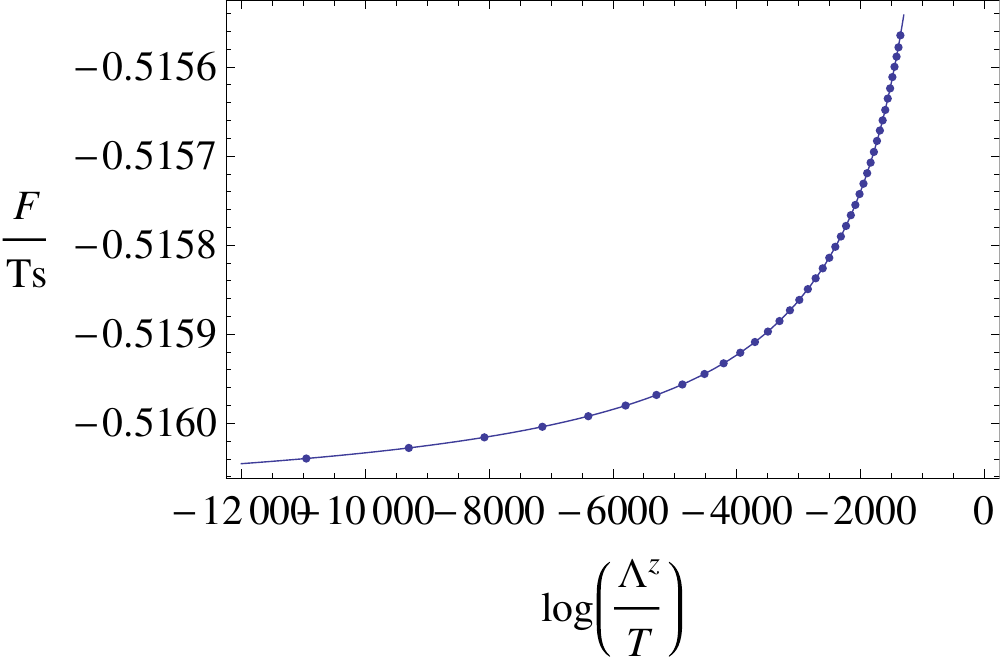}  \includegraphics[scale=0.55]{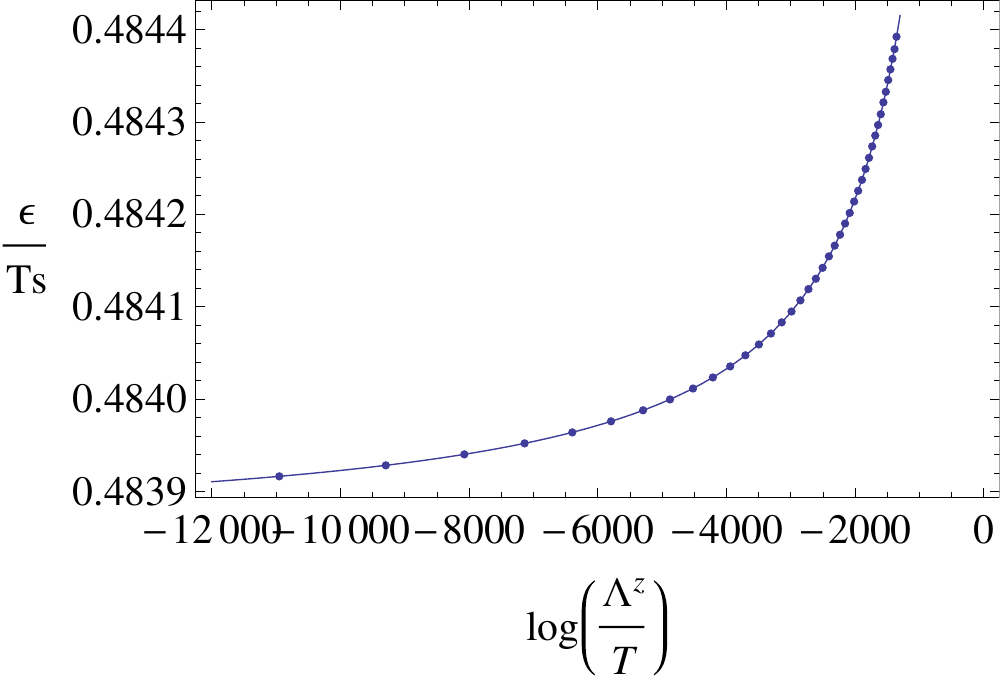} \includegraphics[scale=0.55]{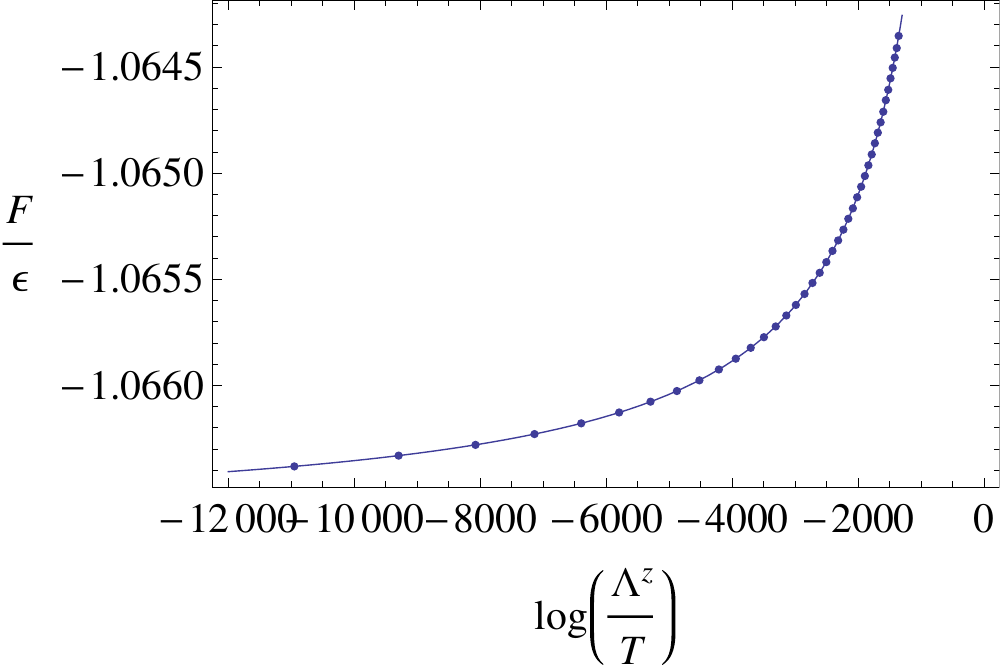}}\\
        \subfloat[For $\tilde{\alpha}=-\frac{1}{4}$ (or $z=4$). Dots are numerical results running $h_0$ from $2.66870$ to $2.66788$, which corresponds to $\log \Lambda^z/T$ from $-24978.2$ to $-1147.85$. Solid line is the fitting function denoted in Table {\ref{table:fitfuncn4}}.]{\includegraphics[scale=0.55]{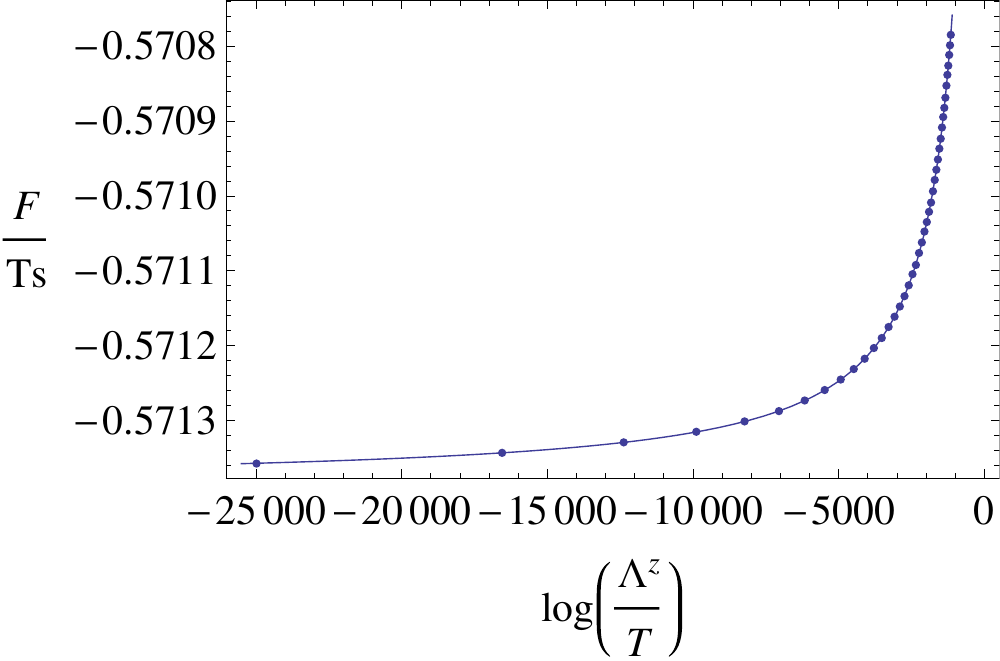} \includegraphics[scale=0.55]{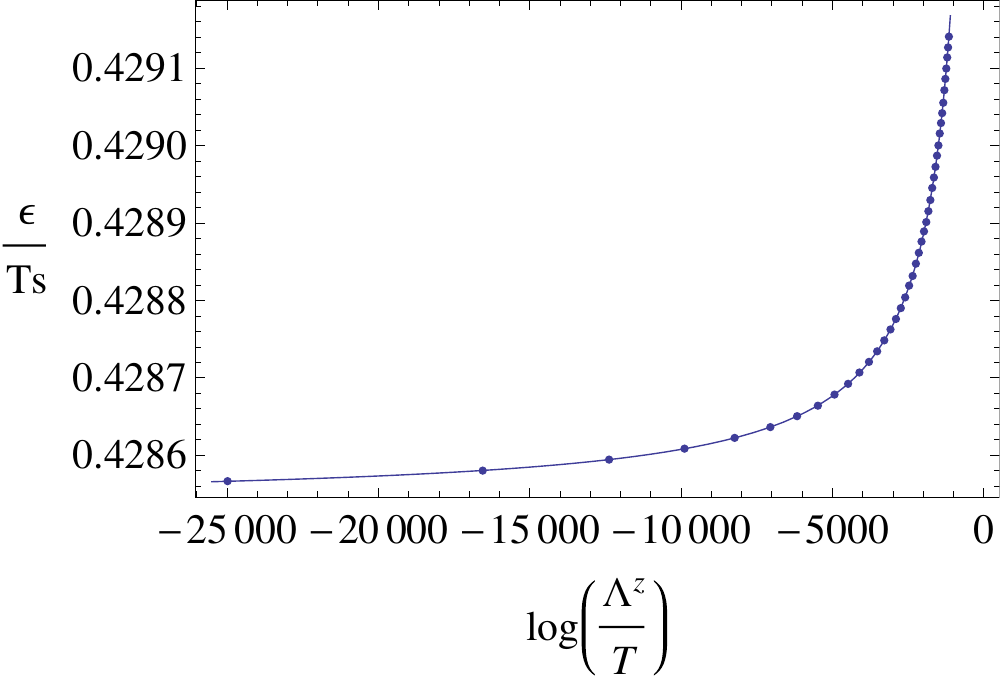} \includegraphics[scale=0.55]{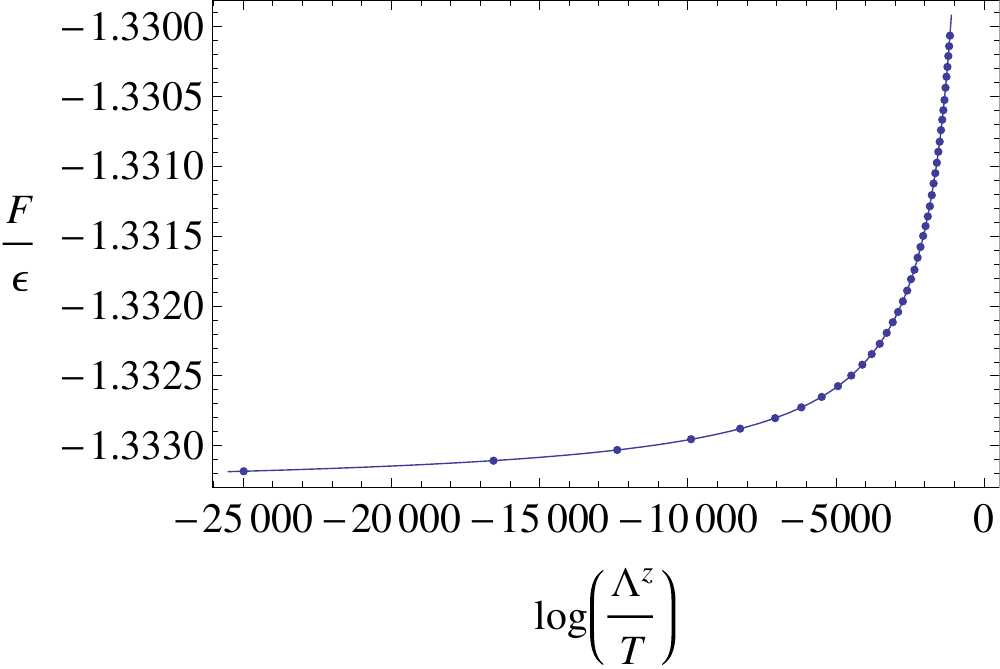}} \\
        \vspace{0.5cm}\\
        \subfloat[For $\tilde{\alpha}=-\frac{3}{10}$ (or $z=4.2$). Dots are numerical results running $h_0$ from $2.89170$ to $2.89088$, which corresponds to $\log \Lambda^z/T$ from $-8863.96$ to $-1051.13$. Solid line is the fitting function denoted in Table {\ref{table:fitfuncn4}}.]{\includegraphics[scale=0.55]{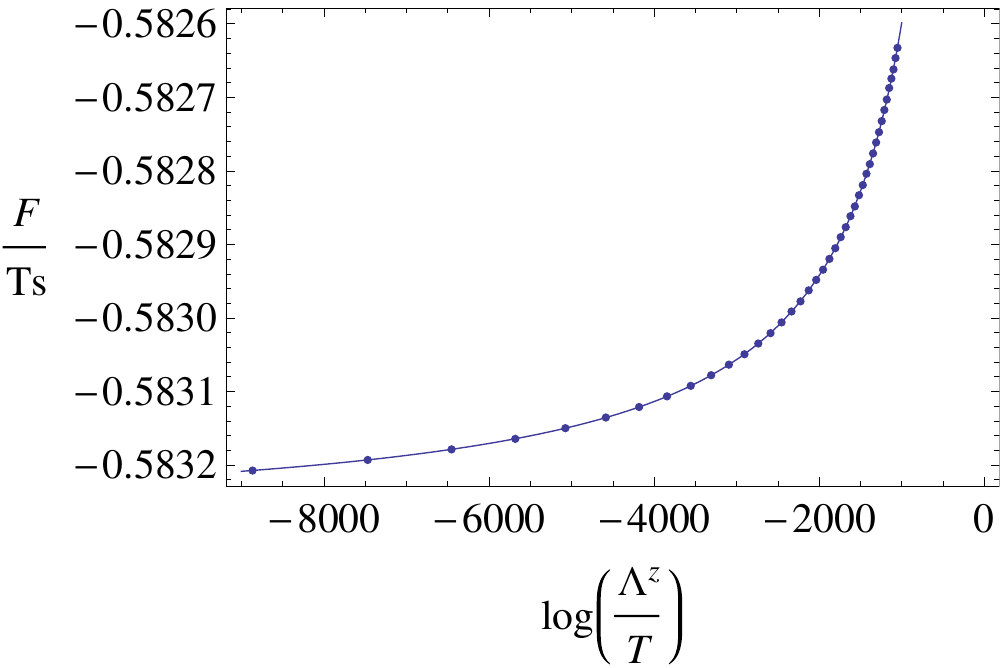} \includegraphics[scale=0.55]{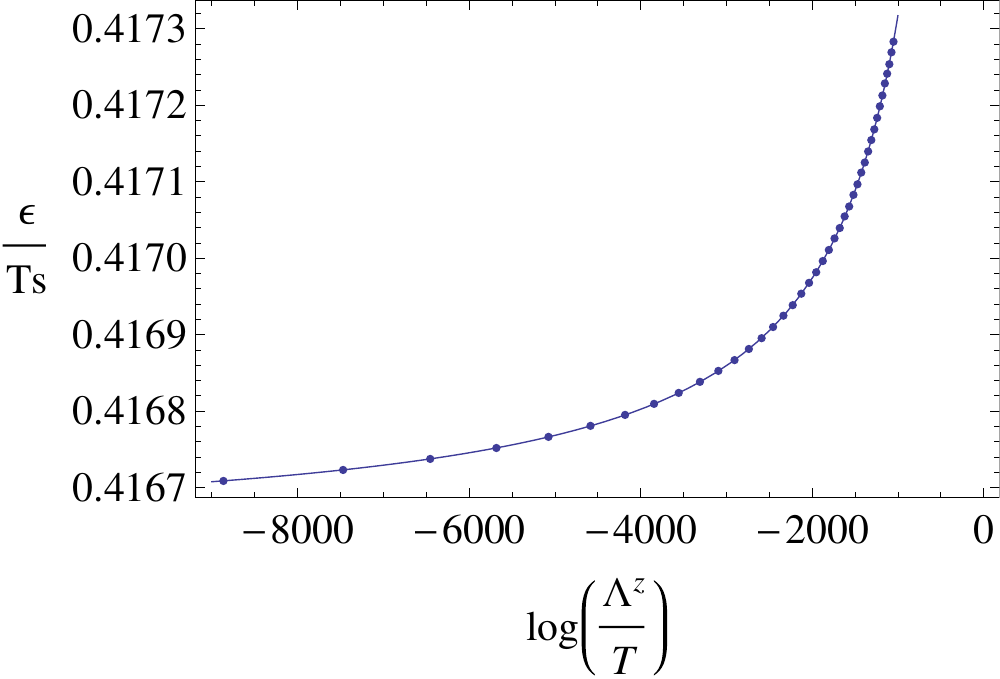} \includegraphics[scale=0.55]{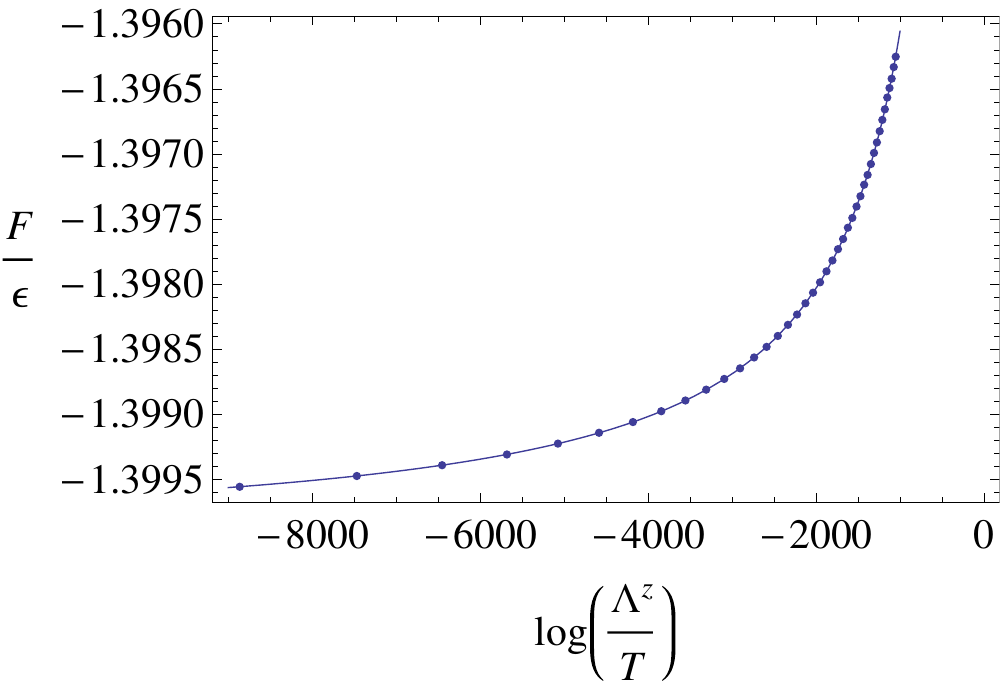}}\\
        \caption{Plots of ${\mathcal{F}}/Ts$ , ${\mathcal{E}}/Ts$ and ${\mathcal{F}}/{\mathcal{E}}$ versus $\log(\Lambda^z/T)$ for negative $\tilde{\alpha}$ for $n=4$.}
        \label{fig:frvsT2}
\end{figure}

\begin{figure}[!h]
        \subfloat[For $\tilde{\alpha}=\frac{1}{4}$ ($or z=2$) and $h_0$ runs from $0.75120$ to $0.75038$.]{\includegraphics[scale=0.85]{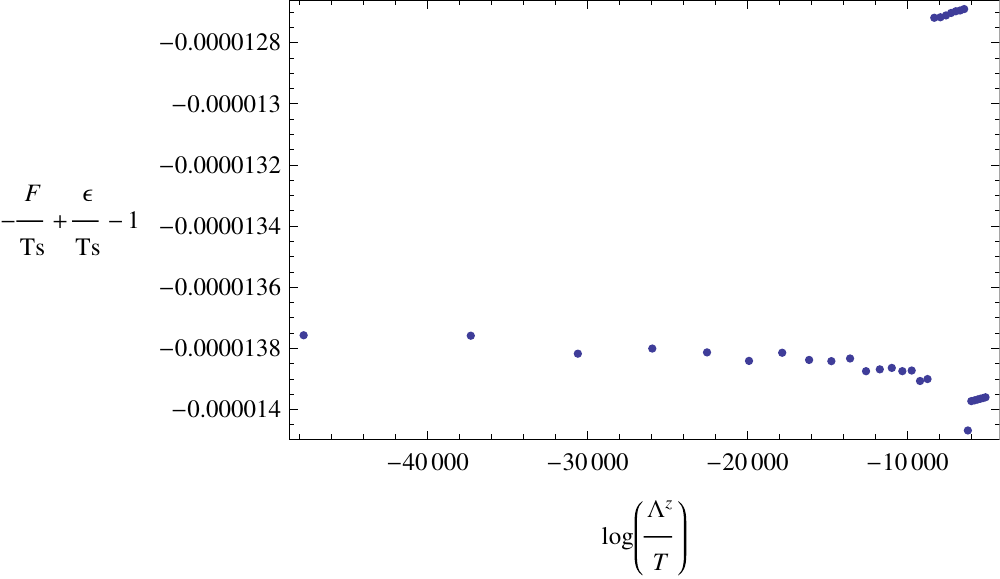}} \; \; \;
        \subfloat[For $\tilde{\alpha}=\frac{1}{10}$ (or $z=2.6$) and $h_0$ from $1.26160$ to $1.26078$.]{\includegraphics[scale=0.85]{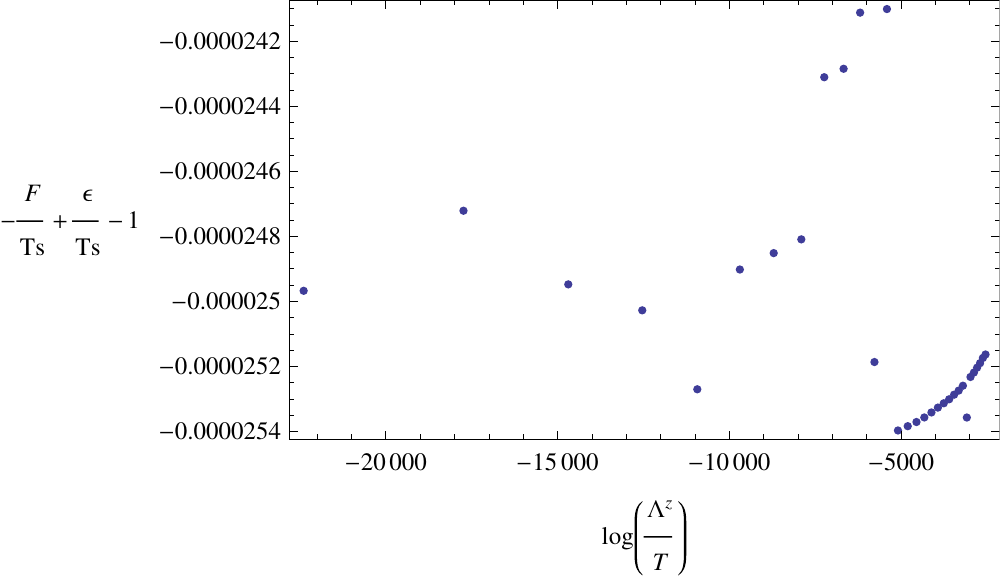}}\\
        \vspace{10pt}\\
        \subfloat[For $\tilde{\alpha}=0$ (or $z=3$) and $h_0$ runs from $1.63430$ to $1.63348$]{\includegraphics[scale=0.85]{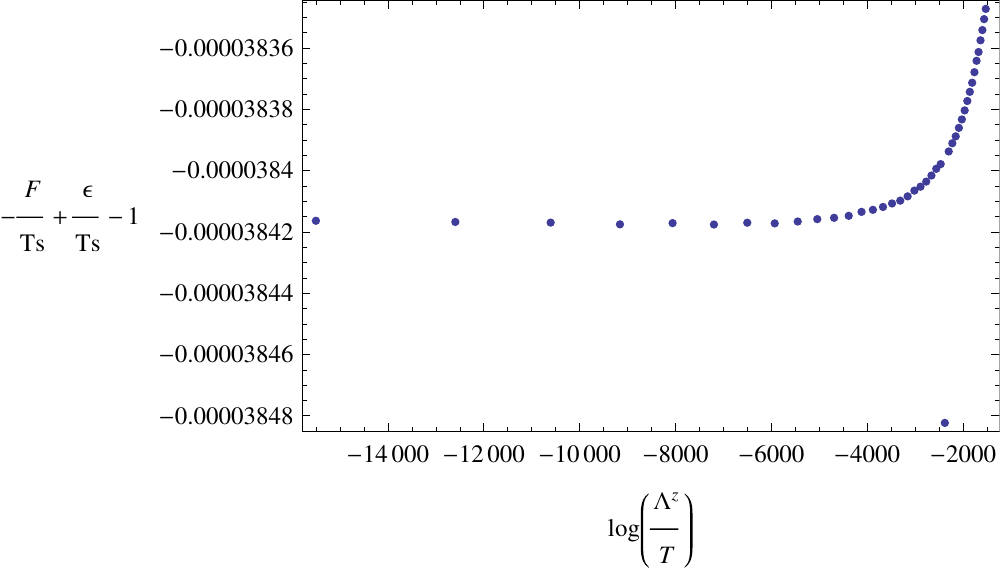}} \; \; \;
        \subfloat[For $\tilde{\alpha}=-\frac{1}{20}$ (or $z=3.2$) and $h_0$ runs from $1.82980$ to $1.82898$]{\includegraphics[scale=0.85]{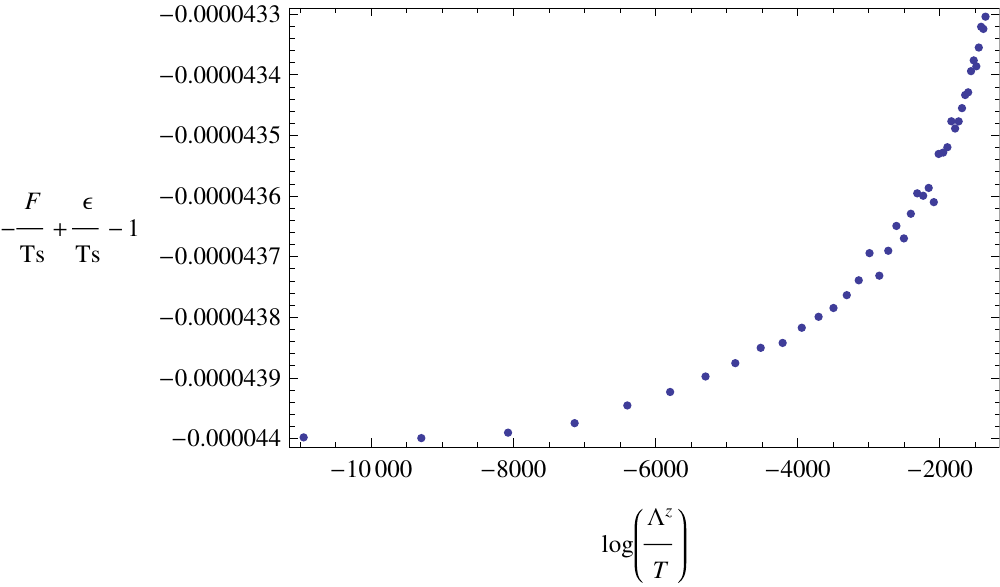}}\\
        \vspace{10pt}\\
        \subfloat[For $\tilde{\alpha}=-\frac{1}{4}$ (or $z=4$) and $h_0$ runs from $2.66870$ to $2.66788$]{\includegraphics[scale=0.85]{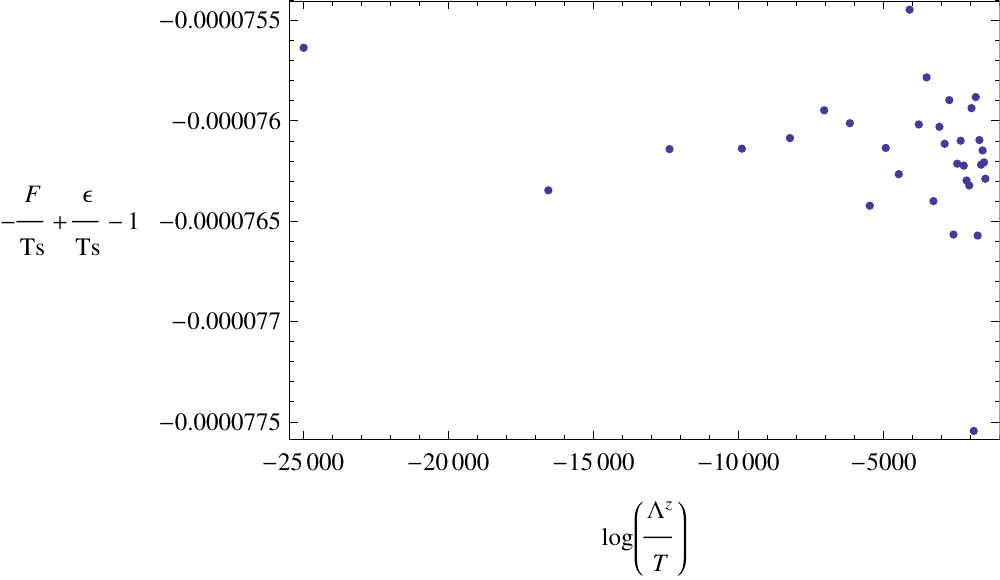}} \; \; \;
        \subfloat[For $\tilde{\alpha}=-\frac{3}{10}$ (or $z=4.2$). $h_0$ runs from $2.89170$ to $2.89088$]{\includegraphics[scale=0.85]{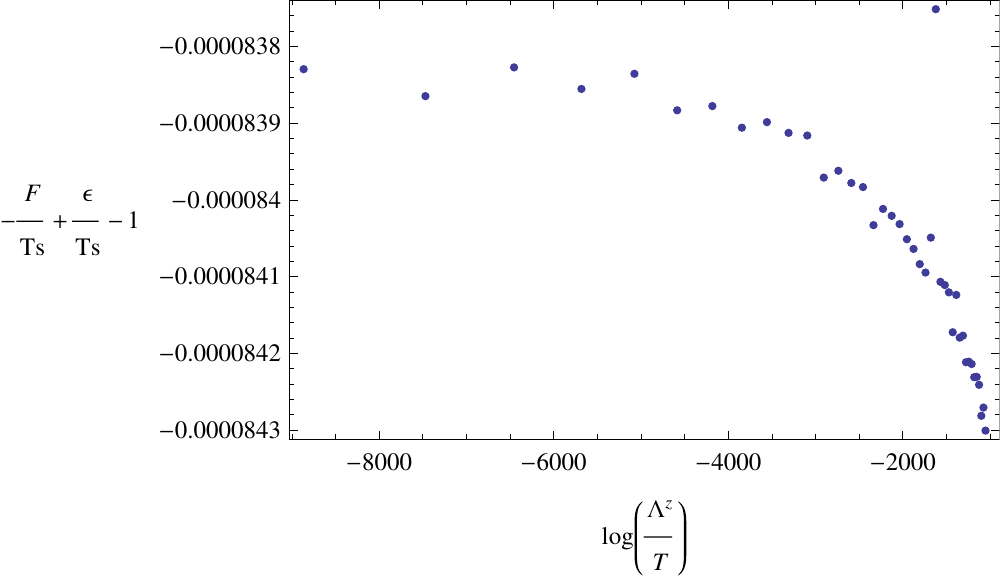}}\\
        \vspace{10pt}\\
        \caption{Plots of errors from $-\mathcal{F}/Ts + \mathcal{E}/Ts-1$}
        \label{fig:errors}
\end{figure}

Collecting  data for each of the values of $\{h_0, \tilde{\alpha}\}$, we  plot $\mathcal{F}/Ts$ and $\mathcal{E}/Ts$ as functions of $\log \Lambda^z/T$ and find their fitting functions. These results are displayed in Figures {\ref{fig:frvsT1}} and {\ref{fig:frvsT2}}, for which $n=4$; the dots are the numerical results and the solid line is the fitting function found. The fitting functions for each $n$ are recorded in Tables {\ref{table:fitfuncn4}}, {\ref{table:fitfuncn5}}, {\ref{table:fitfuncn6}}, {\ref{table:fitfuncn7}}, {\ref{table:fitfuncn8}}, and {\ref{table:fitfuncn9}}. We subsequently check that our numerical results for $\mathcal{F}/Ts$ and $\mathcal{E}/Ts$ with the marginally relevant modes  agree with the integrated first law of thermodynamics in (\ref{frslwthrm}) with an error of less than $10^{-4}$ as shown in Fig. {\ref{fig:errors}}.

\begin{figure}[!t]
    \centering{\includegraphics[scale=0.8]{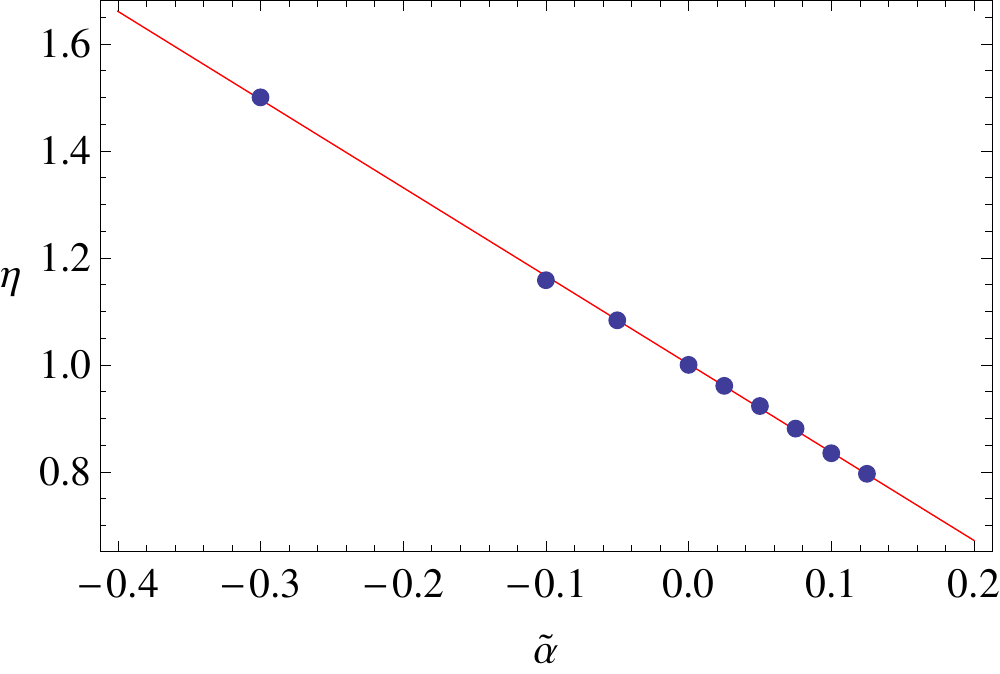}}
    \caption{For $n=4$, red dots are the numerical data and blue solid line is the fitting of the $\eta$ function}
        \label{fig:eta}
\end{figure}

It is straightforward to show using the data in the rightmost columns of tables {\ref{table:fitfuncn4}},
  {\ref{table:fitfuncn5}}, {\ref{table:fitfuncn6}}, {\ref{table:fitfuncn7}}, {\ref{table:fitfuncn8}}, and {\ref{table:fitfuncn9}},
that the values of $\mathcal{F}_0/\mathcal{E}_0 = - \eta(n,\tilde{\alpha})$ are consistent with the expression (\ref{eta}).
We illustrate the   $n=4$ case in Fig. \ref{fig:eta}, where the red dots are the numerical data and the solid line is the  expression for $\eta(4,\tilde{\alpha})$ given in (\ref{eta}). We recover the values of $\mathcal{F}_0/Ts$ and $\mathcal{E}_0/Ts$ in (\ref{tW2}) in the absence of the marginally relevant mode ($\Lambda=0$) at finite temperature.

\begin{table}[!b]
  \begin{tabular}{| c || >{\centering}m{3.6cm} | >{\centering}m{3.6cm} | >{\centering}m{3.6cm} m{0.1cm} |}
    \hline
      ($n=4$) & $\frac{{\mathcal{F}}}{Ts}$ & $ \frac{\mathcal{E}}{Ts} $ & $\frac{\mathcal{F}}{\mathcal{E}}$  & \\[11pt] \hline
    $\tilde{\alpha}=1/4$ or $z=2$ & $ \; -0.40 - \frac{1.21}{\log \Lambda^{z}/T} + \cdots $ & $ \; 0.60 - \frac{1.21}{\log \Lambda^{z}/T} + \cdots $ &  $ \; -0.67 - \frac{3.37}{\log \Lambda^{z}/T} + \cdots $ & \\[11pt] \hline
    $\tilde{\alpha}=1/10$ or $z=2.6$ & $ \; -0.46 - \frac{0.83}{\log \Lambda^{z}/T} + \cdots $ & $ \; 0.54 - \frac{0.83}{\log \Lambda^{z}/T} + \cdots $ & $ \;-0.87 - \frac{2.89}{\log \Lambda^{z}/T} + \cdots$ & \\[11pt] \hline
    $\tilde{\alpha}=0$ or $z=3$ & $ \; -0.50 - \frac{0.76}{\log \Lambda^{z}/T} + \cdots $ & $ \; 0.50 - \frac{0.76}{\log \Lambda^{z}/T} + \cdots $ & $ \; -1.0 - \frac{3.00}{\log \Lambda^{z}/T} + \cdots$ & \\[11pt] \hline
    $\tilde{\alpha}=-1/20$ or $z=3.2$ & $ \; -0.52 - \frac{0.73}{\log \Lambda^{z}/T} + \cdots $ & $ \; 0.48 - \frac{0.74}{\log \Lambda^{z}/T} + \cdots $ & $ \; -1.1 - \frac{3.13}{\log \Lambda^{z}/T} + \cdots$ & \\[11pt] \hline
    $\tilde{\alpha}=-1/4$ or $z=4$ & $ \; -0.57 - \frac{0.69}{\log \Lambda^{z}/T} + \cdots $ & $ \; 0.43 - \frac{0.69}{\log \Lambda^{z}/T} + \cdots $ & $ \; -1.3 - \frac{3.76}{\log \Lambda^{z}/T} + \cdots$ & \\[11pt] \hline
    $\tilde{\alpha}=-3/10$ or $z=4.2$& $ \; -0.58 - \frac{0.69}{\log \Lambda^{z}/T} + \cdots $ & $ \; 0.42 - \frac{0.69}{\log \Lambda^{z}/T} + \cdots $ & $ \; -1.4 - \frac{3.95}{\log \Lambda^{z}/T} + \cdots$ & \\[11pt] \hline
    $\vdots$ & $\vdots$ & $\vdots $ &  $\vdots$ & \\
    \hline
    \end{tabular}
    \caption{fitting functions for $\frac{{\mathcal{F}}}{Ts}$, $\frac{\mathcal{E}}{Ts}$, and $\frac{\mathcal{F}}{\mathcal{E}}$}
    \label{table:fitfuncn4}
\end{table}

We also observed how $\mathcal{F}/Ts$ and $\mathcal{E}/Ts$ behave with the marginally relevant modes according to different values of $n$ and $\tilde{\alpha}$ when $\Lambda^z/T \rightarrow 0$. When $\Lambda \neq 0$ the marginally relevant modes (generated by small values of $\Lambda$) contribute at sub-leading order to $\mathcal{F}/Ts$ and $\mathcal{E}/Ts$ as functions of $\log \Lambda^z/T$, where $\Lambda^z/T \rightarrow 0$. From Figures {\ref{fig:frvsT1}}, \ref{fig:frvsT2} and Tables {\ref{table:fitfuncn4}}, {\ref{table:fitfuncn5}}, {\ref{table:fitfuncn6}}, {\ref{table:fitfuncn7}}, {\ref{table:fitfuncn8}}, and {\ref{table:fitfuncn9}}, we see that for the same dimension $n$, as $z$ increases (or $\tilde{\alpha}$ decreases), this sub-leading contribution  decreases. Furthermore, for the same value of $\tilde{\alpha}$, as $z$ increases (or $n$ increases), this sub-leading contribution also decreases. That is, when $z$ increases due to either decreasing $\tilde{\alpha}$ or increasing $n$ by $z=n-1-2(n-2) \tilde{\alpha}$, this sub-leading contribution decreases.

\afterpage{\clearpage}

\section{Summary and Discussion}
\label{sec:Dscss}

Our main purpose has been to understand how the Gauss-Bonnet coupling constant $\tilde{\alpha}$, defined in (\ref{cplgct}), plays a role in the deformation of  Lifshitz holography in ($n+1$) dimensions.

First recall where $\tilde{\alpha}$ makes a contribution. In the Lifshitz spacetime without the Gauss-Bonnet term ($\tilde{\alpha}=0$), the anisotropy of the spacetime is governed by the massive vector field. This vector field allows a (non-trivial) marginally relevant mode only at $z=(n-1)$  \cite{Park:2012bv}. When the Gauss-Bonnet term is included in the Lifshitz gravity action, the (non-trivial) marginally relevant mode is now restricted by $z=n-1-2(n-2)\tilde{\alpha}$. In other words $z$ is determined not only by the massive vector field but also by the value of $\tilde{\alpha}$. Considering the marginal mode in the pure Lifshitz case (i.e. $\Lambda = 0$)  for  the metric (\ref{nstzmtr}), $\tilde{\alpha}$ contributes to the function $f = \frac{1}{r^{2z}}$ in the metric via $z$ and also modifies the charge of the massive vector field by $q^2 = \frac{z-1}{z}(1-2\tilde{\alpha})$. However, in the deformed Lifshitz spacetime (i.e. considering marginally relevant modes generated by $\Lambda \sim 0$), $\tilde{\alpha}$ modifies both   $f$ and $p$ in the metric with $\Lambda$ and involves the massive vector field, as shown in (\ref{aympksol})-(\ref{aympysol}) or (\ref{aympfsol})-(\ref{aymppsol}).

We found the asymptotic solutions in (\ref{aympksol})-(\ref{aympysol}), and obtained the free energy density in (\ref{fedfn}) and the energy density in (\ref{edfn}) by holographic renormalization. We next derived the near-horizon expansion of the black hole solution  in (\ref{hrsnslf})-(\ref{hrsnslh}), characterized by two parameters $h_0$ and $\tilde{\alpha}$, and calculated thermodynamic quantities such as temperature $T$ and entropy density $s$ in (\ref{thrmdnmqtt}). As our metric solutions contain undetermined parameters $\Lambda$ in the asymptotic solution and  $f_0$ and $p_0$ in the near horizon solution, we numerically integrate the horizon solution towards to boundary and match it with the asymptotic solution plotted towards the horizon by controlling $\Lambda$, $f_0$, and $p_0$ for given $\{h_0, \tilde{\alpha}\}$. After fixing those parameters, we numerically explored   physical quantities such as the free energy density and the energy density.

Numerically we found that there exists a maximum value of $h_0$, which corresponds to high temperature $\Lambda^z/T \rightarrow 0$. The smaller the value of $\tilde{\alpha}$ (or larger $z$) , the larger the maximum value of $h_0$. This explicitly demonstrates that $\tilde{\alpha}$ modifies $h_0$ or the horizon flux of the vector field via $h_0$. The dependence of the maximum value of $h_0$ both on $\tilde{\alpha}$ and $n$ is shown in Figure \ref{fig:hmaxvsn}, and for $n=4$ the maximum value of $h_0$ according to $z$ or $\alpha$ are depicted in Figures \ref{fig:hmaxvsza} (a) and (b) respectively.

Smaller values of $h_0$  correspond to decreased temperature. For a very small value of $h_0$, the zero temperature limit $\Lambda^z/T \rightarrow \infty$ is approached. Due to fixed $\Lambda \sim 0$, our spacetime can be configured  as a slightly deformed Lifshitz spacetime in the UV regime and as an AdS spacetime in IR regime. Hence we expect  holographic renormalization group flow at zero temperature from UV  Gauss-Bonnet-Lifshitz  to IR Gauss-Bonnet-AdS. Such renormalization flow was discovered in the AdS case {\cite{Braviner:2011kz}}, {\cite{Kachru:2008yh}}, but when $\tilde{\alpha} \neq 0$ the situation remains unknown, and an interesting problem for further research.

To investigate the behaviour of physical quantities with the metric solution of the deformed Lifshitz spacetime at high energy scales, which are expected to give   information concerning   marginally relevant operators near the critical regime at finite temperature via holographic duality, we confined ourselves to values of $h_0$ that are a little less than its  maximum value (which corresponds to the high temperature regime $\Lambda^z/T \rightarrow 0$), and obtained the numerical results   shown in Table {\ref{table:results}}. For $\tilde{\alpha} \geq 0$, the value of $z$ becomes less than $n-1$ -- this case appears to have stable properties as no unstable behaviour is observed for the $k$-function and the fitting for $\mathcal{F}/Ts$ and $\mathcal{E}/Ts$. However for $0 > \tilde{\alpha} >  -\frac{1}{2(n-2)}$, the value of $z$ is in the range  $n-1 < z < n$, and instability is expected since the free energy density is bigger than the energy density. Even worse properties emerge for $-\frac{1}{2(n-2)} \lesssim \tilde{\alpha} $, where $\mathcal{F}/Ts$ and $\mathcal{E}/Ts$ both have an oscillatory sharp peak near the horizon and the free energy density is bigger than the energy density. Here ``$\sim$" indicates near the value of $-\frac{1}{2(n-2)} \pm \epsilon$ or $n\pm \epsilon$, where $\epsilon$ is small.

When $\Lambda =0$, we analytically predicted the 0th order expressions for the free energy density and the energy density from  the integrated first law of thermodynamics and  the  trace Ward identity.  Our numerical results are consistent
with these expressions, indicating that the trace Ward identity is valid in the GB case (at least for $z=n-1-2(n-2)\tilde{\alpha}$).

Finally we found the sub-leading behaviour of $\mathcal{F}/Ts$ and $\mathcal{E}/Ts$ as a function of $\log \Lambda^z/T$, which is expected to be contribution of the marginally relevant mode at finite temperature. As the value of $z$ gets bigger due to either  decreasing $\tilde{\alpha}$ or increasing $n$, the factor associated with the sub-leading order (i.e. the magnitude of  the coefficient of $1/\log(\Lambda^z/T)$) becomes smaller as shown in Tables {\ref{table:fitfuncn4}}, {\ref{table:fitfuncn5}}, {\ref{table:fitfuncn6}},{\ref{table:fitfuncn7}}, {\ref{table:fitfuncn8}}, {\ref{table:fitfuncn9}}.

An interesting extension of the methods we have developed would be to find the holographic renormalization flow interpolating   GB-Lifshitz spacetime and   GB-AdS spacetime at zero temperature.  Work on this problem is in progress.

\begin{table}[!h]
  \begin{tabular}{|m{3cm} || m{4cm} | m{4cm} | m{4cm} m{0cm} |} \hline
    & \centering{$\tilde{\alpha} \geq 0$} & \centering{$ 0 > \tilde{\alpha} >  -\frac{1}{2(n-2)}$}   &  \centering{$ -\frac{1}{2(n-2)} \lesssim \tilde{\alpha} $} & \tabularnewline [10pt] \hline
    \textbf{value of $z$} &  \centering{$z \leq n-1$}   &  \centering{$n-1 < z < n$}  &  \centering{$n \lesssim z$} & \tabularnewline [10pt] \hline
    \textbf{$k$ function} &  Increases as the boundary is approached, as shown in Fig. {\ref{fig:mtchk}}-(a),(b),(c)  &  Increases as the boundary is approached, as shown in Fig. {\ref{fig:mtchk}}-(d),(e) &  Decreases as the boundary is approached, as shown in Fig. {\ref{fig:mtchk}}-(f) & \tabularnewline [40pt] \hline
    \textbf{$\mathcal{F}/Ts$ and $\mathcal{E}/Ts$ depending on $\log(r/r_+)$} & No unstable behaviour seen near the horizon in Fig. {\ref{fig:frednst1}}-(a),(b),(c) &  No unstable behaviour seen near the horizon in Fig. {\ref{fig:frednst2}}-(a)  &  Large oscillating peak; unstable behaviour observed near the horizon as shown in Fig. {\ref{fig:frednst2}}-(b),(c),(d) & \tabularnewline [40pt] \hline
    \textbf{Comparing magnitude of $|\mathcal{F}_0/Ts|$ and $|\mathcal{E}_0/Ts|$}  & $ \left|\frac{\mathcal{F}_0}{Ts}\right| \leq \left|\frac{\mathcal{E}_0}{Ts}\right|$ as shown in Table {\ref{table:fitfuncn4}},{\ref{table:fitfuncn5}},{\ref{table:fitfuncn6}},{\ref{table:fitfuncn7}},{\ref{table:fitfuncn8}} {\ref{table:fitfuncn9}} & $\left|\frac{\mathcal{F}_0}{Ts}\right| > \left|\frac{\mathcal{E}_0}{Ts}\right|$ as shown in Table {\ref{table:fitfuncn4}},{\ref{table:fitfuncn5}},{\ref{table:fitfuncn6}},{\ref{table:fitfuncn7}},{\ref{table:fitfuncn8}} {\ref{table:fitfuncn9}} &  $ \left|\frac{\mathcal{F}_0}{Ts}\right| > \left|\frac{\mathcal{E}_0}{Ts}\right|$ as shown in Table {\ref{table:fitfuncn4}},{\ref{table:fitfuncn5}},{\ref{table:fitfuncn6}},{\ref{table:fitfuncn7}},{\ref{table:fitfuncn8}}, {\ref{table:fitfuncn9}} & \tabularnewline [40pt] \hline
    \end{tabular}
    \caption{These results are obtained at high temperature (i.e. $h_0$ is near maximum or $\Lambda^z/T \rightarrow 0$).}
    \label{table:results}
\end{table}

\section*{Acknowledgements}
This work was supported in part by the Natural Sciences and Engineering Research Council of Canada.


\appendix

\section{X,Y, and Z in $\mathcal{F}$, $\mathcal{E}$, and $\mathcal{J}^{\hat{t}}$ in (\ref{fdwtcc})-(\ref{flwtcc})}
\label{sec:test}

\begin{align} \label{XYZ}
X =& z^3 (-1 + 2 z + 2 z^2 - 2 z^3) + z^2 (-6 + 27 z - 34 z^2 - 14 z^3 + 18 z^4) \tilde{\alpha} - z (12 - 101 z + 309 z^2 \nonumber\\
& - 406 z^3 + 144 z^4 + 12 z^5) \tilde{\alpha}^2 + (-8 + 120 z -574 z^2 + 1405 z^3 - 1842 z^4 + 1053 z^5 - 194 z^6) \tilde{\alpha}^3 \nonumber\\
& + (20 - 191 z + 693 z^2 - 1595 z^3 + 2383 z^4 - 1718 z^5 + 408 z^6) \tilde{\alpha}^4 -2 (-1 + 21 z - 150 z^2  \nonumber\\
& + 262 z^3 + 49 z^4  - 295 z^5 + 90 z^6) \tilde{\alpha}^5 + 4 (1 + z)^2 (2 - 17 z + 41 z^2 - 32 z^3 + 4 z^4) \tilde{\alpha}^6, \\
Y =& z^3 (-1 + 4 z - 4 z^2 + 2 z^3) + z^2 (-6 + 31 z - 70 z^2 + 66 z^3 - 30 z^4) \tilde{\alpha} + z (-12 + 95 z - 311 z^2 \nonumber\\
& + 560 z^3 - 502 z^4 + 200 z^5) \tilde{\alpha}^2 + (-8 + 112 z - 490 z^2 + 1209 z^3 - 1934 z^4 + 1689 z^5 - 618 z^6)\tilde{\alpha}^3 \nonumber\\
& + (28 - 215 z + 589 z^2 - 1155 z^3 + 2095 z^4 - 2134 z^5 + 792 z^6) \tilde{\alpha}^4 + (-14 + 38 z + 220 z^2 - 668 z^3 \nonumber\\
&  + 126 z^4 + 654 z^5 - 308 z^6) \tilde{\alpha}^5 + 4 (1 + z)^2 (2 - 17 z + 41 z^2 - 32 z^3 + 4 z^4) \tilde{\alpha}^6),\\
Z =& z^5 (-5 + 24 z - 23 z^2 - 22 z^3 + 18 z^4 + 20 z^5) + z^4 (11 + 23 z - 395 z^2 + 703 z^3 - 40 z^4 - 318 z^5 \nonumber\\
& - 176 z^6) \tilde{\alpha} + z^3 (1 - 140 z + 402 z^2 + 2050 z^3 - 6461 z^4 + 3610 z^5 + 1106 z^6 + 776 z^7) \tilde{\alpha}^2 \nonumber\\
& + z^2 (-37 + 105 z + 1198 z^2 - 7062 z^3 + 2299 z^4 + 25593 z^5 - 29948 z^6 + 6308 z^7 - 3832 z^8) \tilde{\alpha}^3 \nonumber\\
& + 2 z (53 - 351 z + 1059 z^2 - 4788 z^3 + 22016 z^4 - 31125 z^5 - 13341 z^6 + 49514 z^7 - 24299 z^8  \nonumber\\
& + 7982 z^9) \tilde{\alpha}^4 - 2 (-16 + 328 z - 2401 z^2 + 7052 z^3 - 16915 z^4 + 62838 z^5 - 119335 z^6 + 51832 z^7  \nonumber\\
& + 66435 z^8 - 58018 z^9 + 18952 z^10) \tilde{\alpha}^5 + 4 (-28 + 218 z - 1604 z^2 + 3369 z^3 - 1768 z^4 + 22651 z^5  \nonumber\\
&- 76922 z^6 + 67311 z^7 + 11872 z^8 - 31725 z^9 + 12002 z^10) \tilde{\alpha}^6 - 8 (-7 - 46 z - 23 z^2 - 1092 z^3  \nonumber\\
& + 8439 z^4 - 11616 z^5 - 8665 z^6 + 21748 z^7  - 2820 z^8 - 8226 z^9 + 3844 z^10) \tilde{\alpha}^7 + 16 (1 + z)^2 (-2 \nonumber\\
& + 23 z - 322 z^2 + 1396 z^3 - 2240 z^4 + 665 z^5 + 1724 z^6 - 1676 z^7 + 480 z^8) \tilde{\alpha}^8
\end{align}

\section{Table $\mathcal{F}/Ts$ and $\mathcal{E}/Ts$ for $n=5$-$9$}
\label{sec:test}

\begin{center}
\begin{table}[!h]
  \begin{tabular}{| c || m{3.6cm} | m{3.6cm} | m{3.6cm} m{0.1cm} |}
    \hline
      ($n=5$) & \centering{$\frac{{\mathcal{F}}}{Ts}$} & \centering{$\frac{\mathcal{E}}{Ts}$} & \centering{$\frac{\mathcal{F}}{\mathcal{E}}$}  & \\[11pt] \hline
    $\tilde{\alpha}=1/4$ or $z=2.5$ & $ \; -0.38 - \frac{1.29}{\log \Lambda^{z}/T} + \cdots $ & $ \; 0.62 - \frac{1.29}{\log \Lambda^{z}/T} + \cdots $ &  $ \; -0.63 - \frac{3.42}{\log \Lambda^{z}/T} + \cdots $ & \\[11pt] \hline
    $\tilde{\alpha}=1/10$ or $z=3.4$ & $ \; -0.46 - \frac{0.75}{\log \Lambda^{z}/T} + \cdots $ & $ \; 0.54 - \frac{0.75}{\log \Lambda^{z}/T} + \cdots $ & $ \;-0.85 - \frac{2.56}{\log \Lambda^{z}/T} + \cdots$ & \\[11pt] \hline
    $\tilde{\alpha}=0$ or $z=4$ & $ \; -0.50 - \frac{0.67}{\log \Lambda^{z}/T} + \cdots $ & $ \; 0.50 - \frac{0.67}{\log \Lambda^{z}/T} + \cdots $ & $ \; -1.00 - \frac{2.67}{\log \Lambda^{z}/T} + \cdots$ & \\[11pt] \hline
    $\tilde{\alpha}=-1/20$ or $z=4.3$ & $ \; -0.52 - \frac{0.65}{\log \Lambda^{z}/T} + \cdots $ & $ \; 0.48 - \frac{0.65}{\log \Lambda^{z}/T} + \cdots $ & $ \; -1.08 - \frac{2.78}{\log \Lambda^{z}/T} + \cdots$ & \\[11pt] \hline
    $\tilde{\alpha}=-1/6$ or $z=5$ & $ \; -0.56 - \frac{0.63}{\log \Lambda^{z}/T} + \cdots $ & $ \; 0.44 - \frac{0.62}{\log \Lambda^{z}/T} + \cdots $ & $ \; -1.25 - \frac{3.17}{\log \Lambda^{z}/T} + \cdots$ & \\[11pt] \hline
    $\tilde{\alpha}=-3/10$ or $z=5.8$& $ \; -0.59 - \frac{0.60}{\log \Lambda^{z}/T} + \cdots $ & $ \; 0.41 - \frac{0.61}{\log \Lambda^{z}/T} + \cdots $ & $ \; -1.45 - \frac{3.63}{\log \Lambda^{z}/T} + \cdots$ & \\[11pt] \hline
    \vdots & $\; \; \; \; \; \; \; \; \; \;  \; \; \; \; \; \; \; \; \; \; \vdots$ & $\; \; \; \; \; \; \; \; \; \; \; \; \; \; \; \; \; \; \; \; \vdots $ &  $\; \; \; \; \; \; \; \; \; \; \; \; \; \; \; \; \; \; \; \; \vdots$ & \\
    \hline
    \end{tabular}
    \caption{fitting functions for $\frac{{\mathcal{F}}}{Ts}$, $\frac{\mathcal{E}}{Ts}$, and $\frac{\mathcal{F}}{\mathcal{E}}$ in $n=5$}
    \label{table:fitfuncn5}
\end{table}
\end{center}

\begin{center}
\begin{table}[!h]
  \begin{tabular}{| c || m{3.6cm} | m{3.6cm} | m{3.6cm} m{0.1cm} |}
    \hline
      ($n=6$) & \centering{$\frac{{\mathcal{F}}}{Ts}$} & \centering{$\frac{\mathcal{E}}{Ts} $} & \centering{$\frac{\mathcal{F}}{\mathcal{E}}$}  & \\[11pt] \hline
    $\tilde{\alpha}=1/4$ or $z=3$ & $ \; -0.38 - \frac{1.43}{\log \Lambda^{z}/T} + \cdots $ & $ \; 0.63 - \frac{1.41}{\log \Lambda^{z}/T} + \cdots $ &  $ \; -0.60 - \frac{3.65}{\log \Lambda^{z}/T} + \cdots $ & \\[11pt] \hline
    $\tilde{\alpha}=1/10$ or $z=4.2$ & $ \; -0.46 - \frac{0.71}{\log \Lambda^{z}/T} + \cdots $ & $ \; 0.54 - \frac{0.71}{\log \Lambda^{z}/T} + \cdots $ & $ \;-0.84 - \frac{2.40}{\log \Lambda^{z}/T} + \cdots$ & \\[11pt] \hline
    $\tilde{\alpha}=0$ or $z=5$ & $ \; -0.50 - \frac{0.63}{\log \Lambda^{z}/T} + \cdots $ & $ \; 0.50 - \frac{0.63}{\log \Lambda^{z}/T} + \cdots $ & $ \; -1.00 - \frac{2.51}{\log \Lambda^{z}/T} + \cdots$ & \\[11pt] \hline
    $\tilde{\alpha}=-1/20$ or $z=5.4$ & $ \; -0.52 - \frac{0.60}{\log \Lambda^{z}/T} + \cdots $ & $ \; 0.48 - \frac{0.60}{\log \Lambda^{z}/T} + \cdots $ & $ \; -1.08 - \frac{2.61}{\log \Lambda^{z}/T} + \cdots$ & \\[11pt] \hline
    $\tilde{\alpha}=-1/8$ or $z=6$ & $ \; -0.55 - \frac{0.58}{\log \Lambda^{z}/T} + \cdots $ & $ \; 0.45 - \frac{0.58}{\log \Lambda^{z}/T} + \cdots $ & $ \; -1.20 - \frac{2.81}{\log \Lambda^{z}/T} + \cdots$ & \\[11pt] \hline
    $\tilde{\alpha}=-3/10$ or $z=7.4$& $ \; -0.60 - \frac{0.57}{\log \Lambda^{z}/T} + \cdots $ & $ \; 0.40 - \frac{0.57}{\log \Lambda^{z}/T} + \cdots $ & $ \; -1.48 - \frac{3.48}{\log \Lambda^{z}/T} + \cdots$ & \\[11pt] \hline
    \vdots & $\; \; \; \; \; \; \; \; \; \;  \; \; \; \; \; \; \; \; \; \; \vdots$ & $\; \; \; \; \; \; \; \; \; \; \; \; \; \; \; \; \; \; \; \; \vdots $ &  $\; \; \; \; \; \; \; \; \; \; \; \; \; \; \; \; \; \; \; \; \vdots$ & \\
    \hline
    \end{tabular}
    \caption{fitting functions for $\frac{{\mathcal{F}}}{Ts}$, $\frac{\mathcal{E}}{Ts}$, and $\frac{\mathcal{F}}{\mathcal{E}}$ in $n=6$}
    \label{table:fitfuncn6}
\end{table}
\end{center}

\begin{center}
\begin{table}[!h]
  \begin{tabular}{| c || m{3.6cm} | m{3.6cm} | m{3.6cm} m{0.1cm} |}
    \hline
      ($n=7$) & \centering{$\frac{{\mathcal{F}}}{Ts}$} & \centering{$ \frac{\mathcal{E}}{Ts} $} & \centering{$\frac{\mathcal{F}}{\mathcal{E}}$}  & \\[11pt] \hline
    $\tilde{\alpha}=1/4$ or $z=3.5$ & $ \; -0.37 - \frac{1.58}{\log \Lambda^{z}/T} + \cdots $ & $ \; 0.63 - \frac{1.59}{\log \Lambda^{z}/T} + \cdots $ &  $ \; -0.58 - \frac{3.97}{\log \Lambda^{z}/T} + \cdots $ & \\[11pt] \hline
    $\tilde{\alpha}=1/10$ or $z=5$ & $ \; -0.45 - \frac{0.68}{\log \Lambda^{z}/T} + \cdots $ & $ \; 0.55 - \frac{0.68}{\log \Lambda^{z}/T} + \cdots $ & $ \;-0.83 - \frac{2.30}{\log \Lambda^{z}/T} + \cdots$ & \\[11pt] \hline
    $\tilde{\alpha}=0$ or $z=6$ & $ \; -0.50 - \frac{0.60}{\log \Lambda^{z}/T} + \cdots $ & $ \; 0.50 - \frac{0.60}{\log \Lambda^{z}/T} + \cdots $ & $ \; -1.00 - \frac{2.40}{\log \Lambda^{z}/T} + \cdots$ & \\[11pt] \hline
    $\tilde{\alpha}=-1/20$ or $z=6.5$ & $ \; -0.52 - \frac{0.58}{\log \Lambda^{z}/T} + \cdots $ & $ \; 0.48 - \frac{0.58}{\log \Lambda^{z}/T} + \cdots $ & $ \; -1.08 - \frac{2.50}{\log \Lambda^{z}/T} + \cdots$ & \\[11pt] \hline
    $\tilde{\alpha}=-1/10$ or $z=7$ & $ \; -0.54 - \frac{0.56}{\log \Lambda^{z}/T} + \cdots $ & $ \; 0.46 - \frac{0.56}{\log \Lambda^{z}/T} + \cdots $ & $ \; -1.17 - \frac{2.63}{\log \Lambda^{z}/T} + \cdots$ & \\[11pt] \hline
    $\tilde{\alpha}=-3/10$ or $z=9$& $ \;-0.60 - \frac{0.54}{\log \Lambda^{z}/T} + \cdots $ & $ \; 0.40 - \frac{0.54}{\log \Lambda^{z}/T} + \cdots $ & $ \; -1.5 - \frac{3.37}{\log \Lambda^{z}/T} + \cdots $ & \\[11pt] \hline
    \vdots & $\; \; \; \; \; \; \; \; \; \;  \; \; \; \; \; \; \; \; \; \; \vdots$ & $\; \; \; \; \; \; \; \; \; \; \; \; \; \; \; \; \; \; \; \; \vdots $ &  $\; \; \; \; \; \; \; \; \; \; \; \; \; \; \; \; \; \; \; \; \vdots$ & \\
    \hline
    \end{tabular}
    \caption{fitting functions for $\frac{{\mathcal{F}}}{Ts}$, $\frac{\mathcal{E}}{Ts}$, and $\frac{\mathcal{F}}{\mathcal{E}}$ in $n=7$}
    \label{table:fitfuncn7}
\end{table}
\end{center}

\begin{center}
\begin{table}[!h]
  \begin{tabular}{| c || m{3.6cm} | m{3.6cm} | m{3.6cm} m{0.1cm} |}
    \hline
      ($n=8$) & \centering{$\frac{{\mathcal{F}}}{Ts}$} & \centering{$\frac{\mathcal{E}}{Ts}$} & \centering{$\frac{\mathcal{F}}{\mathcal{E}}$}  & \\[11pt] \hline
    $\tilde{\alpha}=1/4$ or $z=4$ & $ \; -0.36 - \frac{1.76}{\log \Lambda^{z}/T} + \cdots $ & $ \; 0.64 - \frac{1.73}{\log \Lambda^{z}/T} + \cdots $ &  $ \; -0.57 - \frac{4.32}{\log \Lambda^{z}/T} + \cdots $ & \\[11pt] \hline
    $\tilde{\alpha}=1/10$ or $z=5.8$ & $ \; -0.45 - \frac{0.66}{\log \Lambda^{z}/T} + \cdots $ & $ \; 0.55 - \frac{0.67}{\log \Lambda^{z}/T} + \cdots $ & $ \;-0.83 - \frac{2.23}{\log \Lambda^{z}/T} + \cdots$ & \\[11pt] \hline
    $\tilde{\alpha}=0$ or $z=7$ & $ \; -0.50 - \frac{0.58}{\log \Lambda^{z}/T} + \cdots $ & $ \; 0.50 - \frac{0.58}{\log \Lambda^{z}/T} + \cdots $ & $ \; -1.00 - \frac{2.33}{\log \Lambda^{z}/T} + \cdots$ & \\[11pt] \hline
    $\tilde{\alpha}=-1/20$ or $z=7.6$ & $ \; -0.52 - \frac{0.56}{\log \Lambda^{z}/T} + \cdots $ & $ \; 0.48 - \frac{0.56}{\log \Lambda^{z}/T} + \cdots $ & $ \; -1.09 - \frac{2.45}{\log \Lambda^{z}/T} + \cdots$ & \\[11pt] \hline
    $\tilde{\alpha}=-1/12$ or $z=8$ & $ \; -0.53 - \frac{0.56}{\log \Lambda^{z}/T} + \cdots $ & $ \; 0.47 - \frac{0.55}{\log \Lambda^{z}/T} + \cdots $ & $ \; -1.14 - \frac{2.53}{\log \Lambda^{z}/T} + \cdots$ & \\[11pt] \hline
    $\tilde{\alpha}=-3/10$ or $z=10.6$& $ \; -0.60 - \frac{0.53}{\log \Lambda^{z}/T} + \cdots $ & $ \; 0.40 - \frac{0.53}{\log \Lambda^{z}/T} + \cdots $ & $ \;-1.51 - \frac{3.35}{\log \Lambda^{z}/T} + \cdots $ & \\[11pt] \hline
    \vdots & $\; \; \; \; \; \; \; \; \; \;  \; \; \; \; \; \; \; \; \; \vdots$ & $\; \; \; \; \; \; \; \; \; \; \; \; \; \; \; \; \; \; \; \; \vdots $ &  $\; \; \; \; \; \; \; \; \; \; \; \; \; \; \; \; \; \; \; \; \vdots$ & \\
    \hline
    \end{tabular}
    \caption{fitting functions for $\frac{{\mathcal{F}}}{Ts}$, $\frac{\mathcal{E}}{Ts}$, and $\frac{\mathcal{F}}{\mathcal{E}}$ in $n=8$}
    \label{table:fitfuncn8}
\end{table}
\end{center}

\begin{center}
\begin{table}[h!]
  \begin{tabular}{| c || m{3.6cm} | m{3.6cm} | m{3.6cm} m{0.1cm} |}
    \hline
      ($n=9$) & \centering{$\frac{{\mathcal{F}}}{Ts}$} & \centering{$ \frac{\mathcal{E}}{Ts} $} & \centering{$\frac{\mathcal{F}}{\mathcal{E}}$}  & \\[11pt] \hline
    $\tilde{\alpha}=1/4$ or $z=4.5$ & $ \; -0.36 - \frac{1.89}{\log \Lambda^{?}/T} + \cdots $ & $ \; 0.64 - \frac{1.89}{\log \Lambda^{z}/T} + \cdots $ &  $ \; -0.56 - \frac{4.59}{\log \Lambda^{z}/T} + \cdots $ & \\[11pt] \hline
    $\tilde{\alpha}=1/10$ or $z=6.6$ & $ \; -0.45 - \frac{0.66}{\log \Lambda^{z}/T} + \cdots $ & $ \; 0.55 - \frac{0.66}{\log \Lambda^{z}/T} + \cdots $ & $ \;-0.83 - \frac{2.19}{\log \Lambda^{z}/T} + \cdots$ & \\[11pt] \hline
    $\tilde{\alpha}=0$ or $z=8$ & $ \; -0.50 - \frac{0.57}{\log \Lambda^{z}/T} + \cdots $ & $ \; 0.50 - \frac{0.57}{\log \Lambda^{z}/T} + \cdots $ & $ \; -1.0 - \frac{2.29}{\log \Lambda^{z}/T} + \cdots$ & \\[11pt] \hline
    $\tilde{\alpha}=-1/20$ or $z=8.7$ & $ \; -0.52 - \frac{0.55}{\log \Lambda^{z}/T} + \cdots $ & $ \; 0.48 - \frac{0.55}{\log \Lambda^{z}/T} + \cdots $ & $ \; -1.09 - \frac{2.40}{\log \Lambda^{z}/T} + \cdots$ & \\[11pt] \hline
    $\tilde{\alpha}=-1/14$ or $z=9$ & $ \; -0.53 - \frac{0.54}{\log \Lambda^{z}/T} + \cdots $ & $ \; 0.47 - \frac{0.54}{\log \Lambda^{z}/T} + \cdots $ & $ \; -1.13 - \frac{2.46}{\log \Lambda^{z}/T} + \cdots$ & \\[11pt] \hline
    $\tilde{\alpha}=-3/10$ or $z=12.2$& $ \; -0.60 - \frac{0.51}{\log \Lambda^{z}/T} + \cdots $ & $ \; 0.40 - \frac{0.51}{\log \Lambda^{z}/T} + \cdots $ & $ \;-1.53 - \frac{3.28}{\log \Lambda^{z}/T} + \cdots $ & \\[11pt] \hline
    \vdots & $\; \; \; \; \; \; \; \; \; \;  \; \; \; \; \; \; \; \; \; \vdots$ & $\; \; \; \; \; \; \; \; \; \; \; \; \; \; \; \; \; \; \; \; \vdots $ &  $\; \; \; \; \; \; \; \; \; \; \; \; \; \; \; \; \; \; \; \; \vdots$ & \\
    \hline
    \end{tabular}
    \caption{fitting functions for $\frac{{\mathcal{F}}}{Ts}$, $\frac{\mathcal{E}}{Ts}$, and $\frac{\mathcal{F}}{\mathcal{E}}$ in $n=9$}
    \label{table:fitfuncn9}
\end{table}
\end{center}

\afterpage{\clearpage}
\newpage
\section{Graph of Entropy Density depending on $\log(\Lambda^z/T)$}
\label{sec:test}
 \begin{figure}[!h]
        \subfloat[For $\tilde{\alpha}=\frac{1}{4}$ ($or z=2$) and $h_0$ runs from $0.75120$ to $0.75038$ in increments of 0.00002. Dots are the numerical results and joined by straight line.]{\includegraphics[scale=0.7]{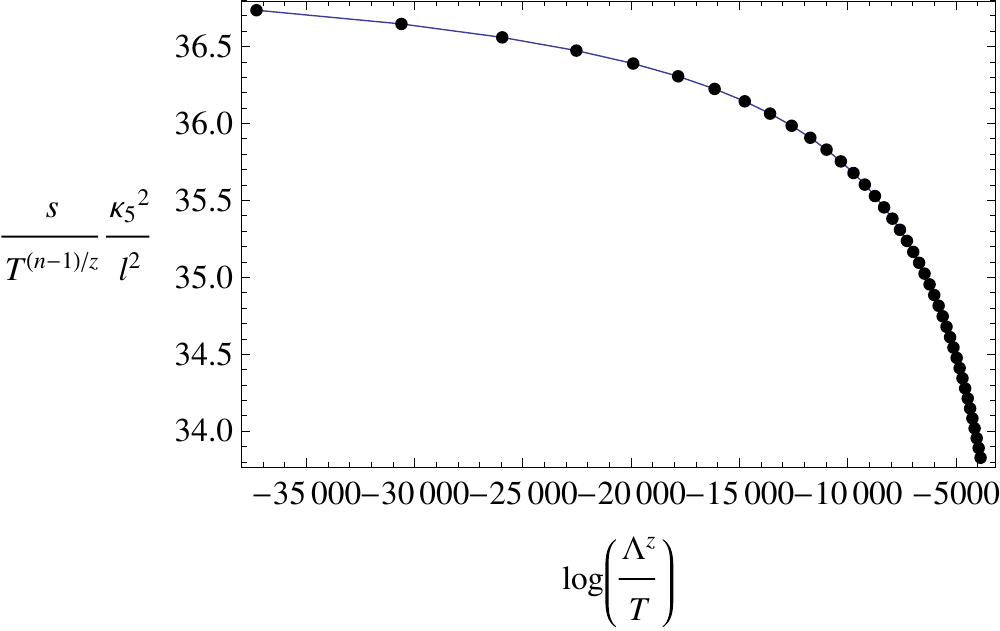}} \; \; \; \; \; \subfloat[For $\tilde{\alpha}=\frac{1}{10}$ (or $z=2.6$) and $h_0$ from $1.26160$ to $1.26078$ in increments of 0.00002. Dots are the numerical results and joined by straight line.]{\includegraphics[scale=0.7]{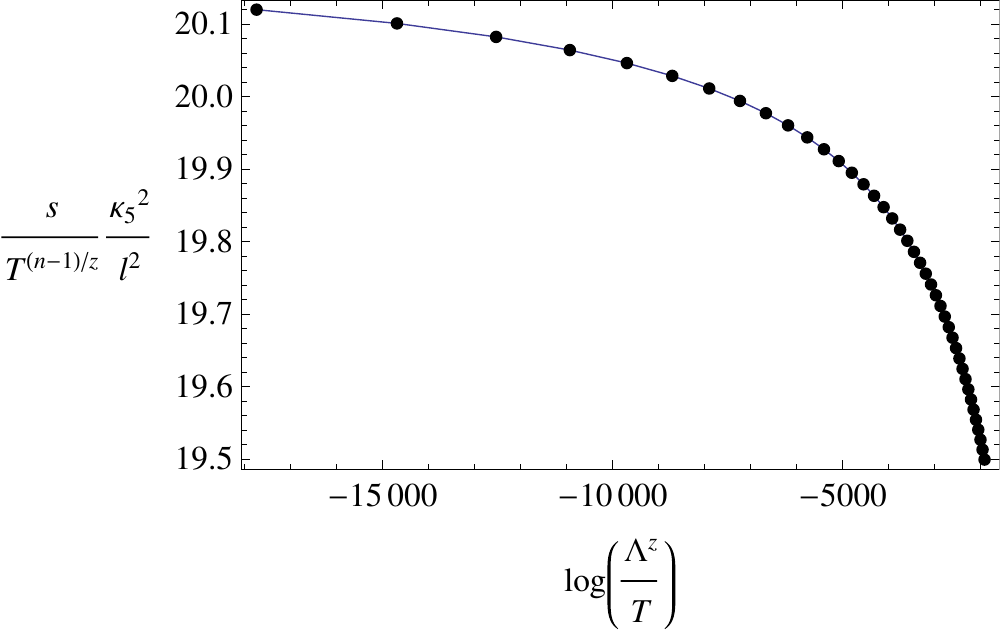}} \\
        \vspace{5pt}\\
        \subfloat[For $\tilde{\alpha}=0$ (or $z=3$) and $h_0$ runs from $1.63430$ to $1.63348$ in increments of 0.00002. Dots are the numerical results and joined by straight line.]{\includegraphics[scale=0.7]{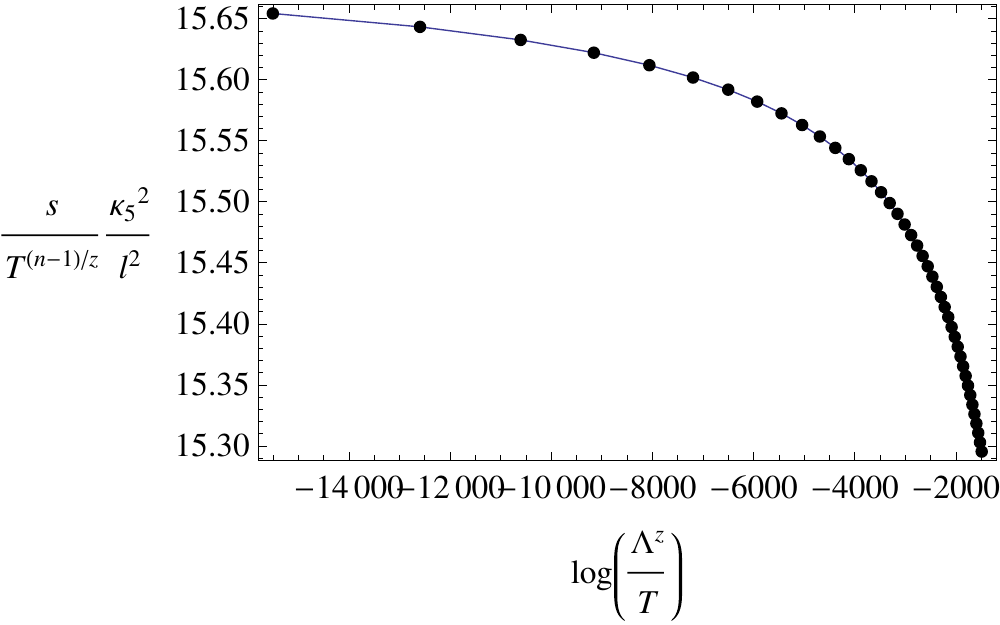}} \; \; \; \; \; \subfloat[For $\tilde{\alpha}=-\frac{1}{20}$ (or $z=3.2$) and $h_0$ runs from $1.82980$ to $1.82898$ in increments of 0.00002. Dots are the numerical results and joined by straight line.]{\includegraphics[scale=0.7]{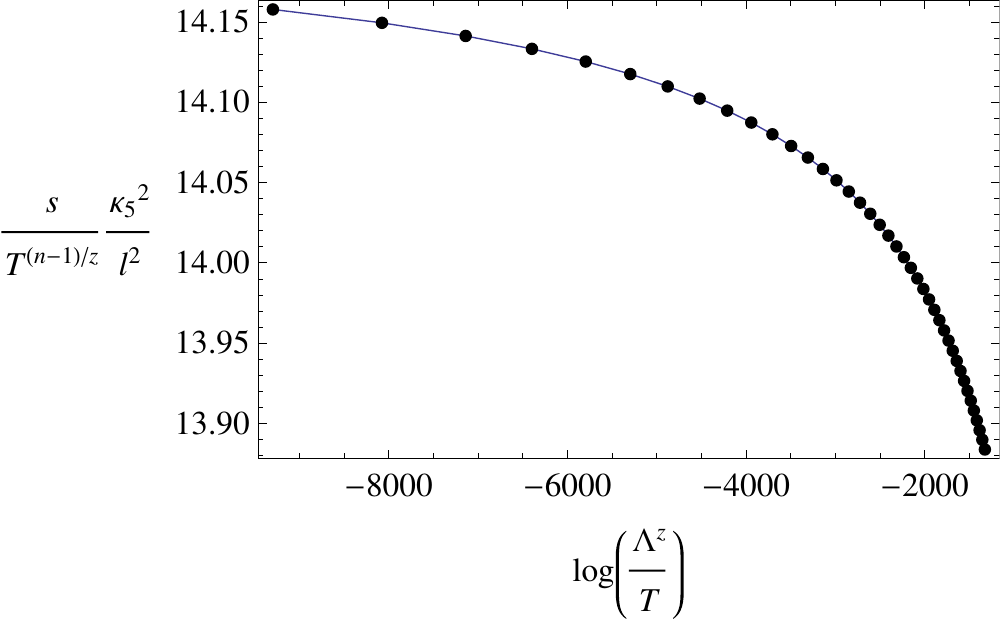}} \\
        \vspace{5pt}\\
        \subfloat[For $\tilde{\alpha}=-\frac{1}{4}$ (or $z=4$) and $h_0$ runs from $2.66870$ to $2.66788$ in increments of 0.00002. Dots are the numerical results and joined by straight line.]{\includegraphics[scale=0.7]{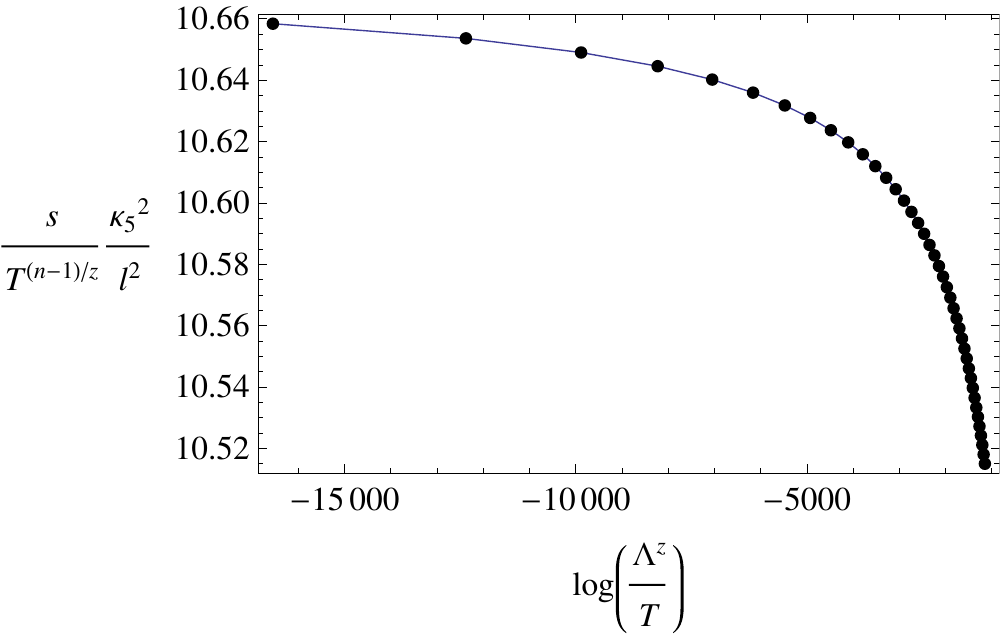}} \; \; \; \; \; \subfloat[For $\tilde{\alpha}=-\frac{3}{10}$ (or $z=4.2$). $h_0$ runs from $2.89170$ to $2.89088$ in increments of 0.00002. Dots are the numerical results and joined by straight line.]{\includegraphics[scale=0.7]{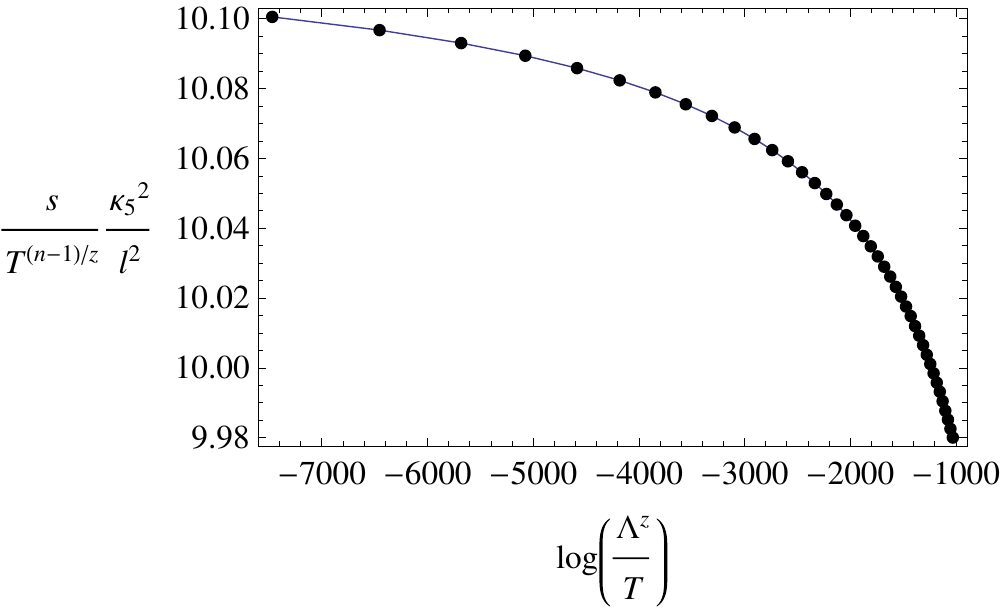}}
        \caption{entropy density per unit temperature on $\log \Lambda^z/T$ for each $\tilde{\alpha}$ for $n=4$}
        \label{fig:ntrp}
\end{figure}



\newpage
\bibliographystyle{JHEP}

\bibliography{GDLHGB}

\providecommand{\href}[2]{#2}\begingroup\raggedright\begin{thebibliography}{10}

\bibitem{Maldacena:1997re}
J.~M. Maldacena, {\it {The Large N limit of superconformal field theories and
  supergravity}},  {\em Adv.Theor.Math.Phys.} {\bf 2} (1998) 231--252,
  [\href{http://xxx.lanl.gov/abs/hep-th/9711200}{{\tt hep-th/9711200}}].

\bibitem{Forkel:2007cm}
H.~Forkel, M.~Beyer, and T.~Frederico, {\it {Linear square-mass trajectories of
  radially and orbitally excited hadrons in holographic QCD}},  {\em JHEP} {\bf
  0707} (2007) 077, [\href{http://xxx.lanl.gov/abs/0705.1857}{{\tt
  arXiv:0705.1857}}].

\bibitem{Gursoy:2007cb}
U.~Gursoy and E.~Kiritsis, {\it {Exploring improved holographic theories for
  QCD: Part I}},  {\em JHEP} {\bf 0802} (2008) 032,
  [\href{http://xxx.lanl.gov/abs/0707.1324}{{\tt arXiv:0707.1324}}].

\bibitem{Hoyos:2006gb}
C.~Hoyos-Badajoz, K.~Landsteiner, and S.~Montero, {\it {Holographic meson
  melting}},  {\em JHEP} {\bf 0704} (2007) 031,
  [\href{http://xxx.lanl.gov/abs/hep-th/0612169}{{\tt hep-th/0612169}}].

\bibitem{Bertoldi:2009dt}
G.~Bertoldi, B.~A. Burrington, and A.~W. Peet, {\it {Thermodynamics of black
  branes in asymptotically Lifshitz spacetimes}},  {\em Phys.Rev.} {\bf D80}
  (2009) 126004, [\href{http://xxx.lanl.gov/abs/0907.4755}{{\tt
  arXiv:0907.4755}}].

\bibitem{Braviner:2011kz}
H.~Braviner, R.~Gregory, and S.~F. Ross, {\it {Flows involving Lifshitz
  solutions}},  {\em Class.Quant.Grav.} {\bf 28} (2011) 225028,
  [\href{http://xxx.lanl.gov/abs/1108.3067}{{\tt arXiv:1108.3067}}].

\bibitem{Hartnoll:2009sz}
S.~A. Hartnoll, {\it {Lectures on holographic methods for condensed matter
  physics}},  {\em Class.Quant.Grav.} {\bf 26} (2009) 224002,
  [\href{http://xxx.lanl.gov/abs/0903.3246}{{\tt arXiv:0903.3246}}].

\bibitem{Horowitz:2010gk}
G.~T. Horowitz, {\it {Theory of Superconductivity}},  {\em Lect.Notes Phys.}
  {\bf 828} (2011) 313--347, [\href{http://xxx.lanl.gov/abs/1002.1722}{{\tt
  arXiv:1002.1722}}].

\bibitem{Kachru:2008yh}
S.~Kachru, X.~Liu, and M.~Mulligan, {\it {Gravity Duals of Lifshitz-like Fixed
  Points}},  {\em Phys.Rev.} {\bf D78} (2008) 106005,
  [\href{http://xxx.lanl.gov/abs/0808.1725}{{\tt arXiv:0808.1725}}].

\bibitem{Mann:2011hg}
R.~B. Mann and R.~McNees, {\it {Holographic Renormalization for Asymptotically
  Lifshitz Spacetimes}},  {\em JHEP} {\bf 1110} (2011) 129,
  [\href{http://xxx.lanl.gov/abs/1107.5792}{{\tt arXiv:1107.5792}}].

\bibitem{Ross:2011gu}
S.~F. Ross, {\it {Holography for asymptotically locally Lifshitz spacetimes}},
  {\em Class.Quant.Grav.} {\bf 28} (2011) 215019,
  [\href{http://xxx.lanl.gov/abs/1107.4451}{{\tt arXiv:1107.4451}}].

\bibitem{Bredberg:2010ky}
I.~Bredberg, C.~Keeler, V.~Lysov, and A.~Strominger, {\it {Wilsonian Approach
  to Fluid/Gravity Duality}},  {\em JHEP} {\bf 1103} (2011) 141,
  [\href{http://xxx.lanl.gov/abs/1006.1902}{{\tt arXiv:1006.1902}}].

\bibitem{Janik:2005zt}
R.~A. Janik and R.~B. Peschanski, {\it {Asymptotic perfect fluid dynamics as a
  consequence of Ads/CFT}},  {\em Phys.Rev.} {\bf D73} (2006) 045013,
  [\href{http://xxx.lanl.gov/abs/hep-th/0512162}{{\tt hep-th/0512162}}].

\bibitem{Rangamani:2009xk}
M.~Rangamani, {\it {Gravity and Hydrodynamics: Lectures on the fluid-gravity
  correspondence}},  {\em Class.Quant.Grav.} {\bf 26} (2009) 224003,
  [\href{http://xxx.lanl.gov/abs/0905.4352}{{\tt arXiv:0905.4352}}].

\bibitem{Cheng:2009df}
M.~C. Cheng, S.~A. Hartnoll, and C.~A. Keeler, {\it {Deformations of Lifshitz
  holography}},  {\em JHEP} {\bf 1003} (2010) 062,
  [\href{http://xxx.lanl.gov/abs/0912.2784}{{\tt arXiv:0912.2784}}].

\bibitem{Park:2012mn}
M.~Park and R.~B. Mann, {\it {Deformations of Lifshitz Holography in
  $(n+1)$-dimensions}},  {\em JHEP} {\bf 1207} (2012) 173,
  [\href{http://xxx.lanl.gov/abs/1202.3944}{{\tt arXiv:1202.3944}}].

\bibitem{Boulware:1985wk}
D.~G. Boulware and S.~Deser, {\it {String Generated Gravity Models}},  {\em
  Phys.Rev.Lett.} {\bf 55} (1985) 2656.

\bibitem{Wheeler:1985nh}
J.~T. Wheeler, {\it {Symmetric Solutions to the Gauss-Bonnet Extended Einstein
  Equations}},  {\em Nucl.Phys.} {\bf B268} (1986) 737.

\bibitem{Cai:2001dz}
R.-G. Cai, {\it {Gauss-Bonnet black holes in AdS spaces}},  {\em Phys.Rev.}
  {\bf D65} (2002) 084014, [\href{http://xxx.lanl.gov/abs/hep-th/0109133}{{\tt
  hep-th/0109133}}].

\bibitem{Cho:2002hq}
Y.~Cho and I.~P. Neupane, {\it {Anti-de Sitter black holes, thermal phase
  transition and holography in higher curvature gravity}},  {\em Phys.Rev.}
  {\bf D66} (2002) 024044, [\href{http://xxx.lanl.gov/abs/hep-th/0202140}{{\tt
  hep-th/0202140}}].

\bibitem{Cvetic:2001bk}
M.~Cvetic, S.~Nojiri, and S.~D. Odintsov, {\it {Black hole thermodynamics and
  negative entropy in de Sitter and anti-de Sitter Einstein-Gauss-Bonnet
  gravity}},  {\em Nucl.Phys.} {\bf B628} (2002) 295--330,
  [\href{http://xxx.lanl.gov/abs/hep-th/0112045}{{\tt hep-th/0112045}}].

\bibitem{Neupane:2002bf}
I.~P. Neupane, {\it {Black hole entropy in string generated gravity models}},
  {\em Phys.Rev.} {\bf D67} (2003) 061501,
  [\href{http://xxx.lanl.gov/abs/hep-th/0212092}{{\tt hep-th/0212092}}].

\bibitem{Jacobson:1993xs}
T.~Jacobson and R.~C. Myers, {\it {Black hole entropy and higher curvature
  interactions}},  {\em Phys.Rev.Lett.} {\bf 70} (1993) 3684--3687,
  [\href{http://xxx.lanl.gov/abs/hep-th/9305016}{{\tt hep-th/9305016}}].

\bibitem{Torii:2005xu}
T.~Torii and H.~Maeda, {\it {Spacetime structure of static solutions in
  Gauss-Bonnet gravity: Neutral case}},  {\em Phys.Rev.} {\bf D71} (2005)
  124002, [\href{http://xxx.lanl.gov/abs/hep-th/0504127}{{\tt
  hep-th/0504127}}].

\bibitem{Zwiebach:1985uq}
B.~Zwiebach, {\it {Curvature Squared Terms and String Theories}},  {\em
  Phys.Lett.} {\bf B156} (1985) 315.

\bibitem{Leith:2007bu}
B.~M. Leith and I.~P. Neupane, {\it {Gauss-Bonnet cosmologies: Crossing the
  phantom divide and the transition from matter dominance to dark energy}},
  {\em JCAP} {\bf 0705} (2007) 019,
  [\href{http://xxx.lanl.gov/abs/hep-th/0702002}{{\tt hep-th/0702002}}].

\bibitem{Nojiri:2005jg}
S.~Nojiri and S.~D. Odintsov, {\it {Modified Gauss-Bonnet theory as
  gravitational alternative for dark energy}},  {\em Phys.Lett.} {\bf B631}
  (2005) 1--6, [\href{http://xxx.lanl.gov/abs/hep-th/0508049}{{\tt
  hep-th/0508049}}].

\bibitem{Charmousis:2002rc}
C.~Charmousis and J.-F. Dufaux, {\it {General Gauss-Bonnet brane cosmology}},
  {\em Class.Quant.Grav.} {\bf 19} (2002) 4671--4682,
  [\href{http://xxx.lanl.gov/abs/hep-th/0202107}{{\tt hep-th/0202107}}].

\bibitem{Davis:2002gn}
S.~C. Davis, {\it {Generalized Israel junction conditions for a Gauss-Bonnet
  brane world}},  {\em Phys.Rev.} {\bf D67} (2003) 024030,
  [\href{http://xxx.lanl.gov/abs/hep-th/0208205}{{\tt hep-th/0208205}}].

\bibitem{Gregory:2009fj}
R.~Gregory, S.~Kanno, and J.~Soda, {\it {Holographic Superconductors with
  Higher Curvature Corrections}},  {\em JHEP} {\bf 0910} (2009) 010,
  [\href{http://xxx.lanl.gov/abs/0907.3203}{{\tt arXiv:0907.3203}}].

\bibitem{Kanno:2011cs}
S.~Kanno, {\it {A Note on Gauss-Bonnet Holographic Superconductors}},  {\em
  Class.Quant.Grav.} {\bf 28} (2011) 127001,
  [\href{http://xxx.lanl.gov/abs/1103.5022}{{\tt arXiv:1103.5022}}].

\bibitem{Park:2012bv}
M.~Park and R.~B. Mann, {\it {Holographic Renormalization of Asymptotically
  Flat Gravity}},  {\em JHEP} {\bf 1212} (2012) 098,
  [\href{http://xxx.lanl.gov/abs/1210.3843}{{\tt arXiv:1210.3843}}].

\bibitem{Taylor:2008tg}
M.~Taylor, {\it {Non-relativistic holography}},
  \href{http://xxx.lanl.gov/abs/0812.0530}{{\tt arXiv:0812.0530}}.

\end{thebibliography}\endgroup

\end{document}